\documentclass[preprint,11pt]{JHEP3}

\JHEPspecialurl{http://jhep.sissa.it/JOURNAL/JHEP3.tar.gz}

\usepackage{epsfig,multicol,amsmath}

\usepackage{epstopdf}
\usepackage{bm}  
\usepackage{amsmath}     
\usepackage{epsfig,epsf}
\usepackage{showkeys}
\usepackage{graphicx}
\usepackage{rotating}
\usepackage{cite}

\def\beq{\begin{equation}}
\def\eeq{\end{equation}}
\def\bea{\begin{eqnarray}}
\def\eea{\end{eqnarray}}

\def\eq#1{{Eq.~(\ref{#1})}}
\def\fig#1{{Fig.~\ref{#1}}}

\newcommand{\as}{\alpha_S}

\newcommand{\Lb}{\left(}
\newcommand{\Rb}{\right)}
\parskip=\itemsep               

\newcommand{\nn}{\nonumber}

\newcommand{\h}{\frac{1}{2}}

\newcommand{\y}{{\cal Y}}

\newcommand{\A}{{\cal A}}

\newcommand{\p}{I\!\!P}

\def\pom{{I\!\!P}}
\def\reg{{I\!\!R}}
\title{ N=4 SYM  and QCD motivated approach to soft interactions at high energies}
\author{\Large  E. Gotsman$^{a}$\thanks{Email:
gotsman@post.tau.ac.il.}\,, E. Levin$^{a,b}$\thanks{Email:
leving@post.tau.ac.il}\,\,and\,\,U. Maor$^{a}$\thanks{Email: maor@post.tau.ac.il.}\, 
\\
a)\,  \,Department of Particle Physics, School of Physics and Astronomy,
Raymond and Beverly Sackler
 Faculty
of Exact Science,  Tel Aviv University, Tel Aviv, 69978, Israel\\
b)\,\,Departamento de F\'\i sica, Universidad T\'ecnica
Federico Santa Mar\'\i a, Avda. Espa\~na 1680,
Casilla 110-V,  Valparaiso, Chile 
\\}

\abstract{ In this paper we  construct a model that satisfies 
 the theoretical requisites of  high
 energy soft interactions, based on two ingredients:(i) the results of N=4 
SYM, which at present  is a unique theory  that allows one
 to deal with a large coupling constant;
 and (ii) the requirement of matching with high energy QCD.
 In accordance with
 these ideas, we assume that the soft Pomeron intercept
 is  rather large, and  the slope of the
 Pomeron trajectory is equal to zero. We derive  analytical formulae 
that sum both enhanced and semi-enhanced diagrams
 for elastic and diffractive amplitudes.
  We  fit  the available experimental data, and predict the value 
for cross sections at the energies  accessible at the LHC. 
 The main corrections to
 the model are studied and evaluated.}

\dedicated{PACS: 13.85.-t, 13.85.Hd, 11.55.-m, 11.55.Bq}
\preprint{TAUP -\\
{\tt [hep-ph]}\\
\today}
\begin{document}
\section{Introduction}

 Thanks to the hard work of experimentalists and theoreticians over the 
past four decades, we know
 that the most economical and reliable method for describing 
soft interactions at high energy, is  the phenomenology based on the
soft Pomeron and secondary Reggeons (see Refs.\cite{COL,SOFT,LEREG} for details
).
Consequently, we believe that the future theory should be a theory of 
Pomeron and 
Reggeons and their interactions. However,  numerous 
attempts\cite{SOFT,ATMP} to build such a theory have failed, 
as one was not able to specify the Pomeron interaction, as well as 
interaction of Pomerons with the target. We are doomed to make  {\it ad hoc}
 assumptions about the vertices of multi-Pomeron interactions
(see Ref. \cite{ATMP,2CH,GLMLAST,KMRS,OS}) which specify the approach, but 
do not make it more theoretically  reliable.
  Such assumptions seem unavoidable, as
 there is no theoretical approach to  non-perturbative QCD and, on the 
other hand,   soft high energy processes appear to be 
 typical examples of non-perturbative physics at long distances.
 These processes are even more difficult
 to calculate, as approximate 
methods such as QCD sum rules and/or effective theories, as well as 
the lattice QCD approach, cannot be employed to determine the high
energy amplitudes.
   The success of any phenomenology does not establish a theory, however
 over the past two years a new approach has been developed (N=4 
SYM),
 which allows one to  study theoretically the regime of the strong
 coupling constant\cite{AdS-CFT}.
\subsection{N=4 SYM}

\FIGURE[ht]{\begin{minipage}{90mm}
{\centerline{\epsfig{file=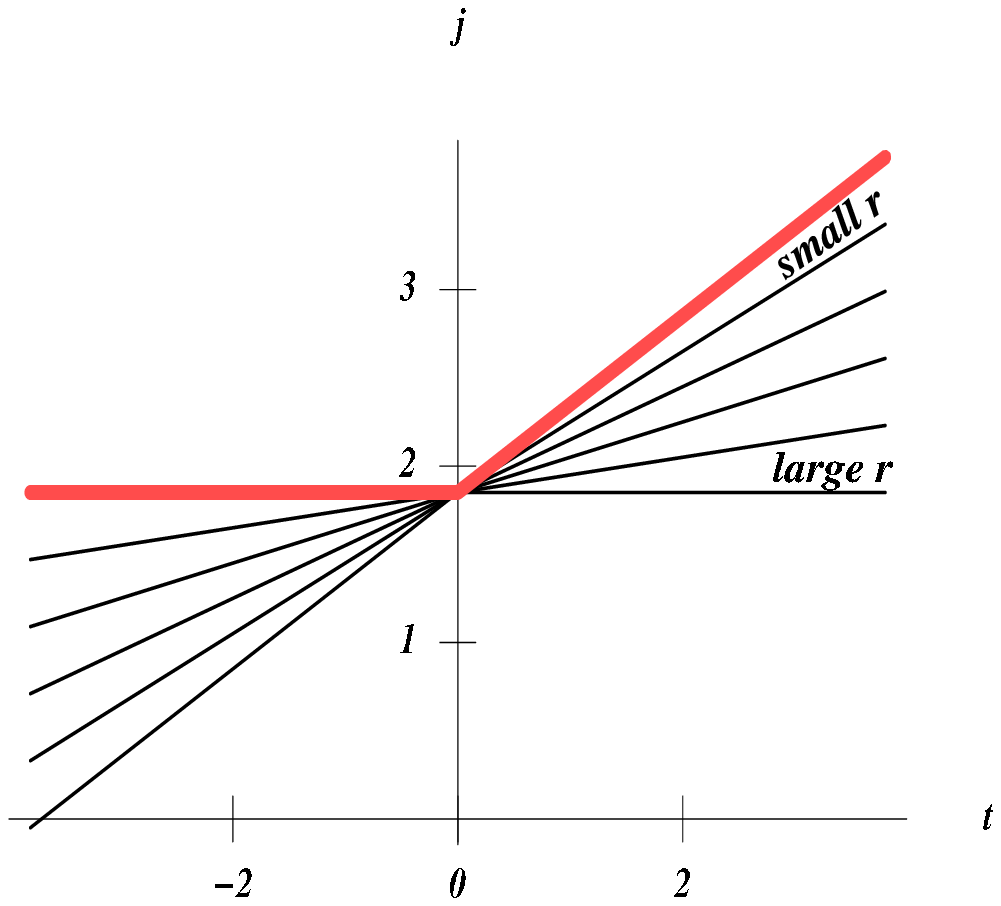,width=85mm}}
\caption{The behaviour of the Pomeron trajectory in N=4 SYM according 
to Ref. \cite{BST}. The figure is taken from Ref.\cite{BST}}
\label{pom4}
}
\end{minipage}
}
 We  use this theory as a guide for the physics phenomena 
 occurring in this regime. The attractive feature of this theory is  that 
N=4 SYM with
 small coupling, leads to normal QCD like physics (see Refs.
\cite{POST,BFKL4}),  with OPE  and linear equations for DIS, as well as 
the BFKL
 equation for the high energy amplitude.
The high energy amplitude reaches the unitarity limit:
 black disc regime, in which
 half of the cross section stems from  elastic scattering, and half 
relates
 to  processes of  multiparticle production.

On the other hand,
making use of the fact that  with the aid of the AdS/CFT
correspondence\cite{AdS-CFT} 
this theory can be solved analytically, and can
be reduced to  weak gravity in $AdS_5$ space.

In the strong coupling limit  the following main features of this theory
cite{BST,HIM,COCO,BEPI,LMKS} are manifest:
 (i) it  has a soft Pomeron, which in this case is 
the reggeized graviton with the  large intercept
 $\alpha(0)_\pom = 2 - 2/\sqrt{\lambda}$ where $\lambda =
 4 \pi N_c \alpha^{YM}_S$ and $\alpha^{YM}_S $ is the
 QCD-like coupling;
 (ii) the main contribution  to the total cross section at high
 energy is due to  the processes of elastic  scattering
 and diffractive dissociation; 
 (iii) the leading Pomeron trajectory has a form shown in \fig{pom4},
 namely, a Regge pole with $ \alpha'_\pom = 0$ in the scattering region
 ($t \,<\, 0$) while $ \alpha'_\pom \,>\, 0$
in the resonance region of positive $t$;
 and (iv) the Pomerons (gravitons)
  interact by means of the triple Pomeron vertex which 
 is small (at least $\propto 2/\sqrt{\lambda}$) .
   The minute value of $\alpha'_\pom$ is not related to
 the small size of the partons in this theory.
   It is related to  small
 values of the fifth coordinate $r$ (see \fig{pom4}).
  The physical meaning of this coordinate is, related to the typical
 size of the colliding particles. 
 It should be
 stressed that all these features are an integral part of the theory and 
therefore,
 for the first time we have  theoretical justification
 for using Reggeon-type phenomenology in high energy scattering.

In this paper we present our approach based on two major assumptions:
 it reproduces the main features of N=4 SYM,
 and it provides a natural matching with the
 perturbative QCD approach.
 For the sake of completeness we discuss the perturbative approach below.

\subsection{Matching with perturbative QCD}

In perturbative QCD the high energy amplitude  has been calculated
 in the leading log approximation, in which
 $\as \ll 1 $ but $\as\, \log s \approx 1$,
 where $\sqrt{s} = W$ denotes the energy in the c.m. 
frame\cite{LONU,BFKL,LI}. 
This amplitude can be written as the exchange of the
 BFKL Pomeron\cite{BFKL},
which has the following form for the cross section for the  scattering of 
one colourless dipole of size $r$ on another dipole of size $R$
\beq \label{I2}
\sigma\Lb r,R;s\Rb\,\,=\,\,\int d^2 b d^2 \rho_1\,d^2 \rho_2 \int^{ +i \infty +
\epsilon}_{-i \infty +\epsilon}\,\,\frac{ d \nu}{2 \pi i}\,V\Lb r, \rho_1;
 \nu\Rb \,e^{ \omega \Lb \nu\Rb \,Y}\,V\Lb R, \rho_2; -\nu\Rb\,
\delta^{(2)}
\Lb \rho_1 - \rho_2 - b\Rb
\eeq
where

\beq \label{I3}
\omega(\nu)\,=\,\as\,\Lb
\,\,2\,\psi(1) \,\,-\,\,\psi(1/2 + i \nu)\,\,-\,\,\psi(1/2 -  i\nu)\Rb
\eeq

and 
 $\psi(z) = d \log \Gamma(z)$, and $\Gamma(z)$ is the Euler gamma 
function.

One can see that the BFKL Pomeron is not a pole in angular momentum,
 but it is a cut.  However, at fixed $\nu $ which is the conjugate variable
 to $ln\Lb r^2/R^2\Rb$, it is a pole.   The position of  
this pole does not depend on momentum transfer (or $b$) and,
 therefore, the corresponding $\alpha'$ is equal to zero.

It turns out that in the wide range of energy
 $1/\as^2\, \gg \, \ln s \,\gg\, 1$ the scattering amplitude
 can be expressed as the
 sum of  BFKL Pomeron exchanges and their interactions
 (see Refs. \cite{GLR,MUQI,MV,B,K,JIMWLK, BRN,BART}.
 Perturbative QCD specifies  the Pomeron vertices,
 and in the LO approach, the only vertex that contributes is
 the triple BFKL Pomeron vertex \cite{BART,BRN}. 

\subsection{Models of high energy soft interaction}

\FIGURE[h]{\begin{minipage}{100mm}
{\centerline{\epsfig{file=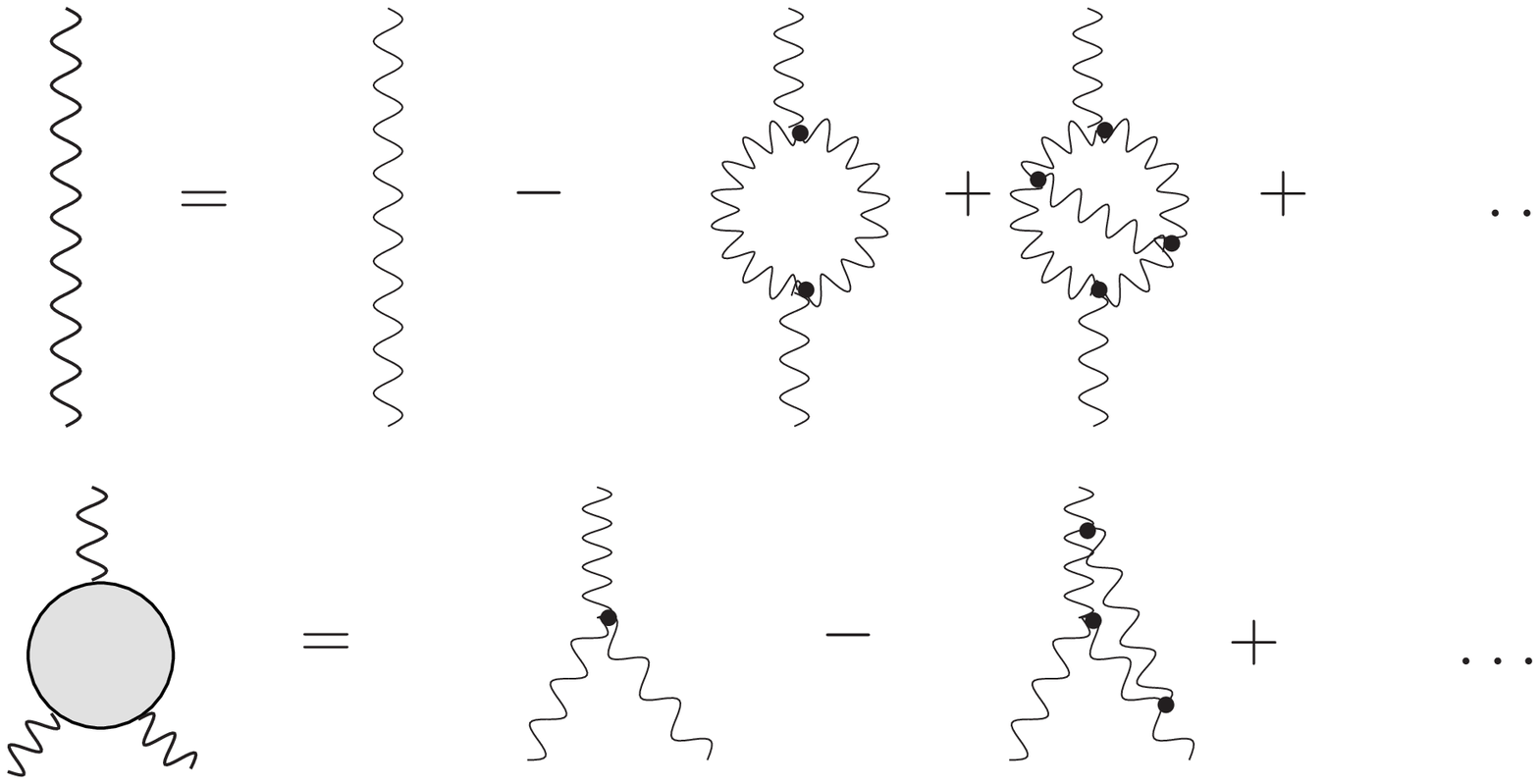,width=95mm}}
\caption{The exact Green's function of the Pomeron as a sum
 of enhanced diagrams and the exact triple Pomeron vertex.}
\label{enhst}}
\end{minipage}
}

We are incapable of  building a theory of  high energy scattering,
 as the problem of the  confinement of quarks and gluons in QCD
 has not been solved, and we do not have a theoretical tool for describing
 the  interactions of quarks and gluons at long distances.
 Therefore, we are doomed to build  models that
 take into account our  ideas of the behaviour
 of QCD at long distances.
 We believe that such models should  include everything
 that we know about QCD at long distances and,
 in particular, should absorb all that
 we  have learned about high energy scattering in N=4 SYM. 
  We can summarize this knowledge as a list of criteria that 
 a model should satisfy, namely,
\begin{enumerate}
\item \quad The model should be built using Pomerons
 and Reggeons as the main ingredients;
\item \quad The intercept of the Pomeron should be rather large.
 In N=4 SYM we 
expect $\Delta_{\p} = \alpha_{\p}(0) - 1 =
1 - 2/\sqrt{\lambda} \approx 0.3 \div 0.4$, since the estimate for 
$\lambda $  from the cross section for  multiparticle production
 as well as from DIS at HERA \cite{LEPO}  is $\lambda \,=\,5 \div 9$;

\item \quad $\alpha'_{\p}(0) \,=\,0$;

\item \quad  A large contribution should come from  processes
 of  diffraction dissociation, since in N=4 SYM at large $\lambda$
 only these processes 
 contribute to the scattering amplitude.
 In other words,  the model should
 include the Good-Walker mechanism \cite{GW} as the main source
 of the diffraction production;
\item \quad  The Pomeron self-interaction should be small
 (of the order of $2/\sqrt{\lambda}$ in N=4 SYM), and much smaller than
 the vertex of interaction of the Pomeron with a hadron, which is
 of the order of $\lambda$;

\item \quad The last requirement follows not from N=4 SYM, but from the 
natural 
matching with perturbative QCD: where the only  vertex that  
contributes is the triple Pomeron vertex.
\end{enumerate}

In this paper we continue to develop a model that satisfies
 all above criteria.
 In our previous paper  (see Ref \cite{GLMM})
we suggested a model that includes the large diffraction from Good-Walker 
mechanism and a Green's function of the Pomeron
 that sums all enhanced diagrams (see \fig{enhst}).
 However, we neglected the
 contribution of other Pomeron diagrams
 (see for example the so called fan diagrams in \fig{fanst}).
 The line of argument
 for justifying this was the following:
 The value of the triple Pomeron vertex turns
 out to be small, and the cross section for
 the diffraction production in the region of large mass 
makes up 10 - 20\% of total cross section of the diffractive production.
 Therefore, we decided in the first approximation to
 neglect the process of diffraction 
of large masses, and calculate them perturbatively.
 In this paper we sum  all 'fan' diagrams together with the enhanced ones.

\FIGURE[ht]{
\centerline{\epsfig{file=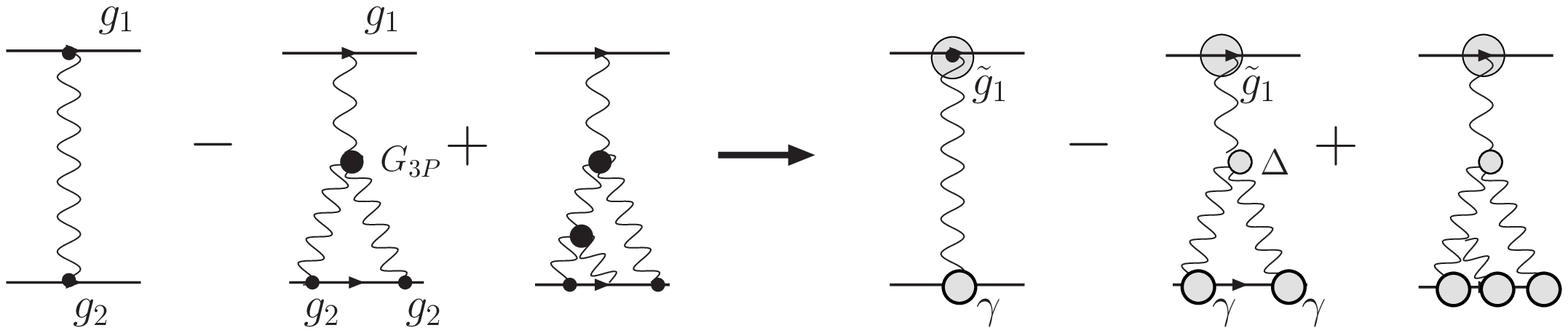,width=120mm}}
\caption{Examples of 'fan' Pomeron diagrams:
 in Reggeon Calculus (the first three diagrams)
 and in the generating function approach (the last three).}
\label{fanst}
}

\section{Main ideas and assumptions}
\subsection{Soft Pomeron}
As we have mentioned  we wish to suggest a procedure which  contains 
the main features of N=4 SYM, and  provides a 
transparent matching  with perturbative QCD,
 which plays the role of the  correspondence principle in our approach.

The good news is that the soft Pomeron is a natural
 ingredient in N=4 SYM. Actually, in N=4 SYM we have an infinite series
 of different Regge poles (see \fig{pom4}).
 Our simplification is to replace this set of poles
 by one pole: the Pomeron, shown in \fig{pom4} in red.
 We feel confident with the assumption regarding 
 the existence of the soft Pomeron,
 since it has been successfully utilized in  high energy phenomenology
for the last forty years,
 and it appears in QCD (see above).  

>From N=4 SYM we   expect that the value of
 $\Delta = \alpha_\pom(0) - 1 =  1 - 2/\sqrt{\lambda} $ could be large.  

\begin{boldmath}
\subsection{$\alpha'_{\pom}$= 0}
\end{boldmath}
Our second ingredient is that the slope of the Pomeron $\alpha'_{\pom}= 
0$.
 As we have discussed, $\alpha'_{\pom}= 0$ is in perfect agreement
 with  N=4 SYM predictions. It  also  agrees
 with the recent fit to  high energy data \cite{GLMM} . 
 This fact might appear strange, as for a long time the widely accepted
 value for  $\alpha'_{\pom}$
was $\alpha'_{\pom} = 0.25 \,GeV^{-2}$ \cite{DL}. 
  Comparing with the parametrization of Ref. \cite{DL}, 
two new ingredients have been introduced:
 a large value of $\Delta \approx 0.3$ instead of $\Delta \approx 0.08$
 in Ref. \cite{DL},
 and sufficiently large shadowing corrections, which were
 considered small in Ref.\cite{DL}.

We believe that a small value of $\alpha'_\pom $, is a signal
 that rather short distances
 contribute to the soft interaction at high energy,
 in agreement with all approaches to the Pomeron
 structure considered above.

It should be stressed that N=4 SYM gives a new possibility
 for  high energy
 assymptotic behaviour: a Regge cut (Pomeron) that
 does not move in the scattering region,
 while it has a slope $\alpha'_\pom$
 in the region of positive $t$.
 In Ref. \cite{GRIBMOV} such a possibility was missed. It is interesting to
 note that in N=4 SYM  this property of
 the Pomeron (graviton) contribution, stems from the
 integration over the fifth coordinate $z$, which
 has the meaning of  integration over the possible size of the hadrons.

\subsection{Good-Walker mechanism}

The third conclusion that we derive from N=4 SYM, is the large 
contribution of 
the diffractive dissociation processes. In our approach the
 diffraction dissociation is taken into account using
 the two channel model 
which we have developed in a number of papers (see Ref. \cite{GLMM} and 
references therein).
In this formalism, diffractively produced hadrons at a given vertex are
considered as a single hadronic state
described by the wave function $\Psi_D$, which is orthonormal
to the wave function $\Psi_h$ of the incoming hadron (proton in the case of
interest), $<\Psi_h|\Psi_D>=0 $.
We introduce two wave functions $\psi_1$ and $\psi_2$ which diagonalize 
the
2x2 interaction matrix ${\bf T}$
\beq \label{2CHM}
A_{i,k}=<\psi_i\,\psi_k|\mathbf{T}|\psi_{i'}\,\psi_{k'}>=
A_{i,k}\,\delta_{i,i'}\,\delta_{k,k'}.
\eeq
In this representation the observed states are written in the form
\beq \label{2CHM31}
\psi_h=\alpha\,\psi_1+\beta\,\psi_2\,,
\eeq
\beq \label{2CHM32}
\psi_D=-\beta\,\psi_1+\alpha \,\psi_2\,,
\eeq
where, $\alpha^2+\beta^2=1$.
Using \eq{2CHM}, we can rewrite the unitarity constraints in the form
\beq \label{UNIT}
Im\,A_{i,k}\left(s,b\right)=|A_{i,k}\left(s,b\right)|^2
+G^{in}_{i,k}(s,b),
\eeq
where $G^{in}_{i,k}$ is the contribution of all non diffractive inelastic
processes,
i.e. it is the summed probability for these final states to be
produced in the scattering of particle $i$ off particle $k$.

A simple solution of \eq{UNIT} has the same structure as in the single 
channel formalism, 
\beq \label{2CHM1}
A_{i,k}(s,b)=i \Lb 1 -\exp\Lb - \frac{\Omega_{i,k}(s,b)}{2}\Rb\Rb,
\eeq
\beq \label{2CHM2}
G^{in}_{i,k}(s,b)=1-\exp\Lb - \Omega_{i,k}(s,b)\Rb.
\eeq
>From \eq{2CHM2} we deduce, that the probability that the initial
projectiles
$(i,k)$ reach the final state interaction unchanged, regardless of the initial
state rescatterings, is
$P^S_{i,k}=\exp \Lb - \Omega_{i,k}(s,b) \Rb$.

The opacities $ \Omega_{i,k}(s,b)$ have to be determined and we will 
discuss this below.
 In general this two channel approach is a particular case
 of the  Good-Walker mechanism\cite{GW}  for  diffractive production,
  and  we can  account for  diffraction in the region of small
 mass using this approach. For the 
region of large values of produced  mass in diffractive dissociation,  
it is necessary 
  to develop a theoretical approach to the Pomeron-Pomeron interaction. 

\subsection{Pomeron -Pomeron vertices}
The general framework for accounting for
 Pomeron interactions was  developed in 70's (see Refs.\cite{GRIBRT,
SOFT} in the framework of Reggeon Calculus. However, being  purely 
phenomenological,  Reggeon Calculus is not
 able to fix the vertices of Pomeron interactions.
 The values of vertices, and their number,  have to be introduced
 in  Reggeon Calculus from the microscopic theory.

Our third ingredient is the assumption that only the  triple
 Pomeron vertex is essential. This assumption is in
 full agreement with N=4 SYM and perturbative QCD: in both
 these theoretical approaches only this vertex contributes.
 We consider this to be  essential, as this
 assumption provides a natural bridge to perturbative QCD.
 To illustrate this matching we consider the sum of the
 `fan' Pomeron diagrams (see \fig{fanst}).

Instead of the generating functional for high density
 QCD (see Ref.\cite{LELU}  ), we can introduce the generating function:
\beq \label{ZF}
Z\Lb Y - Y_0; u \Rb\,\,=\,\,\sum^{\infty}_{n =1}\,P_n\Lb Y - Y_0\Rb\,u^n
\eeq
where $P_n\Lb Y - Y_0 \Rb$ is the probability
 to find $n$-Pomerons at rapidity $Y - Y_0$.
 By fixing the size of the interacting dipoles,
 one can see that \eq{ZF} differs  from the generating
 functional in the dipole approach of high density
 QCD. 

The full set of the `fan' diagrams can be summed using
 the following equation for the generating function
\beq \label{ZFEQ}
\frac{\partial \,Z\,\Lb Y-Y_0; u\Rb}{
\partial \,Y}\,\,= \,\,- \Delta \,u\,(1 - u)\,\,Z\,\Lb Y- Y_0; u \Rb
\eeq
with the amplitude that is given by the following equation 
\beq \label{NF}
 N\left(Y;\{\gamma_i \}\right)\,=  \,- \sum^{\infty}_{n =1} \int
\gamma_n(Y_0) \prod^n_{i=1}\frac{\partial^i}{\partial
u^i }\,\,Z\left(Y, u \right)|_{u=1}\,
 =\,-\,\sum^{\infty}_{n =1} (-1)^n 
\gamma_n(Y_0)\,\,\rho(Y -Y_0) 
\eeq
 If we neglect  dependence on the size of the dipoles, then
both equations coincide with the equation in the dipole approach to
 QCD. 
  These equations sum the set of `fan' diagrams,
 if we replace (see \fig{fanst}
\beq \label{REP}
g_1 \,\,\to\,\,\tilde{g}_1\,=\,\,\frac{g_1\,\Delta}{G_{3\pom}}\,;\,\,\,\,\,\,\,
\,\,\,G_{3\pom}\,\,\to\,\,\Delta
\,;\,\,\,\,\,\,\,\,\,\,g_2\,\,\to\,\,\gamma\,=\,\frac{G_{3\pom}\,g_2}{\Delta}
\eeq
where $\Delta $ denotes the intercept of the Pomeron, and $\gamma$
  the amplitude of the interaction of the `wee' parton
 (colorless dipole) with the target.
 The scattering amplitude is equal to $\tilde{g}_1\,N\Lb \eq{NF}\Rb$.

 Assuming that only the triple Pomeron vertex is essential,
 we achieve the matching between perturbative QCD and our approach.
 We will show below that this matching can be  demonstrated
 for  sets of Pomeron diagrams, more complicated  than the `fan' diagrams.

The theory that includes all the  ingredients that have been
 discussed above, can
be formulated in a functional integral form\cite{BRN},
\begin{eqnarray} \label{FI}
&&Z[\Phi, \Phi^+]\,\,=\,\,\int \,\,D \Phi\,D\Phi^+\,e^S \,\,\,\,\,\,\,\,\mbox{with}\,\,\,\,S \,=\,S_0
\,+\,S_I\,+\,S_E\,,
\end{eqnarray}
where $S_0$ describes the free Pomerons,
 $S_I$ corresponds to their mutual interaction,
and $S_E$ relates to the interaction with the external sources (target and
projectile). Since $\alpha^{\prime}_{\pom}\,=\,0$, $S_0$ has the form
\begin{equation} \label{S0}
S_0\,=\,\int d Y \Phi^+(Y)\,\left\{ -\,\frac{d }{d Y} \,\,
+\,\,\Delta\,\right\} \Phi(Y).
\end{equation}
$S_I$ includes only triple Pomeron interactions and is of the form
\beq \label{SI} 
S_I\,=\,g_{3\pom} \int d Y\,\left\{\Phi(Y)\,
\Phi^+(Y)\,\Phi^+(Y)\,\,+\,\,h.c. \right\}
 \eeq 
 $S_E$ depends on our model for the  interaction of the Pomeron with the 
scattering particles, which  we will specify  later.

This theory, as any theory of Pomeron interactions, is written
in such a way that  the high energy amplitudes satisfy $t$-channel
unitarity. However, $s$-channel unitarity remains a problem.
In Refs.\cite{KLP} it was shown that to  satisfy   $s$-channel unitarity  
we need to add to
 the interaction term ($S_I$), the four Pomeron vertex 
$-\Gamma(2\rightarrow 1) (\Phi^+)^2\,(\Phi)^2$. 

For a better understanding of $s$-channel unitarity
 we reformulate the theory, given by the functional integral of \eq{FI},
 in terms of
the evolution equations for the system of partons.
 As we have mentioned, for perturbative QCD these partons are 
 colourless dipoles, as was shown in Ref.\cite{MUCD}, 
that can decay and merge: one parton to two partons and two partons
into one parton, with probabilities $\Gamma(1 \to 2)$ and $ \Gamma(2 \to 1)$ 
respectively. 
 For such a system of partons, we can write a simple
 evolution equation (Fokker-Planck equation). 
Indeed, let $P_n(y)$ be the probability to find
$n$-parton (dipoles) with rapidity $y$ in the wave function
 of the fastest (parent)
parton (dipole), moving with rapidity $Y\,>\,y$.
For $P_n(y)$, we  write down a recurrence equation 
(see Refs.\cite{GRPO,LELU})
\beq \label{PNEQ}
-\,\frac{\partial\,P_n(y)}{\partial \,y}\,\,=\,\,\Gamma(1 \to 2)\,
\left\{ -\,n\,P_n\,+\, \,(n-1)\,P_{n-1}  \right\}\,\,\,+\,\,\,\Gamma(2 \to 1)
\, \left\{ -\,n\,(n - 1)\,P_n\,+\, \,(n+1)\,n\,P_{n+1}  \right\}.
\eeq
In each bracket the first term on the r.h.s.,
can be viewed as a probability of a dipole annihilation
in the  rapidity range $( y $ to $ y - dy )$ (death term). The second is a
probability to  create one extra dipole (birth term). 
Note the negative sign
in front of $\partial P_n(y)/\partial y$. It appears due to our choice of the
rapidity evolution, which starts at the largest rapidity $y=Y$,
of the fastest dipole and then decreases.
The first two terms are responsible for the process of parton decay,
while the last two terms describe the
contribution of partons merging.

Using \eq{ZF}  we can re-write \eq{PNEQ} in the form

\beq \label{GFEQ}
\,\,-\frac{\partial\,Z(y,\,u)}{\partial\, \y}\,\,
=\,\,-\,\kappa\,u\,(1\,-\,u) 
\,\,\frac{\partial\,Z(y,\,u)}{\partial\, u}\,\,\,+\,\,\,
\,u\,(1\,-\,u) \,\,\frac{\partial^2\,Z(y,\,u)}{\partial^2\, u}.
\eeq
where ${\cal Y}\,\equiv\,\Gamma(2 \to 1)\,(Y - y)\,=\,\,\Delta\,\gamma\,(Y - y)
$ and $\kappa \,\equiv\,1/\gamma$.

\eq{GFEQ} should be added with the initial condition at $Y=Y_0$, as well 
as the 
boundary condition
\beq \label{INC2}
Z(y,\,u\,=\,1)\,\,=\,\,1,
\eeq
which follows from the physical meaning of $P_n$ as a  probability. 
It turns out that \eq{GFEQ} can be solved analytically  with an arbitrary
 initial condition\cite{AMCP, KOLE}. 
 Therefore, formally speaking we could derive the initial
 condition for different processes and obtain the solution using
 the approach of Ref.\cite{KOLE}. However, we chose a different
 and more traditional way to tackle the problem.
 As  was suggested in Ref. \cite{GRIBRT} the solution
 of the Pomeron interaction problem should have two steps.
 Step one,  is to find the correct Green's function of
 the Pomeron, and the exact vertex of the triple Pomeron interaction, 
 by summing all enhanced diagrams (see \fig{enhst}). 
Step  two: to solve the problem of the interaction of new (exact)
 Pomerons. One expects  that the interaction in terms of the
 exact Green's function and vertices, will be much simpler
 at high energy, as the main effect of the interaction
 has been taken into account in  step one.

It is necessary to require that 
there be only one fastest parton (dipole), which is $P_1(y\,=\,Y)\,=\,1$,
while $P_{n>1}(y\,=\,Y)\,=\,0$.  In this case we have the following
 initial condition for the generating function
\beq \label{INC1}
Z(y\,=\,Y)\,=\,u\,.
\eeq
and the solution to \eq{GFEQ} will give the exact Green's
 function of the Pomeron.
However, we prefer to develop a different technique
 for finding the Pomeron Green's function, which makes the calculation 
more  transparent, and leads to  explicit analytical
 formulas for physical observables (see section 3.1).

\subsection{The phenomenological parameters and their typical values}

Unfortunately, even with all assumptions that we have made,
 our approach is still phenomenological, since we have to determine   the 
parameters
 of our interaction  from a fit to the experimental data.
 However, using our main idea that the soft interaction
 stems from rather short distances, we are able to give
 some estimates for these parameters.

First,  we  list all of these parameters:
\begin{enumerate}
\item For description of Good-Walker mechanism of diffraction production
 in the two channel model, we need two phenomenological functions
 $g_1(b)$ and $g_2(b)$, which describe the vertices of interaction
 of the Pomeron with state $1$ and $2$ (see \eq{2CHM31} and \eq{2CHM31}),
 and one number $\beta$ see \eq{2CHM31} and \eq{2CHM31});
\item The Pomeron intercept $\Delta = \alpha(0)\, - \,1$ ;
\item The low energy amplitude of dipole -target interaction $\gamma$;
\end{enumerate}

Since we believe that the short distances contribute
 to the Pomeron structure we expect that
\beq \label{EST}
\Delta\,\,\,\propto\,\,\,\as\,;\,\,\,\,\,\,\,\,\,\,\gamma\,\,\,\propto\,
\,\,\as^2\,;\,\,\,\,\,\,\,\,\,\,
\tilde{g}_1 \,\,\approx\,\,\tilde{g}_2\,\,\propto\,1
\eeq
For     $ \tilde{g}_i(b)$ we use the entire phenomenological assumption 
\beq \label{S}
\tilde{g}_i(b)\,\,\,=\,\,\tilde{g}_i\,S(b)\,\,=\,
\,\frac{\tilde{g}_i}{4 \pi}\,m^3_i\,b\,K_1\Lb m_i\,b\Rb
\eeq 
where $S(b)$ is the Fourier transform of
 the dipole formula for the form factor $1/(1 + q^2/m^2_i)^2$.

\section{Summing the enhanced diagram}

In this section we  sum the enhanced diagrams, and obtain the exact
 Green's function of the Pmeron, as well as the  exact vertex for
 the  triple Pomeron interaction.
  To achieve this we  employ the approximation that has
 been developed  in Ref.\cite{MPSI} (MPSI approximation),
 and the  improved  version as in Refs.\cite{LEPR,LMP}, so as to adjust
 this method for the  summation of the Pomeron loop diagrams,
 for Pomerons with the intercept $\Delta \,>\,0$.
\subsection{Improved MPSI approximation}
To illustrate the method, we calculate the first enhanced diagram of 
\fig{enh1}
\bea
\label{MPSI1}
A\Lb \fig{enh1}\Rb\,
    \,\,&=&  \,\,-\,g_1\,g_2 G^2_{3\pom}\,\,\int^Y_0\,d\,y_1\,\int^{y_1}_0\,d\,y_2\,G(Y - y_1)\,G^2(y_1 - y_2)\,G(y_2 - 0) \,\,\,\nonumber \\
&=&\,\,\,-\,\,g_1\,g_2 G^2_{3\pom}\,\int^Y_0\,d\,y_1\,\int^{y_1}_0\,d\,y_2\,\,e^{\Delta\,(Y
+ y_1 - y_2)}\,\,=\,\,-\,\,\,\frac{g_1\,g_2 G^2_{3\pom}}{\Delta^2}\,\left\{ e^{2\,\Delta\,Y}\,\,+\,\,\,e^{\,\Delta\,Y}\,\,+\,\,\Delta\,Y\,
      e^{\,\Delta\,Y} \right\}\nonumber\\
&=&\,\,\tilde{g}_1\,\tilde{g}_2\,\left\{ \gamma^2\,e^{2\,\Delta\,Y}\,\,+\,\,\gamma^2\,e^{\,\Delta\,Y}\,\,+\,\,\Delta\,\gamma^2\,Y\,
      e^{\,\Delta\,Y} \right\} 
      \eea
      where $G_{3\pom} \,\,=\,\,\Delta\,\gamma$  and $\tilde{g}_i \,\,=\,\,g_i/\sqrt{\gamma}$
(see \fig{enh1gf} for the notation).

\DOUBLEFIGURE[ht]{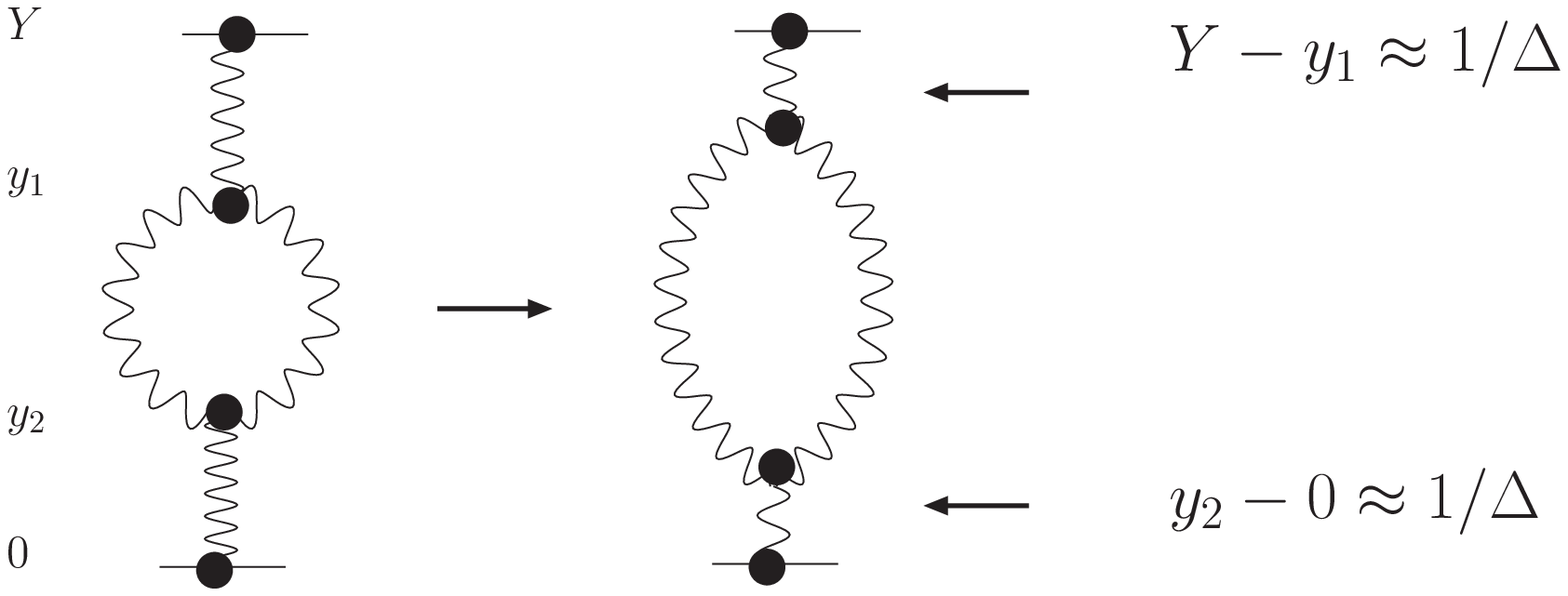,width=70mm,height=30mm}{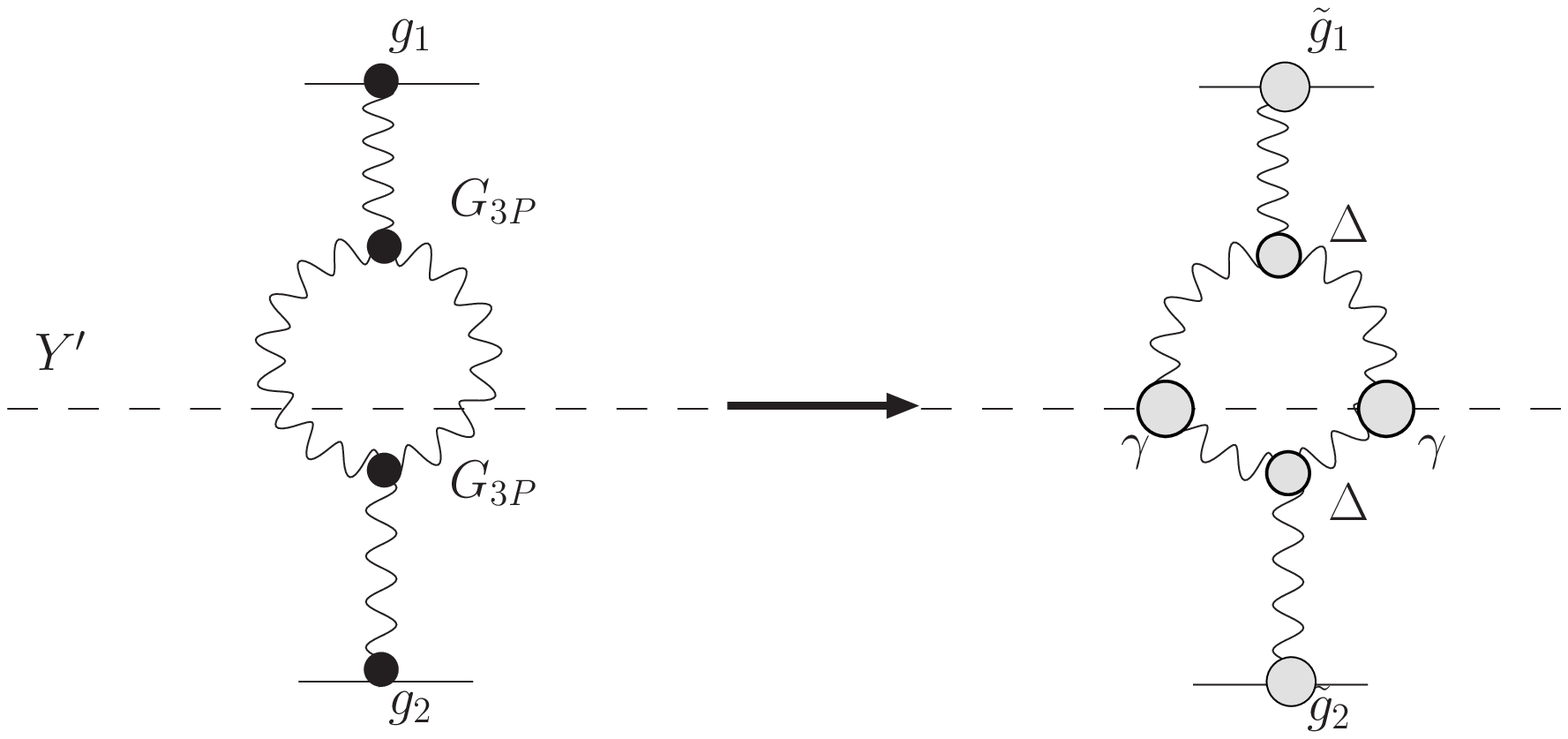,width=70mm,
height=30mm}
{The first enhanced diagram for the Pomeron with  intercept $\Delta \,>\,0$.
\label{enh1}}{The first enhanced diagram in the form suited for the
  MPSI approximation.
\label{enh1gf} }

The main idea of the MPSI approximation is to take into account
 only the first term in \eq{MPSI1}, neglecting other terms, since
 they are suppressed as $\exp[- \Delta Y]$.
 This term is the result of integration for $Y-
 y_1 \approx 1/\Delta$ and $y_2 - 0 \approx 1/\Delta$ (see \fig{enh1}).
 The general expression for the sum of the enhanced diagrams
 in MPSI approximation  is shown in \fig{genmpsi}.
\FIGURE[ht]{\begin{minipage}{80mm}
{\centerline{\epsfig{file=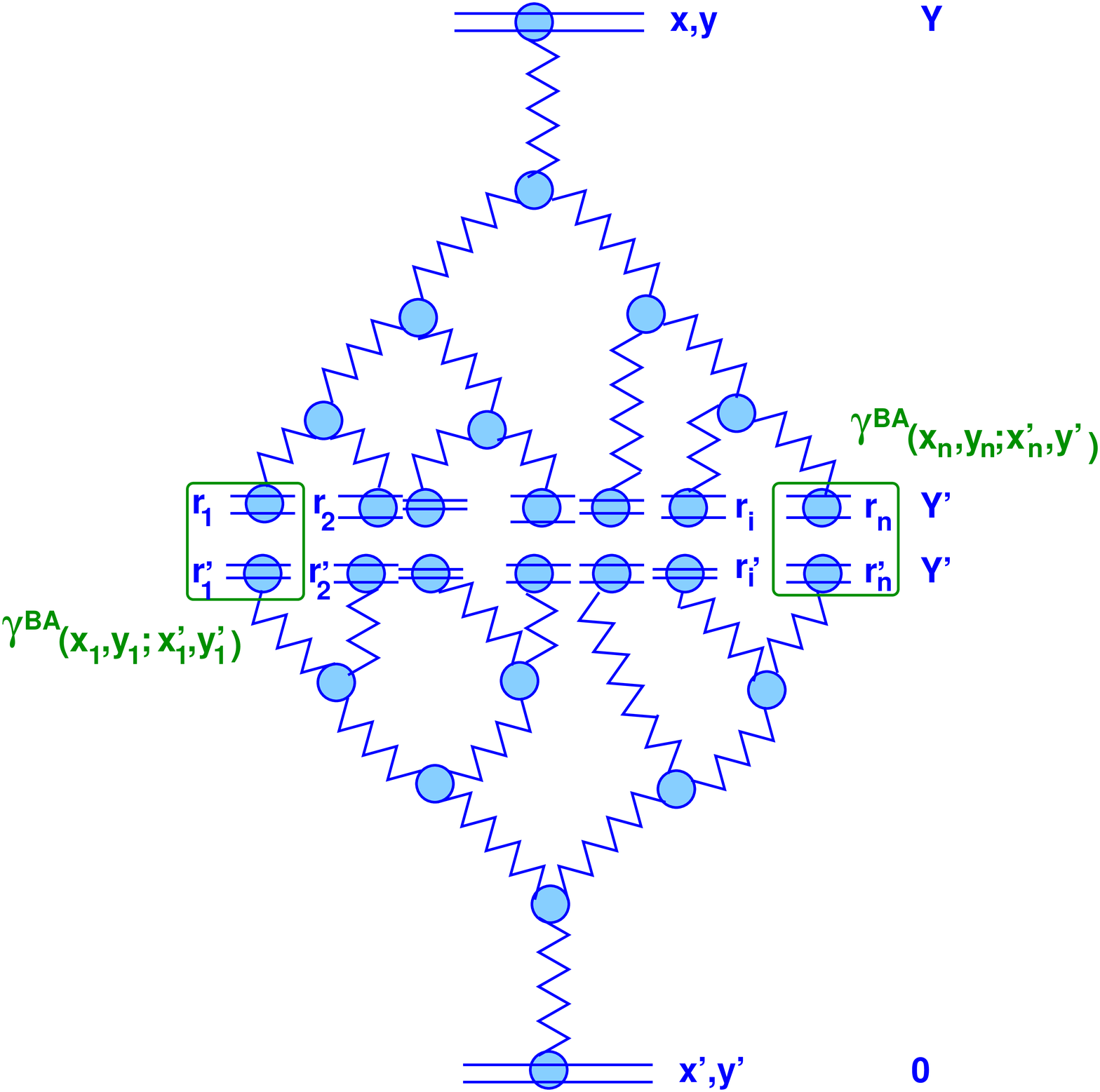,width=75mm}}
\caption{The exact Green's function of the Pomeron as a sum of
 enhanced diagrams in perturbative QCD in the MPSI approximation.}
\label{genmpsi}
}
\end{minipage}
}
One can see that  the MPSI approximation is the
 $t$-channal unitarity constraint adjusted to Reggeon Calculus,
 in the form of the generating functional ( generating function
 in our case for summing Pomeron interactions).
  From this picture, and from our
 knowledge of the cascade described by the `fan' diagrams,  we can 
find
 the answer for the sum of enhanced diagrams only. 
 The physical meaning of the introduced parameters is clear:
 $\gamma$ is the low energy amplitude for two partons (dipoles)
 scattering at  an arbitrary rapidity $Y'$, and $\Delta$ is the
 value of the vertex for the decay of one parton (dipole) to
 two parton (dipoles).
 It should be stressed that the answer does not depend on the
 value of $Y'$, but it should be chosen somewhere in the
 central region for the scattering.

 To find the generating function that describes the sum
 of the `fan' diagrams we need to solve \eq{ZFEQ}.
 This is not a difficult task, and for illustrative purposes we will do 
it in a different way.
 First we introduce a new generating function
\beq \label{MPSI2}
N\Lb Y - Y',\gamma\Rb \,\,\equiv\,\,1\,\,\, -\,\,Z\Lb Y - Y', u =1 -\gamma\Rb
\eeq
>From \eq{NF} one can associate  $N$ with the scattering amplitude,
 when the  variable $\gamma$ is equal to the low energy parton amplitude. 
 If we introduce a new variable
\beq \label{GR}
\gamma_R\,\,\,=\,\,\,\frac{\gamma}{ 1 - \gamma}
\eeq
\eq{ZFEQ} reduces to the form
\beq \label{MPSI3}
  \frac{\partial
\,N\Lb Y- Y'; \gamma_R \Rb}{\partial\,(Y - Y')}\,\,\,\,=\,\,\Delta\,\gamma_R\,\frac{\partial \,N\Lb Y - Y';
\gamma_R\Rb}{\partial\,\gamma_R}
 \eeq
\eq{MPSI3} is the equation for the system of non interacting Pomerons,
 and the general solution has the following form
\beq \label{MPSI4}
 N\Lb Y - Y';
\gamma_R\Rb\,\,\,=\,\,\,\,\sum_{n=1}^{\infty}\,\,\,(-1)^n\,\,C_n\,\,\gamma^n_R\,\,G^n(Y
- 0) 
\eeq

 where the coefficients $C_n$ can be found from the
initial conditions, namely, from the expression for the low energy
amplitude. 
For summing the enhance diagrams  the initial condition 

\beq \label{MPSIIC}
 N\Lb Y - Y'=0 \gamma_R\Rb\,\,=\,\,\gamma\,\,=\,\,\gamma_R/(1 + \gamma_R)
\eeq
  generates $C_n\,\,=\,\,1$ and the solution is
\beq \label{MPSI5} 
N\Lb Y - Y';
\gamma_R\Rb\,\,\,=\,\,\,\frac{\gamma_R\,e^{\Delta\,(Y - Y')}}{1\,\,\,\,+\,\,\,\gamma_R\,e^{\Delta\,(Y - Y')}}
\eeq 

The initial condition of \eq{MPSIIC} has  very simple physics
behind it, and has been discussed in Ref.\cite{LEPR}. 

The main idea of the improved MPSI approximation, is to
 replace $\gamma_R$ in the generating function $
N\Lb
\gamma_R|Y\Rb$ by the low energy amplitude for the dipole-dipole 
interaction, 
(see \fig{genmpsi}).  
It is easy to see that the amplitude in the MPSI approximation has the
 form (in this equation we denote 
$N\Lb
\gamma_R|Y\Rb$ by $N^{MFA}\Lb
\gamma_R|Y\Rb$ where MFA stands for mean field approximation)
\bea \label{MPSI6}
&&N^{MPSI}\Lb Y\Rb\,\,=\\
&&=\,\,\sum^\infty_{n=1}\,\frac{(-1)^n}{n!}\,\Lb\frac{\partial}{\partial \gamma^{(1)}_R}\Rb^n\, N^{MFA}\Lb Y-Y'; \gamma^{(1)}_R\Rb|_{\gamma^{(1)}_R=0}\,\,\Lb\frac{\partial}{\partial\gamma^{(2)}_R}\Rb^n N^{MFA}\Lb Y' - 0; \gamma^{(2)}_R\Rb|_{\gamma^{(2)}_R=0}\,\,\gamma^n_0\,\,\nonumber\\
&&=\,\,\,1\,\,\,-\,\,\exp
\left\{\,-\,\gamma_0 \,\frac{\partial}{\partial
\gamma^{(1)}_{R}}\,\frac{\partial}{\partial
\gamma^{(2)}_R}\,\right\}\,N^{MFA}\Lb Y - Y'\gamma^{(1)}_R \Rb\,
N^{MFA}\Lb  Y'- 0;
\gamma^{(2)}_R\Rb|_{\gamma^{(1)}_R\,=\,\gamma^{(2)}_R \,=\,0}\nonumber
\eea
where $\gamma_0$ is the low energy amplitude for parton scattering
 (we will denote this amplitude as $\gamma$ and hope
 that it will not cause any inconvenience).
\subsection{The range of energy for which our approach is reliable}
 We return to the example of the first enhanced diagram
 (see \eq{MPSI1}). The most dangerous term is the last one.
 It has an extra $Y$ ,  which stems from the region of integration
 $y_1 - y_2 \approx\,1/\Delta$, and it is the first term of
 the renormalization of the  Pomeron intercept.
 One can see  (see Refs. \cite{LEPR,LMP} for details)
 that the renormalized intercept
 $\Delta_R = \Delta - \Delta \gamma^2$
\footnote{To understand, this it is sufficient
 to compare this term with the exchange of one Pomeron,
 which has the form $\tilde{g}_1\,\tilde{g}_2\,\gamma
 \exp\Lb- \Delta Y\Rb$ in our notation.}.
 This   cannot be calculated in the MPSI approximation.
 Therefore, the first estimate for the range of energy
 where we can trust the MPSI approximation, comes
 from the demand that the renormalization of the
 Pomeron intercept should be small. This leads to
\beq \label{ER1}
\Delta\,\gamma\,Y\,\ll\,1;\,\,\,\,\,\mbox{or}\,\,\,
\,\,Y\,\ll\,\frac{1}{\Delta\,\gamma}
\eeq
However,  a more restricted region  can be  obtained from \eq{GFEQ}.
 The term $-\gamma u^2\partial^2 Z/\partial u^2$ describes
 the four Pomeron interaction, and as we have discussed, should
 be considered as small in our approach.
  Solving this equation without this term,
 and taking it into account as a perturbation, it is easy to show that the
 contribution of the four Pomeron interaction turns out to be small for
\beq \label{ER2}
\gamma\,Y\,\ll\,1;\,\,\,\,\,\mbox{or}\,\,\,\,\,Y\,\ll\,\frac{1}{\gamma}
\eeq
The second restriction of our approach, comes from the fact
 that we consider $\alpha'_\pom = 0$.
 Inclusion of a small $\alpha'_\pom$ will destroy our
 approximation for energies which we can find from the condition
\beq \label{ER3}
\alpha'_\pom Y\,>\,1/m^2_i\,\,\,\, \mbox{or}\,\,\,\,\,Y\,>\,\frac{1}{\alpha'_\pom\,m^2_i}
\eeq
Thus, we can trust the MPSI approximation in the region

\beq \label{ER4}
Y \,\,\leq\,\,\mbox{min}\,\left\{ \frac{1}{\gamma}, \frac{1}{\alpha'_\pom\,m^2_i}\right\}
\eeq

\subsection{The Pomeron Green's function and the elastic scattering 
amplitude}
\DOUBLEFIGURE[t]{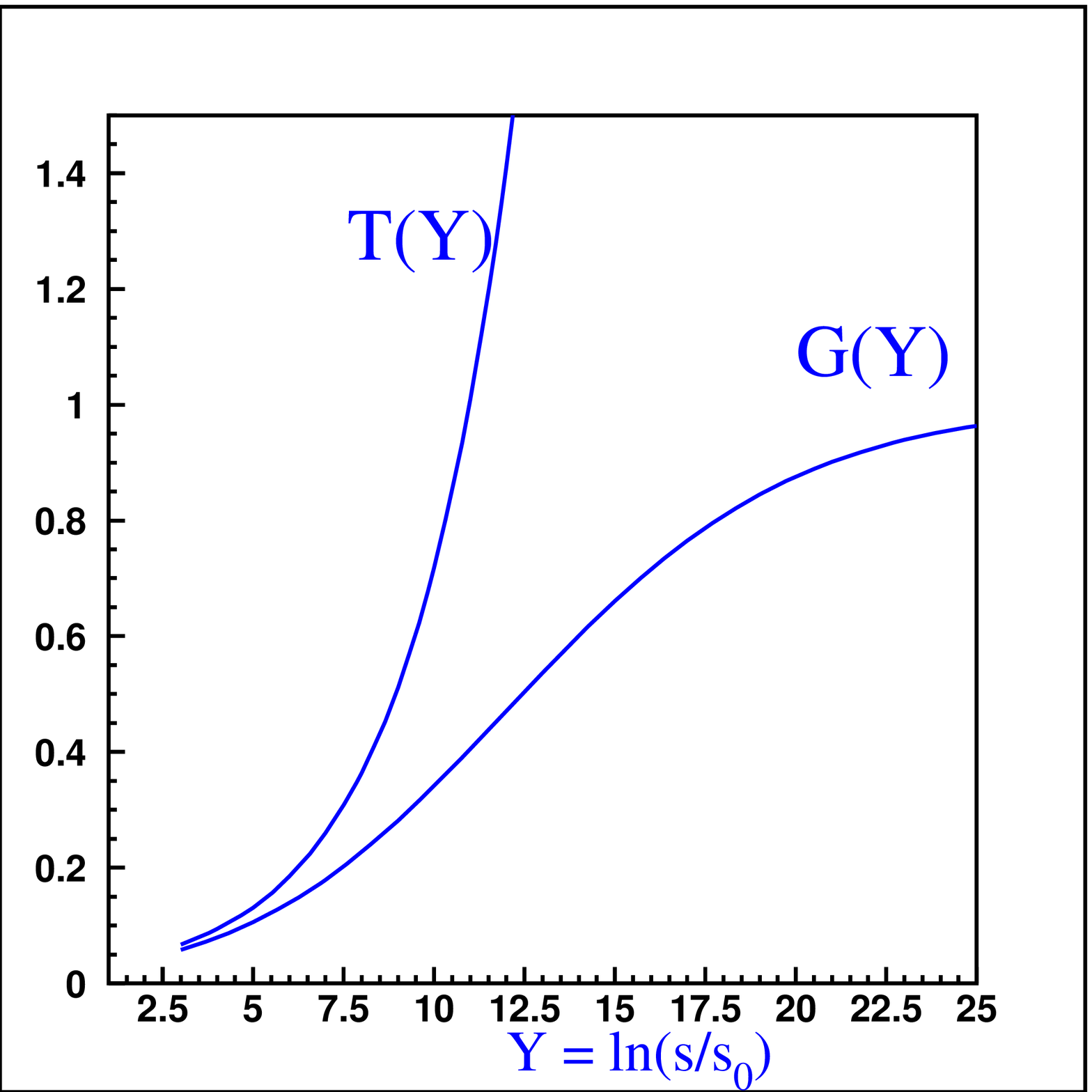,width=60mm,height=50mm}
{ 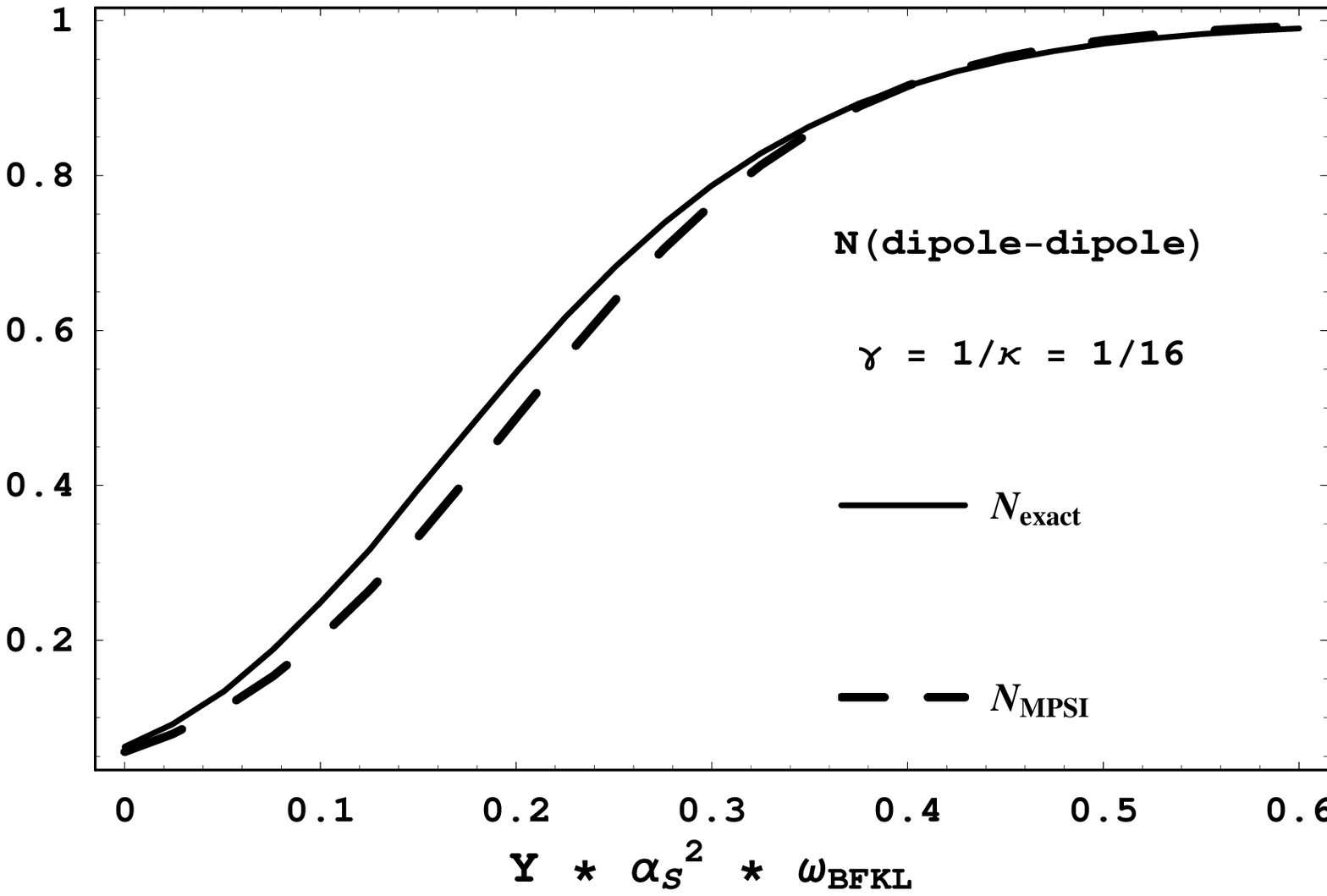,width=80mm,height=55mm}
{The exact Green's function of the Pomeron
 versus $Y = \ln(s/s_0)$ for $s_0 = 1\,GeV^2$, and
 $T\Lb Y\Rb$ for $\Delta = 0.339$ and $\gamma=0.0242$.
The values of parameters have been taken from our
 fit\protect\cite{GLMM}.\label{pomgrf}}
{The comparison of the exact solution of \protect \eq{GFEQ}
  for the Pomeron Green's function $G\Lb Y\Rb $ (see 
Ref.\protect\cite{KOLE,KOLE1}) with $G\Lb Y\Rb $ in
 the improved MPSI approximation (see \protect\eq{ES1}).
 The figure is taken from Ref. \protect\cite{KOLE1}.
\label{compar}}

Using \eq{MPSI6} we can obtain the Green's function of the
 Pomeron in a closed form, namely\cite{GLMM}
\beq \label{ES1}
G\Lb Y\Rb\,\,=\,\,1 \,-\,\\exp\Lb \frac{1}{T\Lb Y\Rb}\Rb\,\frac{1}{T\Lb Y\Rb}\,\Gamma\Lb 0,\frac{1}{T\Lb Y\Rb} \Rb
\eeq
with
\beq \label{ES11}
T\Lb Y \Rb\,\,\,=\,\,\gamma\,e^{\Delta Y}
\eeq
 and $\Gamma\Lb 0, 1/T\Rb$ is the incomplete gamma function
  (see formulae {\bf 8.35} in Ref.\cite{RY}).
Using this function we can write the opacities in the  two channel
 formalism (see \eq{2CHM1}) in the form
\beq \label{ES2}
\Omega_{i,k}\,\,=\,\,\tilde{g}_i(b)\,\tilde{g}_k(b)\,\,G\Lb Y\Rb
\eeq

It should be stressed that the Green's function
 of \eq{ES1} has been found for high energies,
 where the MPSI approach is correct.
 However, the theory given by the functional of \eq{FI} has
 an analytical solution (see Refs. \cite{KOLE,KOLE1}
) at arbitrary values of the energy.
 Comparison with the improved MPSI approximation
 (see \fig{compar}  shows that  this approximation
 describes the lowest energy within accuracy
 of 5 to 10 \%, for $\gamma =1/16$.
 In our fit \cite{GLMM} $\gamma = 0.0242$ which leads to even  better 
accuracy.
 Therefore, we can safely use the improved MPSI approximation 
 starting from $s= 400\,GeV^2$ which was used in our fit.

~

The elastic amplitude is equal
\beq \label{ES3}
a_{el}(b)\,=\,i \Lb \alpha^4 A_{1,1}\,+\,2 \alpha^2\,\beta^2\,A_{1,2}\,+\,\beta^4 A_{2,2}\Rb
\eeq 
where $A_{i,k}$ is given by \eq{2CHM1} with $\Omega_{i,k}$ from \eq{ES2}.

The Pomeron Green's function tends to unity at large values
 of the argument $T$.
 In \fig{pomgrf} we plot this function as well as 
$T\Lb Y\Rb$ for $\Delta = 0.339$ and $\gamma=0.0242$
 these values were found  in our fit (see Ref.\cite{GLMM} and below).
 One can see that $G(Y)$ is quite different from the one Pomeron exchange,
even in the region when $T$ is smaller than 1.

 In the MPSI approximation, the exact triple
 Pomeron vertex (see \fig{enhst}) is equal to the `bare' vertex.

One more comment should be made: \eq{ES1} leads to
\beq \label{ES4}
G\Lb Y\Rb\,\,\,=\,\,\,\sum^\infty_{n=1}\,(-1)^{n + 1}\,n!\,T^n\Lb Y\Rb
\eeq
which is a typical Borel summable asymptotical series.
 Such an amplitude cannot be obtained as a solution
 of the equation after a finite number of integrations.

\subsection{The exact vertex and diffractive production}

\DOUBLEFIGURE[ht]
{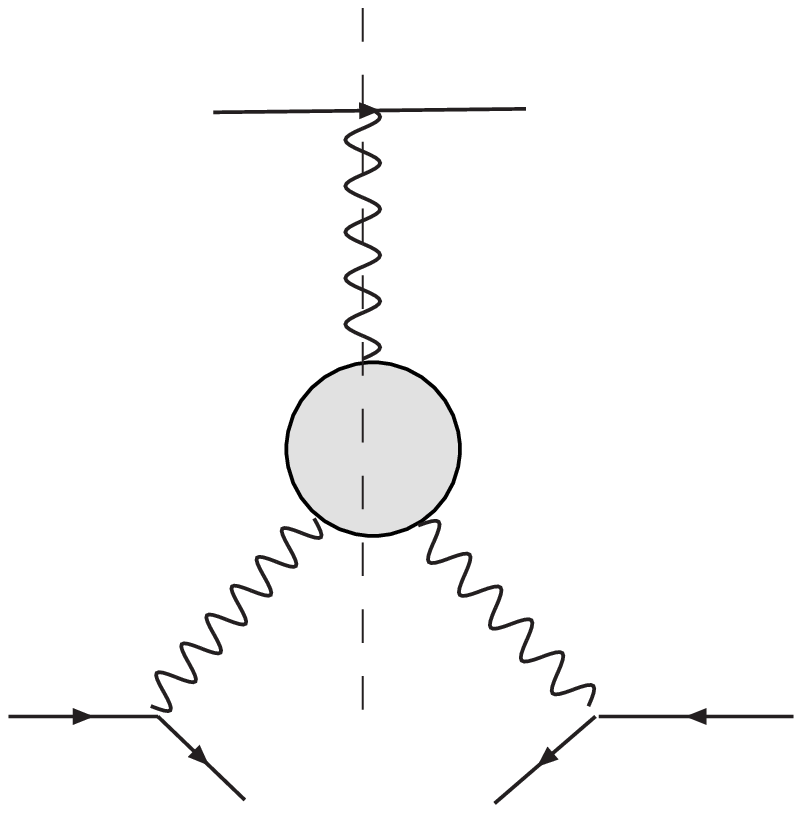,width=50mm,}{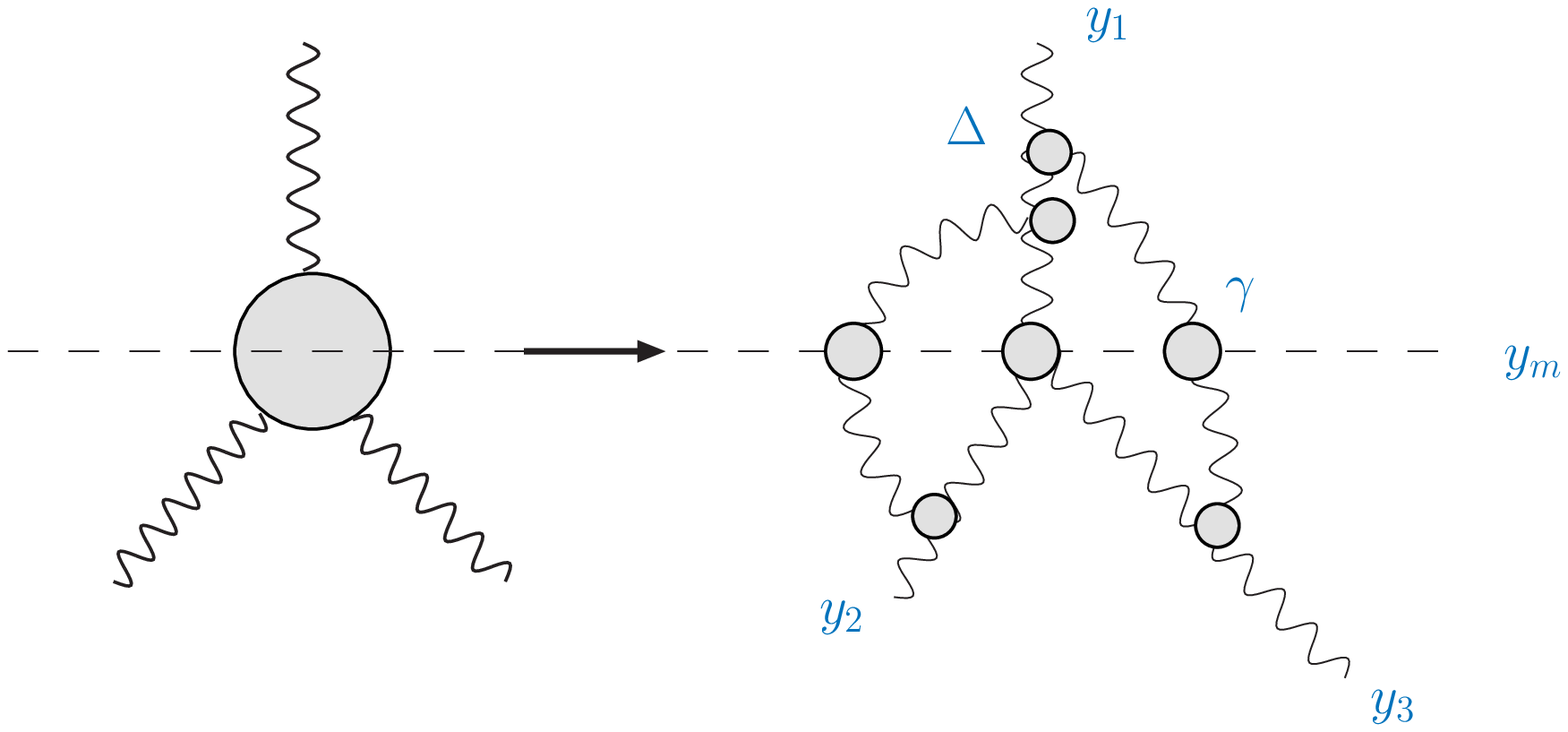,width=100mm}
{The first diagram for single diffraction.
 The wavey lines and the blob denote the exact Green's
 function and exact vertex for Pomerons\protect\cite{GLMM}.
\label{sd1}
}{The diagrams that contribute to the exact vertex.\label{sd2}}

The process of single diffraction has not been taken into
 account in the Green's function of the Pomeron.
In our paper of Ref.\cite{GLMM} we calculated the first
 diagram for the single diffraction dissociation. namely one in \fig{sd1}.
 In this diagram both the Pomeron Green's function, and the vertex are 
exact.  
We found that this contribution leads to a small cross section for  
single
 diffraction in the region of the large mass, and we view this calculation
  as  just the first attempt  to solve
 the problem of the summation of all diagrams.
 We will solve this problem in the next section.
The cross section for  single diffraction has been calculated
 in Ref.\cite{GLMM} using the MPSI approximation.
 Here, we calculate the exact vertex shown in \fig{sd2}. 
We can  use Eq.3.34 of Ref. \cite{GLMM}, where we only need
  to extract the  contribution from the external Green's function of 
the
 Pomeron, and introduce three rapitities as shown in \fig{sd2}. The 
final equation has the form
\bea
&&\Gamma^{MPSI}\Lb y1,y2,y3\Rb\,\,=\label{SD1} \\
&&\,\,\sum^{\infty}_{n=1;m=1}\,\frac{( - 1)^{n + m}}{\,n!\,m! }\,\gamma^{n + m}
\frac{\partial^n\, }{\partial^n\, w^p}
\frac{\partial^m\,}{\partial^n\, \bar{w}^p}\,\,\Gamma^{MFA}_{sd}\Lb  w^p,\bar{w}^p ; 
y1-y_m \Rb |_{w =1; \bar{w}=1} \,\notag\\
&&\times\,\,
\frac{\partial^n\,\Gamma^{MFA}\Lb w^t,y_m - y_2 \Rb }{\partial^n\, w^t}|_{w^t=1}
\frac{\partial^m\,\Gamma^{MFA}\Lb \bar{w}^t,y_m - y_3 \Rb}{\partial^n\, \bar{w}^t}|_
{\bar{w}=1}
\notag\\
&&= \frac{\Delta}{4} \,\frac{1}{T(y_1 - y_2) - T( y_1 - y_3)}\,\left\{ \Gamma_1\Lb 2\,T\Lb y_1 - y_2\Rb\Rb\,\,-\,\,
 \Gamma_1\Lb 2\,T\Lb y_1 - y_3\Rb \Rb\right\}\label{SD2}\\
&&\mbox{with}\,\,\, \Gamma_1\Lb T\Rb\,\,\,=\,\,\,(1/T^3)\times \left\{ T(1+T) - \exp\Lb - 1/T\Rb\,\Lb 1 + 2 T\Rb \,\Gamma\Lb 0,1/T\Rb\right\} \label{SD3}
\eea
where
\beq \label{SD41}
\Gamma^{MFA}\Lb  w,\bar{w} ; y_1 - y_m \Rb\,\,
=\,\,2\,\Delta w\,\bar{w}\,
\frac{1}{\Lb 1\,\,+\,\,(w \,+\,\bar{w})\,(e^{\Delta \Lb y_1 - y_m\Rb} \,
-\,1) \Rb^2}\,\,\,\,\,\,\,
\eeq
and
\beq \label{SD42}
\,\,\,\,\,\,\,\Gamma^{MFA}\Lb  w;  y_m - y_i \Rb\,\,=\,\,\frac{1}{\Lb 1\,\,+\,\,w \,(e^{\Delta \Lb y_m - y_i\Rb} \,
-\,1)\Rb}
\eeq
where $w$ and $\bar{w}$ are the variables that we needed to introduce,
 as has been explained in Ref.\cite{GLMM}.
It is easy to see that
\beq \label{SD5}
\Gamma^{MPSI}\Lb y1,y2,y3\Rb\,\,\,\xrightarrow{y_1 - y_2 \gg 1 ;\,\,y_1 - y_3\,\gg\,1}\,\,\,\frac{\Delta}{8} \,e^{ - \Delta\Lb 2 y_1 - y_2 - y_3\Rb} 
\eeq
and therefore, to a good accuracy we can consider that
 the exact vertex is equal to  the bare one, namely,
$\Gamma^{MPSI}\Lb y1,y2,y3\Rb\,\,=\,\,\Delta$.

The double diffraction contained in the Pomeron Green's function
  has been calculated directly using the unitarity constraint
 (see Ref.\cite{GLMM}).

\section{Summing the full set of diagrams}

\subsection{Elastic scattering}
In this section we sum the full set of the diagrams using
 the improved MPSI approximation.
 In our approach that has been discussed in the previous section,
 the elastic amplitude is written as the sum of
 the eikonal diagrams for each state $i$ in the two channel
 model, with the exact Pomeron Green's function (see \fig{fulset}-A).
 Here we sum all diagrams (see \fig{fulset}-B using the MPSI approximation.
   The diagrams of \fig{fulset}-B  (in the MPSI approximation),
 are shown in \fig{fulset}-C.
\FIGURE[ht]{
\centerline{\epsfig{file=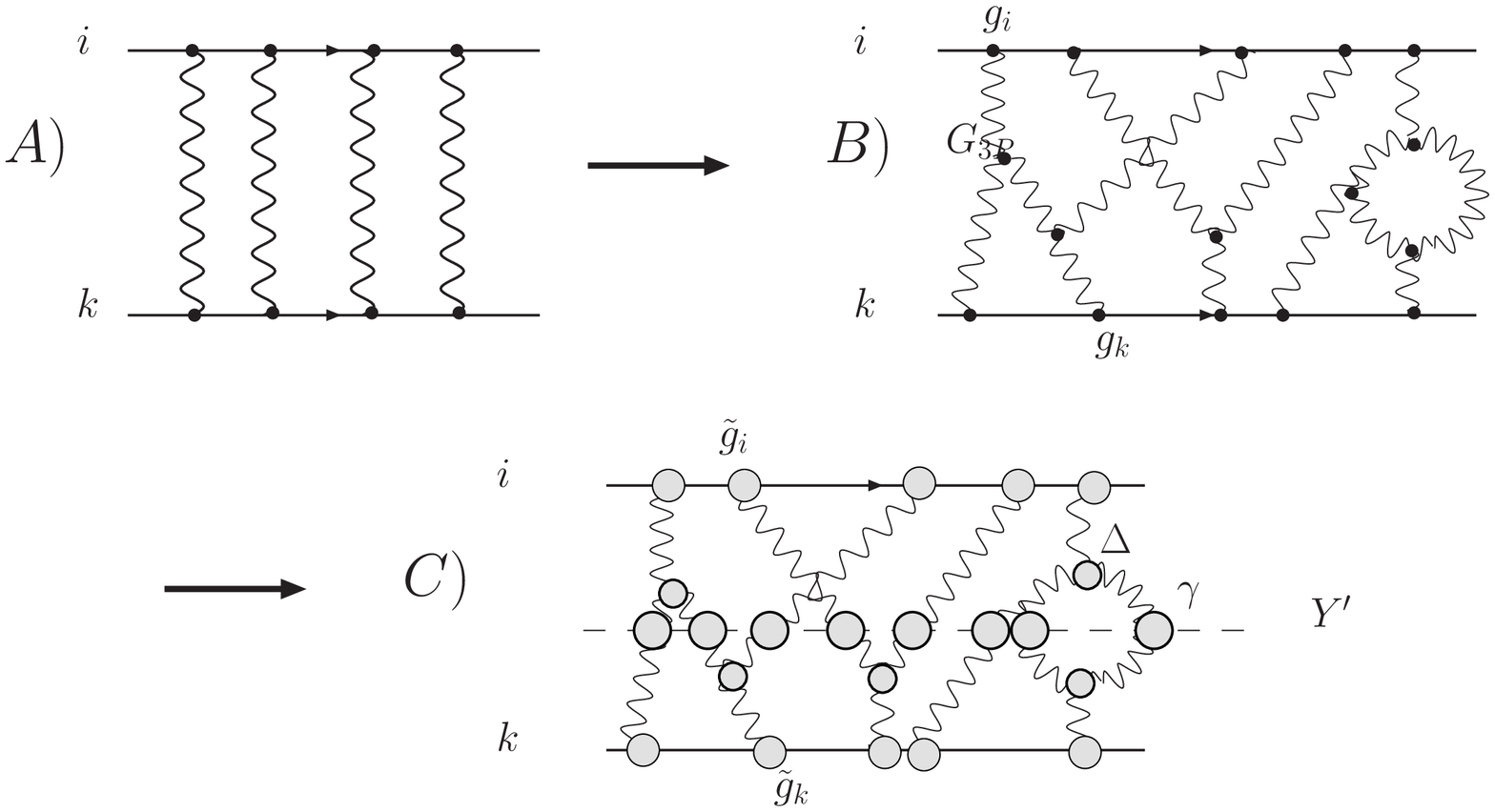,width=120mm}}
\caption{The full set of the diagrams.
  \fig{fulset}-A   the sum of enhanced diagrams in the
 two channel approach, \fig{fulset}-B shows the full set
 of the diagrams which in \fig{fulset}-C is pictured in a way that
 is most suitable to illustrate the MPSI approach.
 The bold wavey line stands for the exact Pomeron
 Green's function that includes all enhanced diagrams.}
\label{fulset}
}
The first step in determining this sum is to find the solution to
 \eq{MPSI3} with the initial condition
\beq \label{FSE1}
N_{in}\Lb \gamma  ; Y - Y'=0 \Rb \,\,=\,\,1\,\,- e^{ - \tilde{g}_i \,\gamma}
\eeq
The meaning of this conditions is very simple and
 stems from \eq{NF}: \eq{FSE1} gives the eikonal formula for
 the scattering amplitude for $Y - Y' =0$.
The general solution of \eq{MPSI3} is  simple, namely,
\beq\label{FSE2}
N^{MFA}\Lb Y-Y'; \gamma_R\Rb\,\,=\,\,N_{in}\Lb
 \gamma = N\Lb Y - Y'; \gamma_R\Rb\Rb
\eeq
where $N\Lb Y - Y'; \gamma_R\Rb$ is given by \eq{MPSI4}.

For the  initial condition of \eq{FSE1}
\beq \label{FSE3}
N^{MFA}\Lb Y-Y'; \gamma_R\Rb\,\,\,=\,\,\,1\,\,\,-\,\exp\left\{-  \tilde{g}_i \frac{\gamma_R\,e^{\Delta\,(Y - Y')}}{1\,\,\,\,+\,\,\,\gamma_R\,e^{\Delta\,(Y - Y')}}\right\}
\eeq
Using the generating function for Laguerre polynomials
 (see Ref.\cite{RY} formula {\bf 8.973(1)}), namely
\beq \label{LPGF}
( 1 - z)^{- \alpha - 1}\, \exp\Lb \frac{x\,z}{z - 1}\Rb\,\,\,=\,\,\,\sum^{\infty}_{n = 0}\,L^{\alpha}_n\Lb x \Rb \,z^n
\eeq
we obtain for \eq{FSE3}

\beq \label{FSE4}
N^{MFA}\Lb Y-Y'; \gamma_R\Rb\,\,\,=\,\,\,-\,\sum^\infty_{n=1}\,L^{-1}_n\Lb\tilde{g}_i\Rb\,\Lb -  \gamma_R e^{\Delta (Y - Y')}\Rb^n
\eeq

Using \eq{FSE4} and \eq{MPSI6} we have for the scattering amplitude
\beq \label{FSE5}
N^{MFA}\Lb Y; \Rb\,\,=\,\,\sum^{\infty}_{n=0}\,n!\,L^{-1}_n\Lb\tilde{g}_i\Rb\,\,L^{-1}_n\Lb\tilde{g}_k\Rb
\Lb - \gamma \,e^{\Delta Y}\Rb^n
\eeq
Introducing $n! = \int^\infty_0 \,\xi^n\,e^{-\xi} \,d \xi$ we
 can re-write \eq{FSE5} in the form
\beq \label{FSE6}
N^{MFA}\Lb Y; \Rb\,\,=\,\,\int^\infty_0 \,d \xi\,\,e^{-\xi} \,d\sum^{\infty}_{n=0}\,\,L^{-1}_n\Lb\tilde{g}_i\Rb\,\,L^{-1}_n\Lb\tilde{g}_k\Rb\,\Lb - \xi \gamma_0\,e^{\Delta Y}\Rb^n
\eeq
Using formula {\bf 8.976(1)} of Ref.\cite{RY}, namely
\beq \label{SUML}
\sum^\infty_{n = 0}\,n!\,z^n\,\frac{L^{\alpha}_n(x)\,L^{\alpha}_n(y)}{\Gamma\Lb n + \alpha +1\Rb}\,\,\,=\,\,\frac{\Lb x\,y\,z\Rb^{- \frac{1}{2}\alpha}}{1 - z}\,\exp\Lb - z \frac{x + y}{1 - z}\Rb\,I_\alpha\Lb 2 \frac{\sqrt{x\,y\,z}}{1 - z }\Rb
\eeq
we derive the final result
\beq \label{FSE7}
N^{MFA}_{i,k}\Lb Y \Rb\,\,=\,\,\int^\infty_0 \,\frac{d \xi}{\xi}\,\,e^{-\xi}\,\frac{\Lb \tilde{g}_i\,\tilde{g}_k\,\xi\,T(Y)\Rb^{\frac{1}{2}}\,}{ 1\,+\,\xi T(Y)}\,\exp\left\{ - \xi \,T(Y)\,\frac{\tilde{g}_i + \tilde{g}_k}{1\,+\,\xi\,T(Y)}\right\}\,J_1\Lb 2\frac{ \sqrt{\tilde{g}_i\,\tilde{g}_i\,\,\xi\,T\Lb Y \Rb}}{1\,+\,\xi\,T(Y)}\Rb
\eeq
where
\beq \label{T}
T\Lb Y\Rb\,\,=\,\,\gamma\,e^{\Delta Y}\,\,\,\,\mbox{where}\,\,\,\,\,\gamma\, \equiv\, \gamma_0.
\eeq

\eq{FSE7} reduces to a simple and elegant formula in the case that
 we can consider $\tilde{g}_i\,T(Y)\,\sim \,1$,
 $\tilde{g}_i\tilde{g}_i\,T(Y)\,>\, \,1$ but $T(Y) \ll 1$.
 Indeed, in this case the integral over $\xi$ can be taken and
\beq \label{FSE9}
A_{i,k}\Lb Y; b\Rb\,\,\,= \,\,1 \,\,\,-\,\,\exp\left\{ - \,
\,\,\h\,\frac{\tilde{g}_i\tilde{g}_k\,T(Y)}{ 1\,+\,T(Y)\,\left[\tilde{g}_i\,\, + \,\,\tilde{g}_k\right]}\right\}
\eeq

In \eq{FSE7} we neglected $b$ dependence.
 In our approach, the   $b$ dependence enters through the 
vertices $\tilde{g}_i(b)$. For \eq{FSE9} it is easy to write
 the expression that takes into account the correct
 impact parameter behaviour, namely, it has the form 
\beq \label{FSE91}
A_{i,k}\Lb Y; b\Rb\,\,\,= \,\,1 \,\,\,-\,\,\exp\left\{ - \,\,\h\,\int d^2 b'\,
\,\,\,\frac{\Lb \tilde{g}_i\Lb\vec{b}'\Rb\,\tilde{g}_k\Lb\vec{b} - \vec{b}'\Rb\,T(Y)\Rb}{ 1\,+\,T(Y)\,\left[\tilde{g}_i\Lb\vec{b}'\Rb + \tilde{g}_k\Lb\vec{b} - \vec{b}'\Rb\right]}\right\}
\eeq
To obtain the elastic amplitude we need to substitute the amplitudes
$A_{i,k}$  in \eq{ES3}.

\FIGURE[ht]{\begin{minipage}{70mm}
{\centerline{\epsfig{file=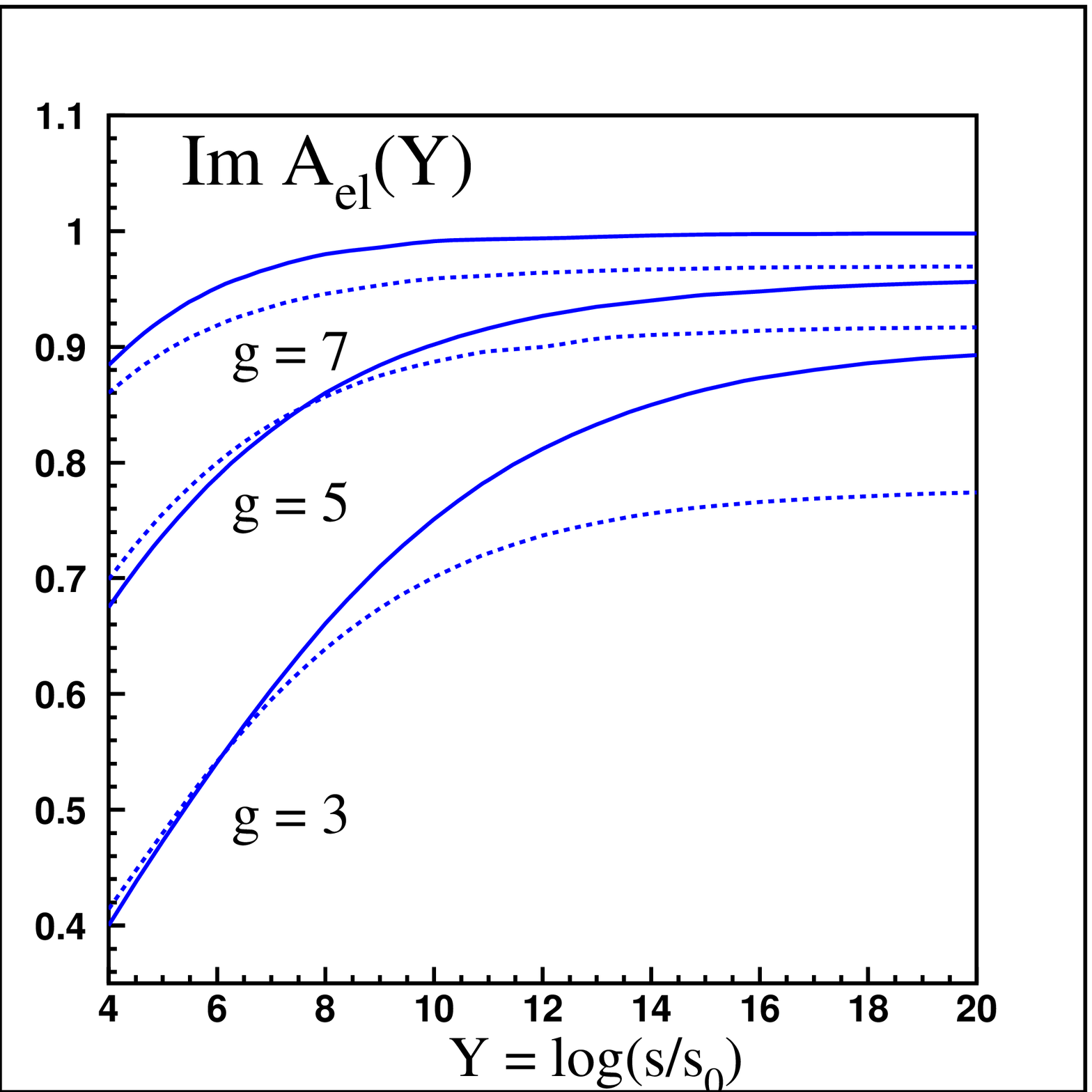,width=60mm}}
\caption{ Comparison of the exact imaginary part
 of the elastic amplitude, given by \protect\eq{FSE7},
with the approximation for $\tilde{g}\,
 T\Lb Y\Rb \approx 1$ while $\Delta \,T\Lb Y\Rb  \,\ll\, 1$.
 The values of $\gamma = 0.0242$ and $\Delta = 0.339$ were taken
 from the fit \cite{GLMM} using the sum of enhanced diagrams (see section 3)}
\label{compar1}
}
\end{minipage}
}

For the nucleus-nucleus scattering \eq{FSE9} is the generalization
 of the Glauber formula, and we intend
publishing our estimates for the cross section of
 nucleus-nucleus scattering elsewhere.

\eq{FSE9} is very simple and we intend to use this
 approximation for the description of the experimental data.
We plot in \fig{compar1}  the comparison of
  $Im A_{el}\Lb Y\Rb$ given by  the correct \eq{FSE7} ,
 with the approximate \eq{FSE9} for $\gamma=0.0242$  
 and $\Delta  = 0.339$ at the values of $\tilde{g}_1=\tilde{g}_2 = 3$
 and $5$. The values of parameters $\gamma$ and $\Delta$ are
  taken from our fit\cite{GLMM} with the amplitude that
 takes into account the emhanced diagrams only (see section 3).
 Note that at large $\tilde{g}$, the accuracy of the
 approximate solution is about $ 5\%$, while at low values
 of $\tilde{g}$ the errors could be as large as $12\%$. (see 
\fig{compar1}).

\eq{FSE91} can be derived by direct summation of the Pomeron
 diagrams, without assuming the MPSI approximation (see Ref. \cite{GLMA}).

Using the fact that \eq{FSE91} sums the net set of the diagrams
 not only in the MPSI approximation, we can also  sum
 more complicated diagrams in which the 'bare' 
 Pomerons in \fig{fulset}-C are replaced by the exact Pomeron
 of \eq{ES1} (see \fig{simsetim}, where we  show  examples
 of the diagrams that we sum, as well as  examples
 of the diagrams that we still need to calculate.
  In the selection of these diagrams
 we used parameters: $\tilde{g}_{i,k}G\Lb T\Lb Y\Rb\Rb \geq\,1$ and 
 $\Delta^2 G\Lb T\Lb Y\Rb\Rb \ll 1$. In the fit of Ref.
 \cite{GLMM} $\tilde{g} \,\geq \,5.8$ and $\Delta = 0.339$ and,
 therefore, we can obtain the scattering amplitude using
 this re-summation procedure within the accuracy
 $\Delta^2/\tilde{g}_i \leq 0.1/5.8 = 0.02$.
  We need only to replace $T\Lb Y\Rb$ in  \eq{FSE91}  by $G\Lb T\Lb Y 
\Rb\Rb$.

\FIGURE[ht]{
\centerline{\epsfig{file=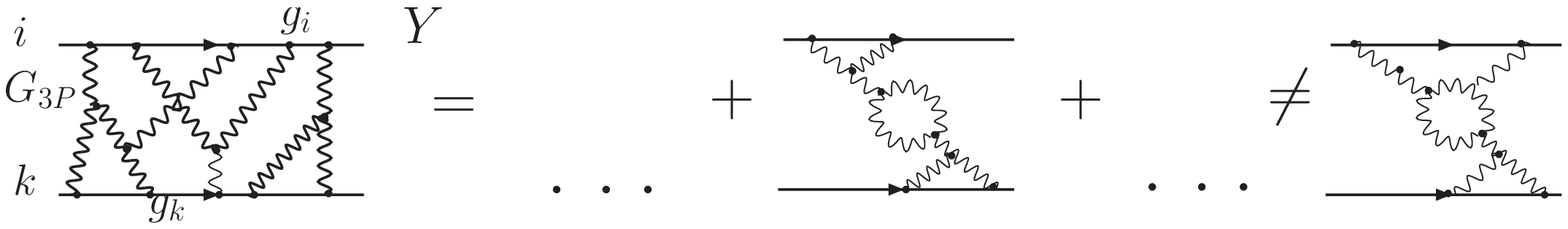,width=160mm}}
\caption{The set of diagrams that is selected and summed using the fact 
that
 $\tilde{g}_{i} G\Lb T\Lb Y\Rb\Rb \geq\,1$ while
 $\Delta^2 G\Lb T\Lb Y\Rb\Rb \ll 1$.}
\label{simsetim}
}

\subsection{Diffractive production}
\begin{figure}[h]
\begin{minipage}{11cm}
\begin{center}
\includegraphics[width=0.70\textwidth]{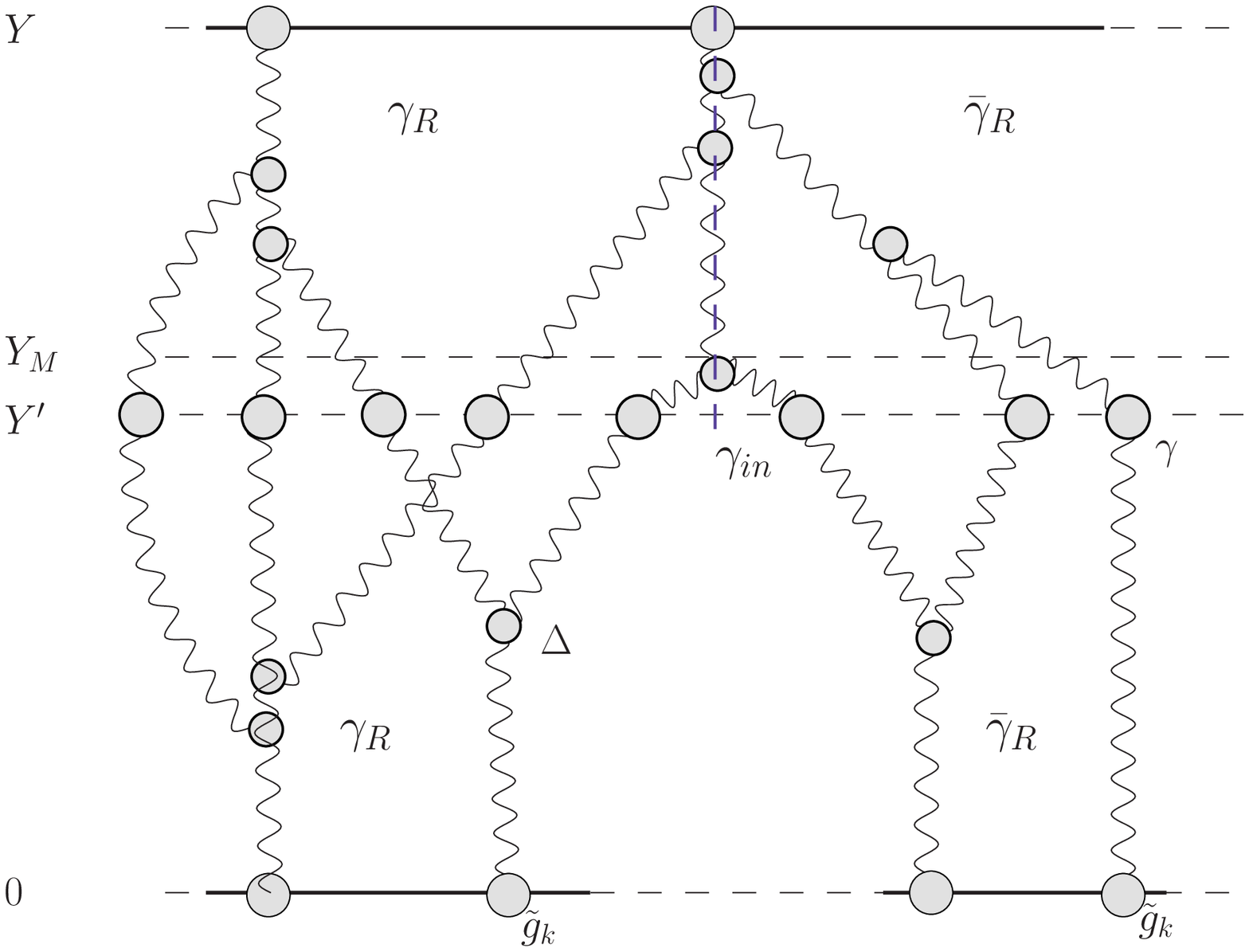}
\end{center}
\end{minipage}
\begin{minipage}{6cm}
\caption{The MPSI approximation for the cross section for  single 
diffractive production of  mass ($M^2$,$
Y - Y_M = \ln (M^2/s_0)$). The dashed lines show the cut Pomerons.  
All other notations, are as in
\protect\fig{fulset}.}
\label{sdst}
\end{minipage}
\end{figure}

 To calculate the single diffraction process it is necessary 
 to introduce three different variables :
 $\gamma_R$,$\bar{\gamma}_R$ and $\gamma_{in}$ to describe  Pomerons in
 the amplitude, in the complex
conjugate amplitude, and the amplitude of the cut Pomeron,
 respectively (see Ref.\cite{LEPR} for details).
 The difference to the calculation of the elastic amplitude lies mostly
 in the fact that we need to take into account all three amplitudes
 in \eq{MPSI6}.
 For the case of single diffraction, we have only one cut Pomeron 
 at $Y=Y_m$ that
decays in one $\gamma_R$ Pomeron, and one $\bar{\gamma}_R$ Pomeron.
  Therefore, the cut Pomeron in \fig{sdst} has the following form\cite{LEPR}
\beq \label{FSSD1}
N^{MFA}_{\mbox{cut Pomeron}}\Lb  \gamma_R,\bar{\gamma}_R ; Y - Y_M \equiv Y_m = \ln(M^2/s_0) \Rb\,\,
=\,\,2\,\Delta_\pom \gamma_R\,\bar{\gamma}_R\,
\frac{e^{\Delta_\pom Y_m}}{\Lb 1\,\,+\,\,(\gamma_R \,+\,\bar{\gamma}_R)\,\,e^{\Delta_\pom Y_m}  \Rb^2}
\eeq

Choosing $Y' = Y_M$  we have the $N^{MFA}\Lb  \gamma_R,\bar{\gamma}_R ;
 Y - Y_M \equiv Y_m  = \ln(M^2/s_0)\Rb$ for the full set of the diagrams
 in the form
\bea \label{FSSD2}
&&N^{MFA}\Lb  \gamma_R,\bar{\gamma}_R ; Y - Y_M = \ln(M^2/s_0) \equiv Y_m \Rb\,\,\,=\,\,\\
&&
N^{MFA}_{\mbox{cut Pomeron}}\Lb  \gamma_R,\bar{\gamma}_R ; Y - Y_M = \ln(M^2/s_0) \equiv Y_m \Rb
\,\exp\Lb - \frac{\tilde{g}_i \gamma_R \,e^{\Delta Y_m}}{1\,+\,\gamma_R\,e^{\Delta Y_m}}\Rb\,\,
\exp\Lb - \frac{\tilde{g}_i \bar{\gamma}_R \,e^{\Delta Y_m}}{1\,+\,\bar{\gamma}_R\,e^{\Delta Y_m}}\Rb\,\,
\nonumber
\eea
Using \eq{FSSD1} we can reduce \eq{FSSD2} to the form
\bea \label{FSSD3}
&&N^{MFA}\Lb  \gamma_R,\bar{\gamma}_R ; Y - Y_M =  \ln(M^2/s_0)
 \equiv Y_m\Rb\,\,\,=\,\,\\
&&
\,\,2\,\Delta_\pom \gamma_R\,\bar{\gamma}_R\,\int\,t\,d t \,e^{ - t\,\Lb  1 + (\gamma_R + \bar{\gamma}_R)\exp\Lb \Delta Y_m\Rb\Rb}\,\exp\Lb - \frac{\tilde{g}_i \gamma_R \,e^{\Delta Y_m}}{1\,+\,\gamma_R\,e^{\Delta Y_m}}\Rb\,\,
\exp\Lb - \frac{\tilde{g}_i \bar{\gamma}_R \,e^{\Delta Y_m}}{1\,+\,\bar{\gamma}_R\,e^{\Delta Y_m}}\Rb\,\,
\nonumber
\eea
Using \eq{LPGF} and \eq{FSE4} we can expand $N^{MFA}$ with respect
 to powers of $\gamma_R \,\exp\Lb \Delta Y_m\Rb$ and $\bar{\gamma}_R \,\exp\Lb 
\Delta Y_m\Rb$, namely
\bea \label{FSSD4}
&&N^{MFA}\Lb  \gamma_R,\bar{\gamma}_R ; Y - Y_M =  \ln(M^2/s_0) \equiv Y_m\Rb\,\,\,=\,\,\\
&& 2 \Delta \tilde{g}_i\,e^{- \Delta Y_m}\,\,\sum_{N = n + \bar{n}}\,\sum^{n}_{k=0}\,\sum^{\bar{n}}_{\bar{k}=0}\frac{t^{k + \bar{k}}}{k!\,\bar{k}!}\,(-1)^N\,\,\gamma^n\,\bar{\gamma}^{\bar{n}}\,\Lb - e^{\Delta\,Y_m}\Rb^N\,L^{-1}_{n - k - 1}\Lb - \tilde{g}_i\Rb\,\,L^{-1}_{\bar{n} - \bar{k} - 1}\Lb - \tilde{g}_i\Rb\,\nonumber
\eea
where $n$, $k$, $\bar{n}$ and $\bar{k}$ are integer numbers.

Using formulae {\bf 8.972(1)} and {\bf 9.211(2)} of Ref. \cite{RY}
 we obtain
\beq \label{FSSD5}
L^{\alpha}_n( - g)\,\,=\,\,\frac{1}{n!}\,e^{-g}\,(-g)^{- \frac{\alpha}{2}}\,\int^\infty_0\,d \xi \,e^{-\xi}\,\xi^{n +  \frac{\alpha}{2}}
\,\,J_{\alpha}\Lb 2 \sqrt{g\,\xi}\Rb
\eeq
Substituting \eq{FSSD5}  in \eq{FSSD4} we have
\bea \label{FSSD6}
&&N^{MFA}\Lb  \gamma_R,\bar{\gamma}_R ; Y_m \Rb\,\,\,=\,\,2\,\Delta \tilde{g}_i\,e^{- \Delta Y_m}\,\,\sum_{N = n + \bar{n}}\,\sum^{n}_{k=0}\,\sum^{\bar{n}}_{\bar{k}=0}\frac{t^{k + \bar{k}}}{k!\,\bar{k}!}\,(-1)^N\,\,\gamma^n\,\bar{\gamma}^{\bar{n}}\,\Lb - e^{\Delta\,Y_m}\Rb^N\, \\
&&\times\,\,\frac{1}{(n-k-1)!}\,e^{-\tilde{g}_i}\,(-\tilde{g}_i)^{ \frac{1}{2}}\,\int^\infty_0\,d \xi \,e^{-\xi}\,\xi^{n +  \frac{\alpha}{2}}\,\,J_{1}\Lb 2 \sqrt{\tilde{g}_i\,\xi}\Rb\,\,\nonumber\\
&&\times\,\,\frac{1}{(\bar{n}-\bar{k}-1)!}\,e^{-\tilde{g}_i}\,(-\tilde{g}_i)^{ \frac{1}{2}}\,\int^\infty_0\,d \bar{\xi}\, \,e^{-\bar{\xi}}\,\bar{\xi}^{n +  \frac{\alpha}{2}}\,\,J_{1}\Lb 2 \sqrt{\tilde{g}_i\,\bar{\xi}}\Rb\,\nonumber
\eea
The generating function that describes the low cascade is equal to
\beq \label{FSSD7}
N^{MFA}\Lb  \gamma_R,\bar{\gamma}_R ; Y -Y_m \Rb\,\,=\,\,N^{MFA}\Lb  \gamma_R ; Y -Y_m |\eq{FSE4}\Rb\,\times 
N^{MFA}\Lb  \bar{\gamma_R }; Y -Y_m |\eq{FSE4}\Rb
\eeq
and can be re-written in the following form
\bea \label{FSSD8}
&&N^{MFA}\Lb  \gamma_R,\bar{\gamma}_R ; Y -Y_m \Rb\,\,=\,\,
\,\,\sum_{N = n + \bar{n}}\,\sum^{n}(-1)^N\,\,\gamma^n\,\bar{\gamma}^{\bar{n}}\,\Lb - e^{\Delta\,(Y - Y_m)}\Rb^N\, \\
&&\times\,\,\frac{1}{n!}\,e^{-\tilde{g}_k}\,(-\tilde{g}_k)^{ \frac{1}{2}}\,\int^\infty_0\,d \xi \,e^{-\xi}\,\xi^{n +  \frac{\alpha}{2}}\,\,J_{1}\Lb 2 \sqrt{\tilde{g}_k\,\xi}\Rb\,\,\nonumber\\
&&\times\,\,\frac{1}{\bar{n}!}\,e^{-\tilde{g}_k}\,(-\tilde{g}_k)^{ \frac{1}{2}}\,\int^\infty_0\,d \bar{\xi}\, \,e^{-\bar{\xi}}\,\bar{\xi}^{n +  \frac{\alpha}{2}}\,\,J_{1}\Lb 2 \sqrt{\tilde{g}_k\,\bar{\xi}}\Rb\,\nonumber
\eea
\eq{MPSI6} should be replaced by a more general equation, namely,
\bea \label{FSSD9}
&&N^{MPSI}_{SD}\Lb Y_m;Y - Y_m\Rb\,\,\,=\,\,
 \,\,\sum^\infty_{n=1}\,\frac{(-1)^n}{n!}\,\,\,\,\sum^\infty_{\bar{n}=1}\,\frac{(-1)^{\bar{n}}}{\bar{n}!}\,\,\Lb  \gamma_0\,T_{SD}\Rb^{n + \bar{n}} \, \nonumber\\
&&
 \times\,\,\Lb\frac{\partial}{\partial \gamma^{(1)}_R}\Rb^n\,
\Lb\frac{\partial}{\partial \bar{\gamma}^{(1)}_R}\Rb^{\bar{n}} \,\,N^{MFA}\Lb Y_m; \gamma^{(1)}_R, 
\bar{\gamma}^{(1)}_R\Rb|_{\gamma^{(1)}_R= \bar{\gamma}^{(1)}_R=0}\nonumber\\
&&\times\,\,\Lb\frac{\partial}{\partial\gamma^{(2)}_R}\Rb^n\,\,\Lb\frac{\partial}{\partial \bar{\gamma}^{(2)}_R}\Rb^{\bar{n}}  N^{MFA}\Lb Y - Y_m; \gamma^{(2)}_R, 
 \bar{\gamma}^{(2)}_R\Rb|_{\gamma^{(2)}_R=\bar{\gamma}^{(2)}_R=0}\,\,\,\,
\eea
Using \eq{FSSD5} we can sum over $k$ and $\bar{k}$ reducing \eq{FSSD6}
 to the following equation
\bea \label{FSSD10}
&&N^{MFA}\Lb  \gamma^{(1)}_R,\bar{\gamma}^{(1)}_R ; Y_m \Rb\,\,=\,\, 2 \Delta \gamma^{(1)}_R\,\bar{\gamma}^{(1)}\,e^{ - \Delta Y_m} \,T^2_{SD}\,\,\tilde{g}_i\,e^{- 2 \tilde{g}_i} \,\,\sum^\infty_{n=1}\,\frac{(-1)^n}{(n - 1)!}\,\,\,
\sum^\infty_{\bar{n}=1}\,\frac{(-1)^{\bar{n}}}{(\bar{n} - 1)!} \\
&& \times\,\int t\, dt\,\int\,d \xi_1\,\Lb \xi_1 + t\Rb^{ n -1}\,\int d \bar{\xi}_1\,\Lb \bar{\xi}_1 + t\Rb^{ \bar{n} -1}\,\Lb \xi_1\,\bar{\xi}_1\Rb ^{- \frac{1}{2}}\,\,e^{ - \xi_1 - \bar{\xi}_1}\,J_{1}\Lb 2 \sqrt{\tilde{g}_i\,xi_1}\Rb\,
\,J_{1}\Lb 2 \sqrt{\tilde{g}_i\,\bar{\xi}_1}\Rb\nonumber
\eea
Using \eq{FSSD10} we can reduce \eq{FSSD9} to the form
\bea \label{FSSD11}
&&N^{MPSI}_{SD}\Lb Y_m;Y - Y_m\Rb\,\,\,=\,\,\,\, 2 \Delta\, \tilde{g}_i\gamma^2_0\,\,e^{ - \Delta Y_m} \,T^2_{SD}\Lb Y,Y_m\Rb\,\,\int\,t dt \,e^{-t - \tilde{g}_i - \tilde{g}_k}\,\int d \xi_1\,d \xi_2\,e^{ - \xi_1 - \xi_2}\,\frac{\xi_2}{\sqrt{\xi_1\,\xi_2}}\,\\
&&\times\sum^\infty_{n = 1}\,\frac{1}{( n - 1)!}\,\Lb t \,+\,\xi_1\Rb^{ n -1}\,\xi^{n - 1}_2\,\,J_{1}\Lb 2 \sqrt{\tilde{g}_i\,\xi_1}\Rb\,\,J_{1}\Lb 2 \sqrt{\tilde{g}_k\,\xi_2}\Rb\,\times\left\{ \xi_1 \to \bar{\xi}_1; \xi_2 \to \bar{\xi}_2; n \to \bar{n} \right\}\,\times \Lb - \gamma_0\,T_{SD}\Rb^{ n + \bar{n}}\nonumber
\eea
Summing over $n$ and $\bar{n}$ we obtain
\bea \label{FSSD12}
&&N^{MPSI}_{SD}\Lb Y_m;Y - Y_m\Rb\,\,\,= \\
&& 2 \Delta\, \,e^{ - \Delta Y_m} \,T^2_{SD}\Lb Y,Y_m\Rb\,\tilde{g}^{-1}_k\,e^{- 2 \tilde{g}_i - 2 \tilde{g}_k}\,\int \,t \,d t
\int d \xi_1 d \xi_2 d \bar{\xi}_1 d \bar{\xi}_2\,e^{ - t( \xi_2 + \bar{\xi}_2)\,\gamma\,T_{SD} - t}\,e^{- \xi_1 - \xi_2 - \bar{\xi}_1 - \bar{\xi}_2}\nonumber\\
&&\times\frac{\xi_2 \bar{\xi}_2}{\sqrt{\xi_1\,\xi_2}}\,\frac{1}{\sqrt{\bar{\xi}_1\,\bar{\xi}_2}}\,
\,\,e^{ - \gamma\,\xi_1\,\xi_2\,T_{SD}}\,e^{ - \gamma\,\bar{\xi}_1\,\bar{\xi}_2\,T_{SD}}\,\,J_{1}\Lb 2 \sqrt{\tilde{g}_i\,\xi_1}\Rb\,\,J_{1}\Lb 2 \sqrt{\tilde{g}_i\,\bar{\xi}_1}\Rb\,J_{1}\Lb 2 \sqrt{\tilde{g}_k\,\xi_2}\Rb\,\,J_{1}\Lb 2 \sqrt{\tilde{g}_k\,\bar{\xi}_2}\Rb\nonumber
\eea
and after interation over $t$ we have 
\bea \label{FSSD13}
&&N^{MPSI}_{SD}\Lb Y_m;Y - Y_m\Rb\,\,\,=\\
&&
 2 \Delta\, \,e^{ - \Delta Y_m} \,T^2_{SD}\Lb Y,Y_m\Rb\,\tilde{g}^{-1}_k\,e^{- 2 \tilde{g}_i + 2 \tilde{g}_k}\,
\int \frac{d \xi_1 d \xi_2 d \bar{\xi}_1 d \bar{\xi}_2}{\Lb 1 + (\xi_2 + \bar{\xi}_2)\,\gamma T_{SD}\Rb^2}
\,e^{- \xi_1 - \xi_2 - \bar{\xi}_1 - \bar{\xi}_2}\\
&&\times\,\,
\frac{\xi_2 \bar{\xi}_2}{\sqrt{\xi_1\,\xi_2}}\,\frac{1}{\sqrt{\bar{\xi}_1\,\bar{\xi}_2}}\,
e^{ - \gamma\,\xi_1\,\xi_2\,T_{SD}}\,e^{ - \gamma\,\bar{\xi}_1\,\bar{\xi}_2\,T_{SD}}\,\,J_{1}\Lb 2 \sqrt{\tilde{g}_i\,\xi_1}\Rb\,\,J_{1}\Lb 2 \sqrt{\tilde{g}_i\,\bar{\xi}_1}\Rb\,J_{1}\Lb 2 \sqrt{\tilde{g}_k\,\xi_2}\Rb\,\,J_{1}\Lb 2 \sqrt{\tilde{g}_k\,\bar{\xi}_2}\Rb
\nonumber
\eea
Introducing new variables $ y = \xi_1 \xi_2$ and $\bar{y} =
 \bar{\xi}_1 \bar{\xi}_2$ and  integrating over them, we
obtain the final formula for the single diffraction amplitude $A_{i,k}\Lb Y, Y_m\Rb$ in the form
\bea \label{SDF} 
A^{SD}_{i;k, l}\Lb Y, Y_m\Rb\,\,&=&\,\, 2 \Delta\, \,e^{ - \Delta Y_m} \,T^2_{SD}\Lb Y,Y_m\Rb\,\tilde{g}_i\,\sqrt{\tilde{g}_k\tilde{g}_l}\,e^{ - 2 \tilde{g}_i -\tilde{g}_k - \tilde{g}_l}\,\int\frac{d \xi_2 \,d \bar{\xi}_2\,\sqrt{\xi_2 \bar{\xi}_2}}{\Lb
1 \,+\,(\xi_2 + \bar{\xi}_2) \cdot T_{SD}\Rb^2}e^{ - \xi_2 - \bar{\xi}_2}
\,\\
&\times &\,\,\,\,J_{1}\Lb 2 \sqrt{\tilde{g}_k\,\xi_2}\Rb
J_{1}\Lb 2 \sqrt{\tilde{g}_l\,\bar{\xi}_2}\Rb\,\,
\left\{1\,-\,\exp\Lb \frac{-\tilde{g}_i}{1\,+\,\xi_2\,T_{SD}}\Rb\right\}\,\,
\left\{1\,-\,\exp\Lb \frac{-\tilde{g}_i}{1\,+\,\bar{\xi}_2\,T_{SD}}\Rb\right\}\nonumber
\eea
where  
\beq \label{TSD}
T_{SD}\Lb Y; Y_m\Rb\,\,\,=\,\,\,\gamma\,\Lb e^{\Delta Y_m}\, - \,1\Rb\,e^{ \Delta (Y - Y_m)}
\eeq
In \eq{SDF} we took into account that in the two channel model,
 the low cascade could 
 be  initiated by different states ($k$ and $l$ ).
As in  the case of the elastic amplitude, $\tilde{g}_i$ in \eq{SDF}
 should be replaced by $\tilde{g}_i\Lb\vec{b}'\Rb$ and
 by $\tilde{g}_{k,l}\Lb \vec{b} - \vec{b}'\Rb$ and should
 be integrated over $d^2 b'$.

The total  cross section of the diffractive production
 can be wriiten as a sum of two terms: the Good -Walker term which is equal to
\beq \label{FSSDGW}
\sigma^{GW}_{SD}\,\,=\,\,\int\,\,d^2b\, \Lb \alpha\beta\{-\alpha^2\,A^{el}_{1,1}+(\alpha^2-\beta^2)\,A^{el}_{1,2}+\beta^2 \,A^{el}_{2,2}\}\,\Rb^2
\eeq
where $A_{i,k}$ are given by \eq{FSE9},
 and the term which describes the diffractive production in the region of large mass, namely
\bea\label{FSSDLM}
&&\sigma^{\mbox{Large mass}}_{SD}\,\,=\,\,\,\,2 \int d Y_m \int d^2 b \,\\
&&\left\{\,\alpha^6\,A^{SD}_{1;1,1}\,e^{- \Omega_{1 1}\Lb Y;b\Rb}\,\,+\,\alpha^2\beta^4 A^{SD}_{1; 2, 2}\,e^{- \Omega_{1 2}\Lb Y;b\Rb} + 2\,\alpha^4\,\beta^2 \,A^{SD}_{1;1,2}\,e^{- \h\Lb \Omega_{1 1}\Lb Y;b\Rb + \Omega_{1 2}\Lb Y; b \Rb\Rb} \right. \nn\\
&&\left.\,\,+\,\,\beta^2\,\alpha^4 \,A^{SD}_{2; 1, 1}\,e^{- \Omega_{1 2}\Lb Y;b\Rb}\,\,+\,\,2\,\beta^4\alpha^2
\,A^{SD}_{2; 1,2}\,e^{- \h\Lb \Omega_{1 2}\Lb Y;b\Rb + \Omega_{2 2}\Lb Y; b \Rb\Rb}\,\,+\,\,\beta^6\,\,A^{SD}_{2; 2,2}\,e^{- \Omega_{2 2}\Lb Y;b\Rb} \right\} \nonumber
\eea

To find the cross section for  double diffraction we
  use the  $s$-channel unitarity constraints as
 was suggested in Ref.\cite{GLMM}
\beq \label{FSDD1}
2 A^{el}_{i,k}(Y;b)\,\,\,=\,\,| A^{el}_{i,k}(Y;b)|^2 \,+ 2\,A^{SD}_{i;k,k}(Y; b )\,\,+\,\,A^{DD}_{i,k}\,\,+\,\,A^{in}_{i,k}(Y;b)
\eeq
It has been shown in Refs.\cite{LEPR,KL,BORY} that the inelastic
 amplitude $A^{in}_{i,k}(Y;b) \,=\,A^{el}_{i,k}(2T(Y);b)$.
 Therefore, from \eq{FSDD1} the amplitude for  double diffraction
 production is equal to
\bea\label{FSDD2}
&&A^{DD}_{i,k}\Lb Y; b\Rb \,\,=\,\,\\
&&
2 A^{el}_{i,k}\Lb T\Lb Y\Rb; b \Rb\,\,\,-\,\,| A^{el}_{i,k}\Lb T\Lb Y\Rb; b \Rb|^2 \,- 2\,\int \,d Y_m A^{SD}_{i;k,k}\Lb T_{SD}\Lb Y, Y_m; b \Rb; b \Rb\,\,-\,\,A^{el}_{i,k}\Lb 2 T\Lb Y \Rb;b\Rb\nonumber
\eea
Finally, the  cross section of the double diffractive
 production is the sum of the Good-Walker contribution, which has the form
\beq \label{FSDD3}
\sigma^{G W}_{DD}\,\,=\,\,\int \,d^2 b \,\, \alpha^2\,\beta^2\left\{ A^{el}_{1,1}\, -\,2
\,A^{el}_{1,2}\,+\, A^{el}_{2,2} \right\}^2
\eeq
with $A^{el}_{I,k}$  given by \eq{FSE9}, and the term which is determined 
by the Pomeron interaction, and which contributes to the production of 
large masses. 
namely,
\beq \label{FSDD4}
\sigma^{\mbox{Large mass}}_{DD}\,\,=\,\,\int\,d^2 b\,\left\{\alpha^4\,A^{DD}_{1,1}\,e^{- \Omega_{1 1}\Lb Y;b\Rb}\,+ \,2 \alpha^2\,\beta^2 A^{DD}_{1,2}\\,e^{- \Omega_{1 2}\Lb Y;b\Rb},\,+\,\,\beta^4\,A^{DD}_{2,2}\,e^{- \Omega_{12 2}\Lb Y;b\Rb}\,\right\}
\eeq

\section{Corrections to our approach}
\subsection{$\alpha'_\pom \neq 0$ and the Pomeron  as a  fixed branch 
point}
In constructing our model we have made two assumptions that 
considerably simplify our approach.
 Firstly, we considered the case of $\alpha'_\pom =0$ for
 summation of the enhanced diagrams, in spite of the fact
 that $\alpha'_\pom \approx 0.02 \,GeV^{-2}$, is the value
 obtained from our fit of the experimental data. Secondly, we  replaced  
the Pomeron
Regge pole by a fixed square root singularity,
  that appears in the  N=4 SYM.
  Both these assumptions are not  principle in nature,
 but have been made to facilitate the simplicity
 and transparency of our approach.
 In this section we will   expand our formalism so as to assess the 
consequence of our assumptions.  

\FIGURE[t]{\begin{minipage}{80mm}
{\centerline{\epsfig{file=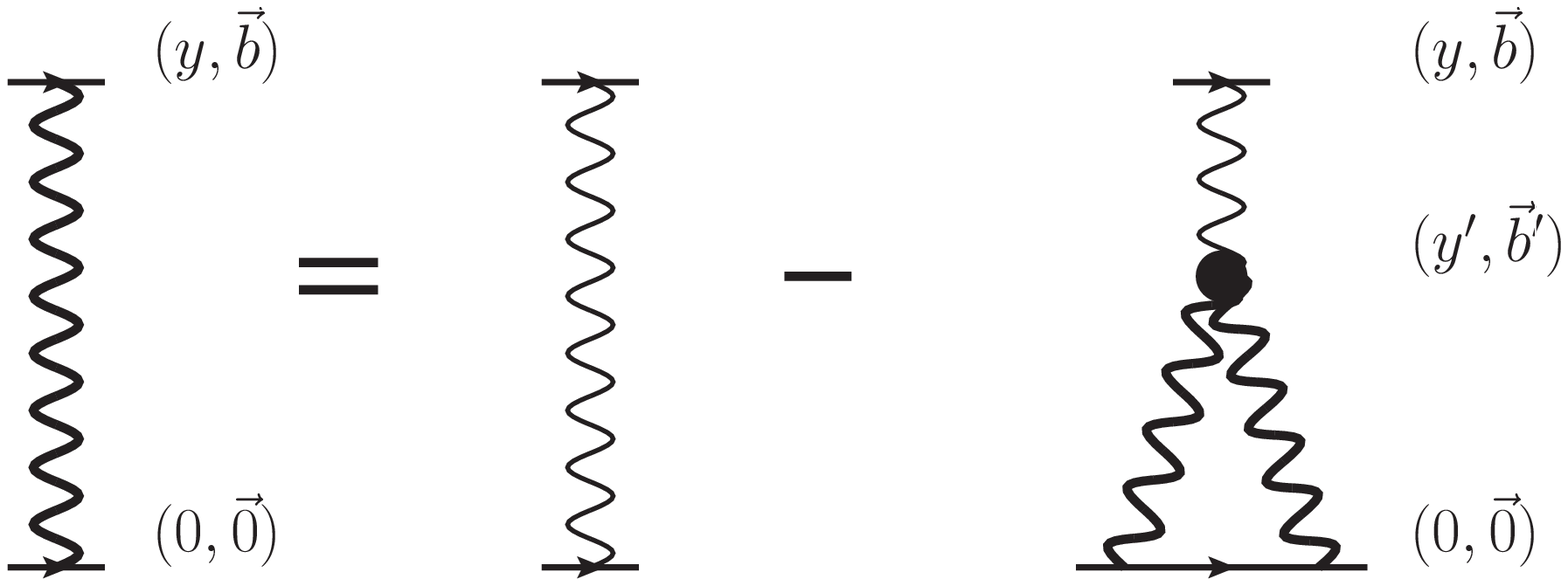,width=75mm}}
\caption{The graphic form of the equation that sums `fan' diagrams. The
 Green's function of
 the exact Pomeron is depicted as bold wavey  line, while the 
 `bare' Pomeron is shown by  thin wavey line.}
\label{eqpom}
}
\end{minipage}
}
We  know  how to include the impact parameter dependence in the 
generating function formalism, \eq{ZFEQ} should be
 replaced by a new equation \cite{GRPO}, which has the following form
\bea \label{COR1}
&&\frac{\partial \,Z\,\Lb Y-Y_0;b; u\Rb}{
\partial \,Y}\,\,= \\
&&\alpha'_\pom\,\nabla^2_b \,Z\,\Lb Y-Y_0;b; u\Rb\, 
- \Delta \,u\,(1 - u)\,\,Z\Lb Y- Y_0; b; u \Rb\nn.
\eea

\eq{COR1} is written for the generating function that sums the `fan' diagrams.
 For the  scattering amplitude, this sum is the solution of the simple
 equation shown in \fig{eqpom}, and it is of the form
\beq \label{COR2}
G\Lb y; b\Rb \,\,=\,\,G_0\Lb y, b \Rb \,\,-\,\,G_{3\pom}\,
\int^y_0 d y' \,\int\,
d^2 b'\,G_0\Lb y - y',\vec{b} - \vec{b}'\Rb\,G^2\Lb y'-0, \vec{b}'\Rb.
\eeq
 $G_0\Lb y,b\Rb$ is the Green's function of the `bare' Pomeron
\beq \label{COR3}
G_0\Lb y,b\Rb\,\,=\,\,\frac{1}{4\pi\alpha'_\pom\,y}\,e^{ \Delta y\,\,-\,\,\frac
{b^2}{4\, \alpha'_\pom y}}\,\,\,\,\,\xrightarrow{\alpha'\,\to\,0}\,\,\,
\,\,\delta^{(2)}\Lb b \Rb,
\eeq
which is   the  Fourier transform of $G\Lb y, 
q\Rb\,=\,\exp\Lb 
\Delta y - \alpha'_\pom\,q^2\Rb$, where  $q^2 = - t$ is the
 momentum transfer squared.
 $G_0$ satisfies the equation
\beq \label{COR4}
\frac{\partial G_0\Lb y, b \Rb}{\partial y} \,-\,\alpha'_\pom\,\nabla^2_b\,G_0
\Lb y, b \Rb\,\,=\,\,\Delta\,G_0\Lb y, b \Rb.
\eeq
Using \eq{COR4},  \eq{COR2}  can be rewritten in the form
\bea \label{COR5}
&&\frac{\partial G\Lb y, b \Rb}{\partial y} \,-\,\alpha'_\pom\,\nabla^2_b\,G\Lb
 y, b \Rb\,\,=\,\,\Delta \,G\Lb y, b \Rb\,\,-\,\,\Delta\, G^2\Lb y, b \Rb,
\eea
where we substitute $G_{3 \pom} = \Delta$ as in \eq{ZFEQ}.

This equation is the same as \eq{COR1} for the generating function.
Therefore, in the framework of the MPSI approximation \eq{COR1},
 together with the 
obvious property of 
\beq \label{COR6}
G_0\Lb y, b \Rb\,\,=\,\,\int d^2 b' \,G_0\Lb y - y', \vec{b} - \vec{b}' \Rb\,
G_0\Lb y', \vec{b}' \Rb,
\eeq
lead to the Green's function of the exact Pomeron that includes
 the impact  parameter dependence.

\FIGURE[t]{\begin{minipage}{75mm}
{\centerline{\epsfig{file=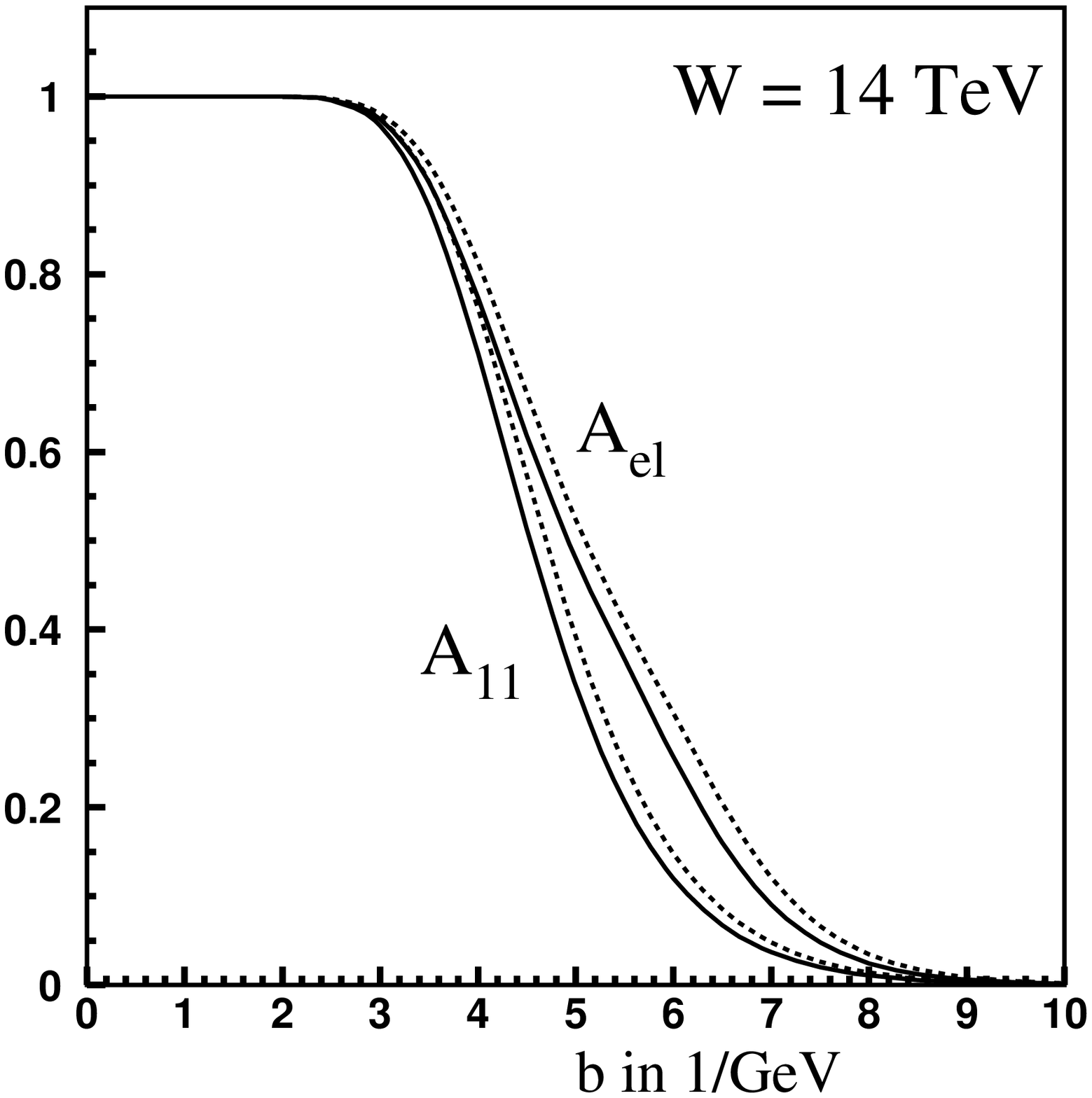,width=75mm}}
\caption{The scattering asmplitude $A_{11}$ and $A_{el}$ for $\alpha'_\pom=0$ (solid lines)
 and $\alpha_\pom = 0.01\,1/GeV^2,$(dashed lines). All other parameters are taken from Ref.\cite{GLMM}.}
\label{ampl}
}
\end{minipage}
}

As we have  discussed,  the solution to \eq{ZFEQ} can be  found
 in the form $Z\Lb   G_0\Lb y \Rb f(u)\Rb$ in which $f(u)$ 
 that follows from the equation, and the
form of $Z$ stems from the initial condition $Z\Lb  y=0, u\Rb = u$.
The solution is
\beq \label{COR7}
Z \Lb y, u\Rb\,\,=\,\,\frac{u}{ u\, \, +\, (1 - u)\, G_0\Lb y \Rb}.
\eeq
We  suggest that the solution of \eq{COR1} has the same form as
 \eq{COR7}, namely,
\beq \label{COR8}
Z \Lb y, u; b \Rb\,\,=\,\,\frac{u}{ u\, \, +\, (1 - u)\, G_0\Lb y; b \Rb}.
\eeq
One can see that this form of $Z \Lb y, u; b \Rb$ satisfies the
 initial condition $Z \Lb y=0, u; b \Rb\,=\,u\,\delta^{(2)}
 \Lb \vec{b}\Rb$ and $Z \Lb y, u=1; b \Rb\
\,=\,\,1$.  Inserting \eq{COR8} one sees that it
 does not satisfy \eq{COR1}. We consider 

\beq \label{COR9}
\frac{\partial \,Z\,\Lb Y-Y_0;b; u\Rb}{
\partial \,Y}\,\,
-
\eeq
$$ \,\,\alpha'_\pom\,\nabla^2_b \,Z\,\Lb Y-Y_0;b; u\Rb\, 
+ \,\Delta \,u\,(1 - u)\,\,Z\,\Lb Y- Y_0; b; u \Rb
$$
$$
\,=\,\,\,\frac{b^2}{4\, \alpha'^2_\pom y^2}\,
G^2_0\Lb y,b,\Rb\frac{d^2 Z\Lb G_0\Lb y, b\Rb; u\Rb}{\Lb d G_0\Rb^2}\,\,
\,\propto \,\,\,\frac{b^2}{4\, \alpha'^2_\pom y^2}\,G^{-1}_0 .
$$

>From \eq{COR9} we  see that 
 if $b^2 < \Lb \alpha'_\pom\,y\Rb^2$,
the corrections are small. 
 Recall that the Pomeron contribution is
dominant at $b^2 \leq 4 \alpha'_\pom\,y$.
 It is also  small at large values of $y$
since 
$G_0 \,\gg\,1$. 

Having \eq{COR8} as the solution for the generating function,
 we can  find the exact Green function  ($ G\Lb y, b\Rb$)
 for the Pomeron in the  MPSI approximation which is equal to \eq{ES1} 
with 
\bea \label{COR10}
T\Lb Y\Rb\,\, &\longrightarrow &\,\,T\Lb Y; b \Rb\,\,=\\
 &=& \,\, \gamma\,G_0\Lb y,b,\Rb\,\, =\,\,
\gamma\,\frac{1}{4 \pi \alpha'_\pom\,y}\,e^{ \Delta y\,\,-\,\,\frac{b^2}{4\, \alpha'_\pom y}}\nn.
\eea
 Instead of \eq{ES2} for $\Omega_{i k}$ we have
\beq \label{COR11}
\Omega_{i k}\Lb  y, b\Rb\,\,=\,\,\int \,d^2  b'\tilde{g}_i,\tilde{g}_k\,S_{i k}\Lb b'\Rb\,
 G\Lb y, \vec{b} - \vec{b}'\Rb.
\eeq
In \fig{ampl} we plot the b space amplitudes for the highesr LHC energy 
with
 $\alpha'_\pom = 0$ and $\alpha'_\pom=0.01\,GeV^{-2}$.
   The difference is negligibly small and,
 therefore, we neglect the restriction on the kinematic
 region that followed from taking $\alpha'_\pom =0$ (see \eq{ES3}
 and \eq{ES4}). Note,  our formalism ( due to the MPSI approximation)
is only valid for W $\leq$ 100 TeV \cite{GLMM}, this is far beyond the LHC 
energy range.

\eq{COR2}  shows that a method to include  the energy behaviour of
 the `bare' Pomeron, is different from the one for a Regge pole.
 As we have mentioned in N=4 SYM the leading singularity in the 
 angular momentum plane is not a pole but a fixed cut,
 with an energy behaviour 
\beq 
\label{COR12}
G_0\Lb y, b \Rb\,\,\propto\,\,y^{- j_0}\,e^{\Delta y}
 \eeq
which satisfies the equation
\beq \label{COR13}
\frac{ \partial G_0\Lb y, b \Rb}{\partial y}\, \,=\,\,\Delta\,G_0\Lb y, b \Rb\,\,-\,\,j_0/y\,G_0\Lb y, b \Rb\,\,\to\,\,\Delta\,G_0\Lb y, b \Rb
\eeq

>From \eq{COR13} one can see that we can neglect the contribution
of the factor, $y^{-j_0}$  by 
 differentiating {\bf over $y$}.  Therefore, 
\eq{COR2}   
  can be reduced to \eq{COR5} which has solution of \eq{COR8} with $G_0$
from \eq{COR12}.

\subsection{Low energy description and the  `threshold effect'.}

\FIGURE[ht]{
\centerline{\epsfig{file=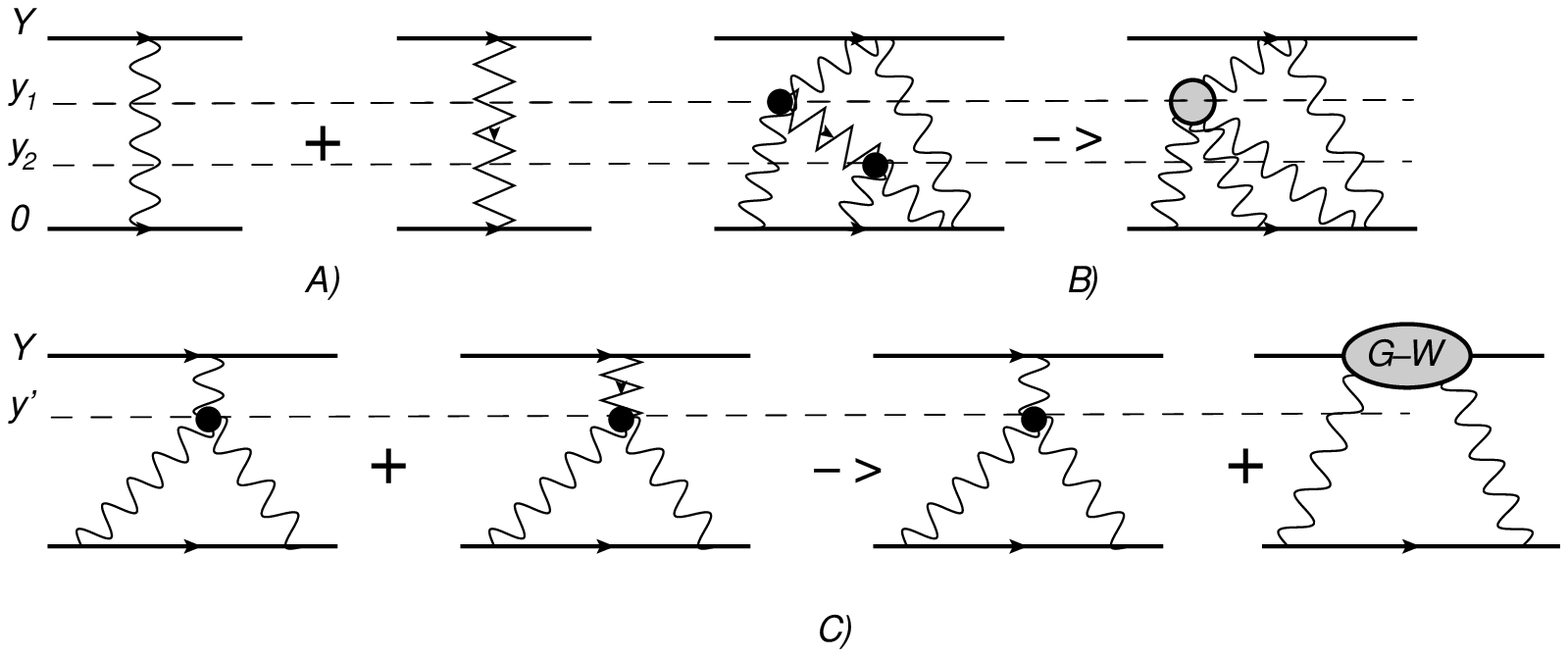,width=150mm}}
\caption{Contributions of secondary  Reggeons denoted by zigzag lines.
 Wavey lines denote the Pomeron.
 \protect\fig{lowen}-A shows the contribution to  the scattering amplitude due 
to exchange
 of a Pomeron and Reggeon. \protect\fig{lowen}-B and \protect\fig{lowen}-C 
illustrates
 the fact that the exchange of secondary  Reggeons can be reduced to
 the inclusion of the vertices
for Pomeron-Pomeron interactions,
 or can be included in the Good-Walker mechanism.}
\label{lowen}
}

\FIGURE[h]{\begin{minipage}{90mm}
{\centerline{\epsfig{file=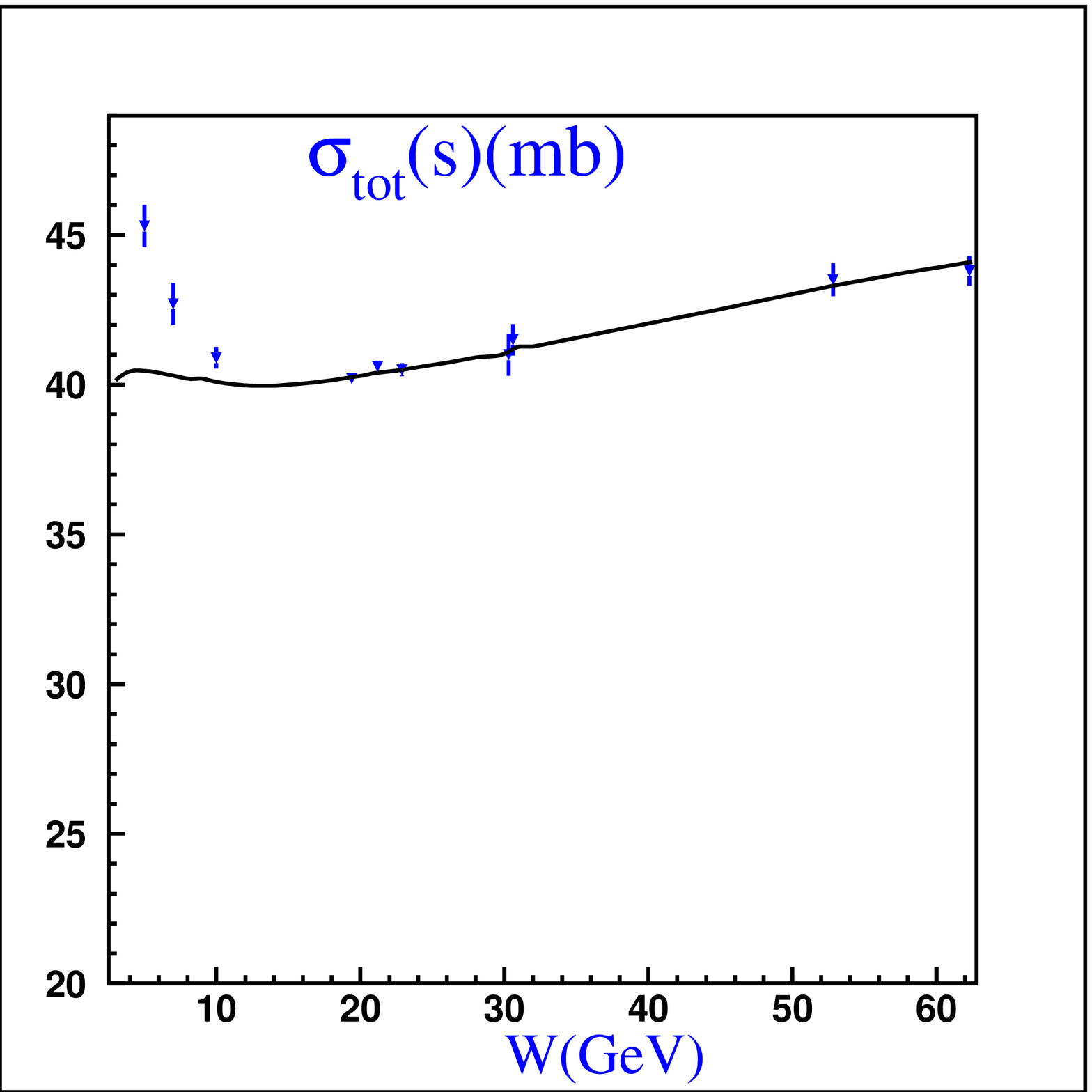,width=85mm,height=75mm}}
\caption{The total cross section ($\sigma_{tot} = 1/2\Lb\sigma_{tot}(pp)
 + \sigma_{tot}(p \bar{p})\Rb$) at low energies in the framework of
 our approach with the parameters determined
 by  high energy data. The curve
 illustrates our parametrization. }
\label{sigttle}
}
\end{minipage}
}

\subsubsection{Low energy behaviour of the scattering amplitude}

As we shall see  in the next section, most of our data base consists of
lower energy points from ISR and Sp${\bar p}$S/SppS (W $\approx$ 20 -70   
GeV), where the contribution of the secondary Regge exchanges are
important.
 A secondary Reggeon has an energy behaviour $\exp\Lb \Delta_\reg 
 (Y - 0)\Rb$. 
 The sum $ \pom + \reg$ describes the energy behaviour of
 the  elastic scattering amplitude
 without screening corrections.
 This  sum replaces the single  Pomeron exchange in the definition of 
$\Omega_{ik}$.
  Inserting this sum everywhere in the more complicated diagram (see 
\fig{lowen}),
 one can see that the integrations over rapidities reduces  the contributions
 of the secondary Reggeons.  By introducing  new vertices
 for the Pomeron-Pomeron interactions (  see \fig{lowen}-B) 
the
integration over $y_1 - y_2$ can be replaced by a
 new $\pom \to 3 \pom$ vertex),  or it 
can be absorbed into G-W mechanism (see \fig{lowen}-C 
Since in our approach the vertices, other than the triple
 Pomeron vertex are considered to be small,
 we arrive at the conclusion that for lower energies
  we  only need to replace the single Pomeron
 exchange by $\pom + \reg$, in the definition of $\Omega_{ik} $.

In \fig{sigttle} we  compare our prediction for lower energies
 with the experimental data and obtain
a satisfactory description to within 10\%.
 The conclusion is very simple: we do not need an
 additional source to describe the lower
 energy behaviour of the amplitude.
\subsubsection{The `threshold' effect}
As a consequence of the above conclusion there is no requirement to 
introduce a Pomeron threshold. i.e.
  we do not need to assume that the Pomeron contributes to the 
scattering amplitude 
  for $Y > y_0$ with $y_0 \approx 1.5 \div 2.5$. 

Nevertheless, it is interesting to estimate the influence of the
 threshold cutoff on the 
value of the scattering amplitude, and especially on the value
 of the survival probability.
 We calculate the simplest enhanced diagram of \fig{enh1}
 introducing this cutoff, 
 assuming that $Y- y_1 > y_0$ as well as $y_2 > y_0$. We obtain 
\beq \label{TE1}
A\Lb \mbox{\fig{enh1}}\Rb \,\,=\,\,\frac{\tilde{g}^2 \gamma^2}{\Delta^2}\,
e^{ - 2 \Delta \Lb Y - y_0\Rb}\,\,+\,\,{\cal O}\Lb \gamma\,
 e^{ - \Delta Y}\Rb .
\eeq
First, we would like to draw the readers attention to the fact, 
 that without a cutoff
 the typical $Y- y_1 \approx 1/\Delta$ and $y_2 \approx 1/\Delta$.
 Therefore, we need to introduce a cutoff only if $y_0 > 1/\Delta$.
 Second,
 we need to change $\gamma \to \gamma\,\exp\Lb \Delta \,y_0\Rb$ and
 multiply the Pomeron
 exchange by $\Theta\Lb Y - y_0\Rb$. Doing so in our parametrization
we obtain the following values for $\langle|S^2_{enh}|\rangle$ for
 $y_0 = 0, 1.5, 2.3$. For the
 Tevatron energy (W = 1.8 TeV) we  obtain    0.285, 0.7 ,0.99.
  For LHC ($W= 14\,TeV$)
  the corresponding values for $\langle|S^2_{enh}|\rangle$  are  0.06, 
0.12, 0.19. $\langle|S^2_{enh}|\rangle$ is the survival probability 
initiated by the Pomeron interactions.
 The conclusion from this exercise is very simple:
 in our approach we do not need to
 introduce a threshold for  Pomeron exchange.

~

~

~

~

~

~

\section{The fit to  the data and its phenomenology }

\subsection{ The main formulae of the fit}
The main formulae, that we use, have been given in  sections 
4.1(\eq{FSE91} and
  4.2
(\eq{SDF},\eq{FSSDGW}.\eq{FSSDLM} and \eq{FSDD4}).
 However, we change these formulae so as  to take into
 account the enhanced diagrams.
 As has been discussed in section 4.1 this procedure reduces
 to the following substitution in our basic formulae:
\bea \label{FI1}
T\Lb Y\Rb\,\mbox{$\Lb\eq{ES11}\Rb$}\,\,&\longrightarrow
 &\,\,G\Big(T\Lb Y\Rb\Big)\mbox{$\Lb\eq{ES1}\Rb$};\nn\\
T_{SD}\Lb Y; Y_m\Rb\mbox{$\Lb\eq{TSD}\Rb$}\,\,&\longrightarrow
&\,\,\frac{1}{\gamma}\,G\Big(T\Lb Y_m\Rb\Big)\,G\Big
(T\Lb Y - Y_m\Rb\Big);
\eea

We also prefer to  use  $g_i$ and $G_{3\pom}$,
instead of $\tilde{g}_i$ and $\gamma$ in
 the main formulae of \eq{FSE91},
considering only $G\Big(T\Lb Y\Rb\Big)$ as a function of $\gamma$.
 Finally, this formula has the form
\beq \label{FI0}
A_{i,k}\Lb Y; b\Rb\,\,\,= \,\,1 \,\,\,-
\,\,\exp\left\{ - \,\,\h\,\Omega^{i,k}_{\pom}\Lb Y; b\Rb\right\}
\eeq
\beq \label{FIMF}
\Omega^{i,k}_{\pom}\Lb Y; b\Rb\,\,\,= \,\,\, \int d^2 b'\,
\,\,\,\frac{ g_i\Lb\vec{b}'\Rb\,g_k\Lb\vec{b} -
 \vec{b}'\Rb\,\Big( 1/\gamma\, G
\Lb T(Y)\Rb\Big)}{ 1\,+\,\Lb
 G_{3\pom}/\gamma\Rb G\Big(T(Y)\Big)\,\left[g_i\Lb
\vec{b}'\Rb + g_k\Lb\vec{b} - \vec{b}'\Rb\right]}
\eeq

Note that $g_i $ in \eq{FIMF} have dimension of inverse momentum (see 
\eq{S})
 as well as $G_{3\pom}$, while $\gamma$ is dimensionless.
 Actually $\gamma^2 \,=\,\int d^2 k \,G^2_{3 \pom}$, but because we do not 
know
 the dependence of $G_{3\pom}$ with respect to transverse momenta of
Pomerons,
  we consider $\gamma$ and $G_{3\pom}$ as independent parameters of the 
fit.

In \eq{SDF} we have to replace all $\tilde{g}_i$ by $g_i$ and multiply it 
by
 factor $G_{3\pom}/\gamma^2$ in addition to the substitution of \eq{FI1}.

For completeness of presentation we list below
 the formulae for physical observables (see Re.\cite{GLMM} for
details).
The amplitudes for the observable processes have the form
\beq \label{EL}
a_{el}(s,b)=
i\{\alpha^4A_{1,1}+2\alpha^2\beta^2A_{1,2}+\beta^4\A_{2,2}\},
\eeq
\beq \label{SD}
a_{sd}(s,b)=
i\alpha\beta\{-\alpha^2A_{1,1}+(\alpha^2-\beta^2)A_{1,2}+\beta^2A_{2,2}\},
\eeq
\beq \label{DD}
a_{dd}=
i\alpha^2\beta^2\{A_{1,1}-2A_{1,2}+A_{2,2}\}.
\eeq`
\par
The corresponding cross sections are given by
\beq \label{XST}
\sigma_{tot}(s)=2\int d^2 b \,a_{el}\Lb s,b\Rb,
\eeq
\beq \label{XSEL}
\sigma_{el}(s)=\int d^2 b \,|a_{el}\Lb s,b\Rb|^2,
\eeq
\beq \label{XSSD}
\sigma_{sd}(s)=\int d^2 b \,|a_{sd}\Lb s,b\Rb|^2,
\eeq
\beq \label{XSDD}
\sigma_{dd}(s)=\int d^2 b \,|a_{dd}\Lb s,b\Rb|^2.
\eeq

\subsection{The strategy of our fitting procedure}
Our Pomeron model, as well as the KMR \cite{KMRS} and Ostapchenko 
\cite{OS} models have the same following  ingredients:

1) A bare non-screened Pomeron exchange amplitude.

2) s-channel unitarity is enforced by eikonal rescatterings of the 
colliding projectiles.

3) These rescatterings proceed through elastic and diffractive states 
according to the G-W mechanism.

4) t-channel unitarity is maintained through the Pomeron interactions. 

In our model the Pomeron is specified by nine  parameters, the Tevatron 
data on its own is not sufficient to determine the parameters. 
Consequently, we have also to include ISR - SP${\bar p}$S/SppS lower 
energy
(W $\approx$ 20-70 GeV) data. This data has small errors which facilitate
a reasonably reliable fit.  To reduce the number of Reggeon
parameters, we define
$\sigma_{tot} = \frac{1}{2}(\sigma_{tot}(pp) + \sigma_{tot}(p{\bar p})$.
The inclusion of the Regge sector of our fit requires five additional 
parameters. i.e. we have fourteen parameters in all.

Our data base has  58 experimental data points,
which include the p-p and $\bar{p}$-p total cross sections,
integrated elastic cross sections, integrated single and double
 diffraction cross sections, $B_{el}$, to which we have added a 
consistency check of the  CDF data
$\frac{d\sigma_{el}}{dt}$ (-t $\leq 0.5 GeV^2)$,
$\frac{d^2 \sigma_{sd}}{dt dM^2/s}$ ( t = 0.05 GeV$^2$) and $B_{sd}$.

The data points were fitted to determine the 14 free parameters of our 
model. We fit simultaneously  the entire data base. The only minor tuning 
which we employ is discussed in section 6.2.

As was mentioned in the previous section,  most of the experimental
 data is   available at low energies ($\sqrt{s} \approx 20 - 70 \,GeV$)
 where the secondary Reggeon contributions are essential.
 this data has small errors.
 We deal with the secondary reggeon in the same way as in Ref.\cite{GLMM} 
adding
 $\Omega_{\reg}$ to $\Omega_\pom$ in \eq{FI0}.

Using the parameters of the fit (see Table 1), we find that
 the contributions
 of the large mass diffraction  to the single diffractive
 cross section as well
 as to the double
 diffractive cross section,  are rather small.
 Therefore, we neglect these contributions
and use \eq{FSSDGW} and \eq{FSDD3} in the fit.
 After determining the parameters of the fit  
 we use them to describe the large mass diffraction.

\FIGURE[ht]{
\begin{tabular}{c c}
\epsfig{file=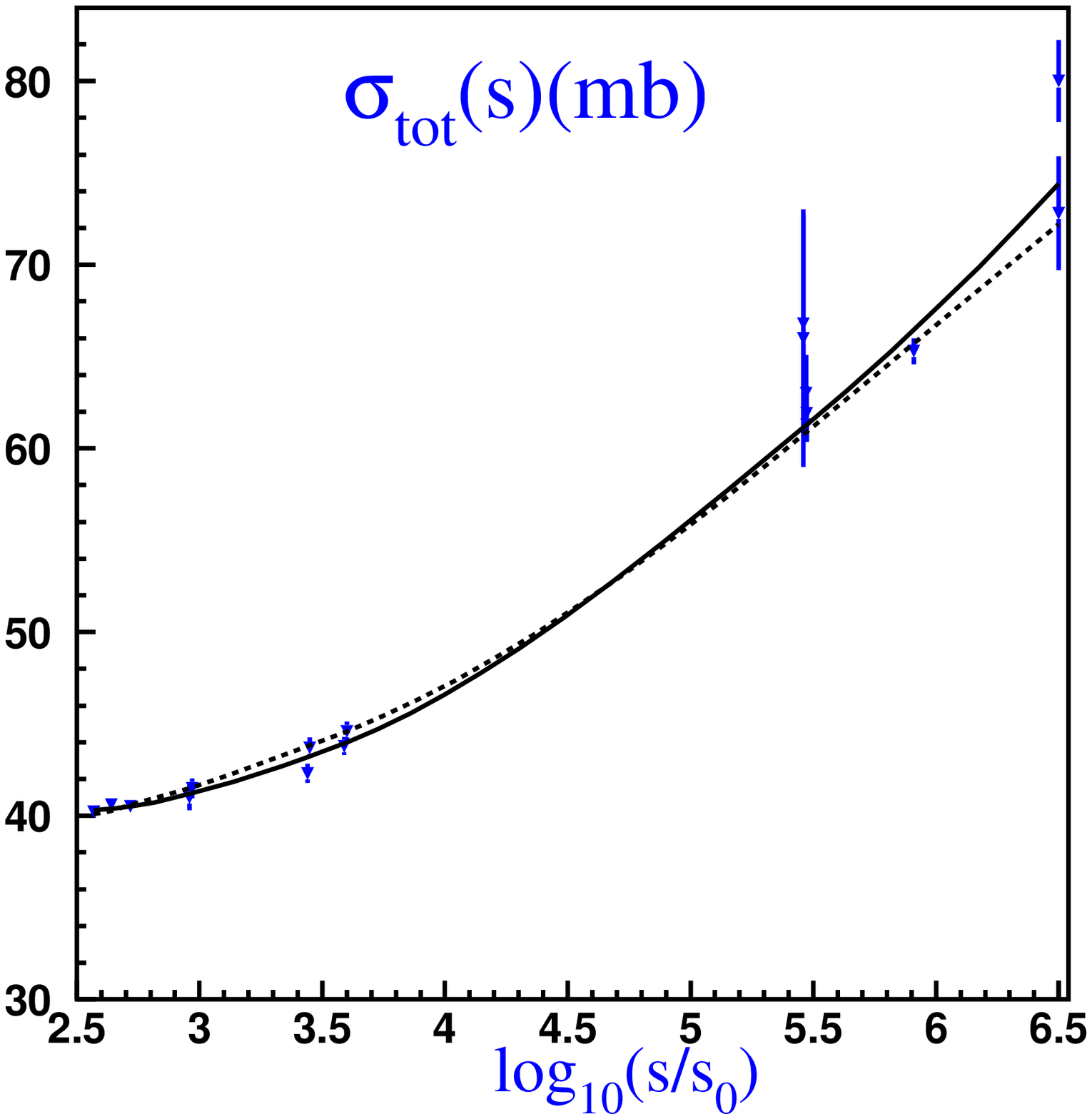,height=50mm,width=65mm}
 &\epsfig{file=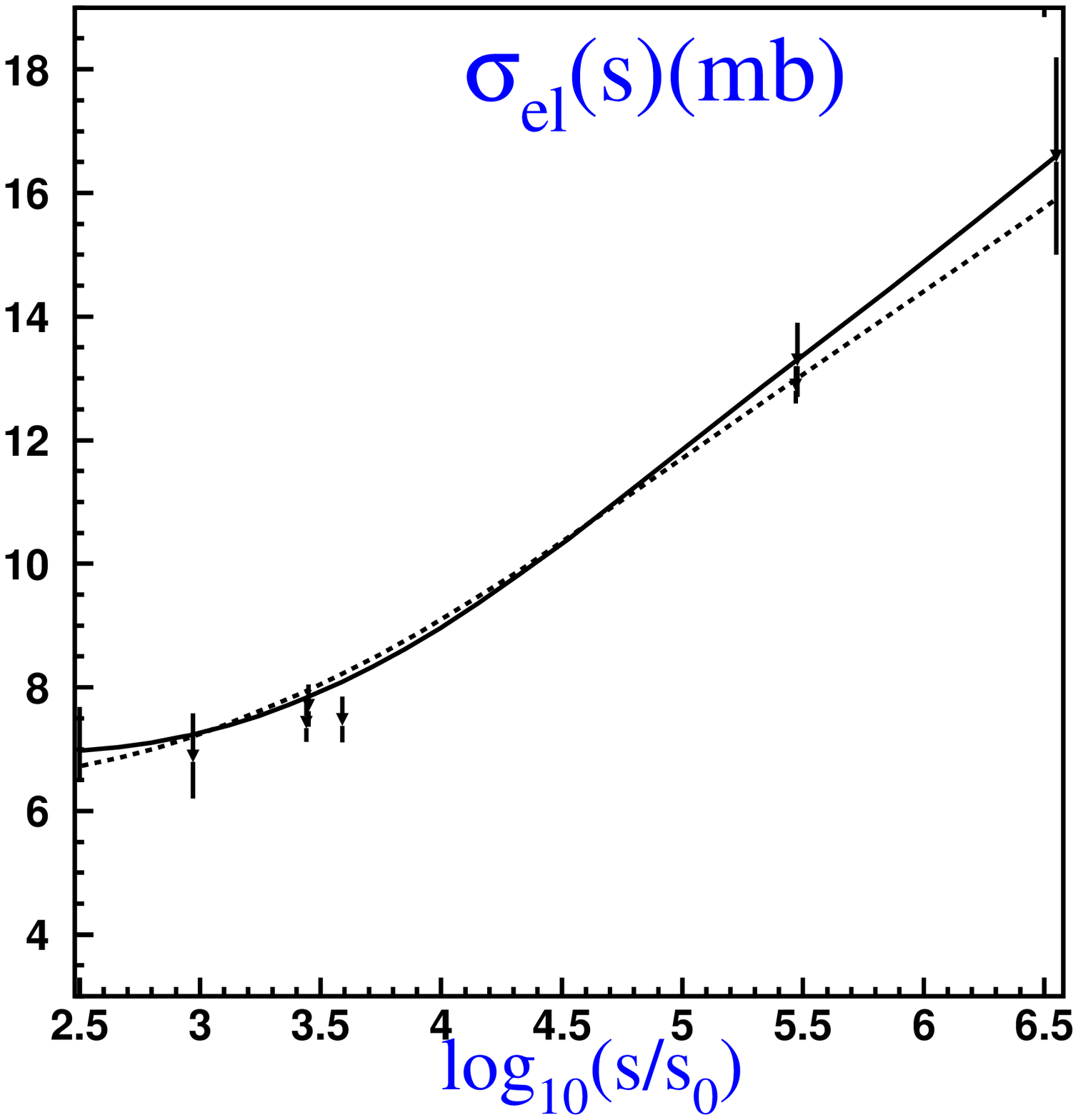,height=50mm,width=65mm}\\
\fig{fit}-a & \fig{fit}-b\\
\epsfig{file=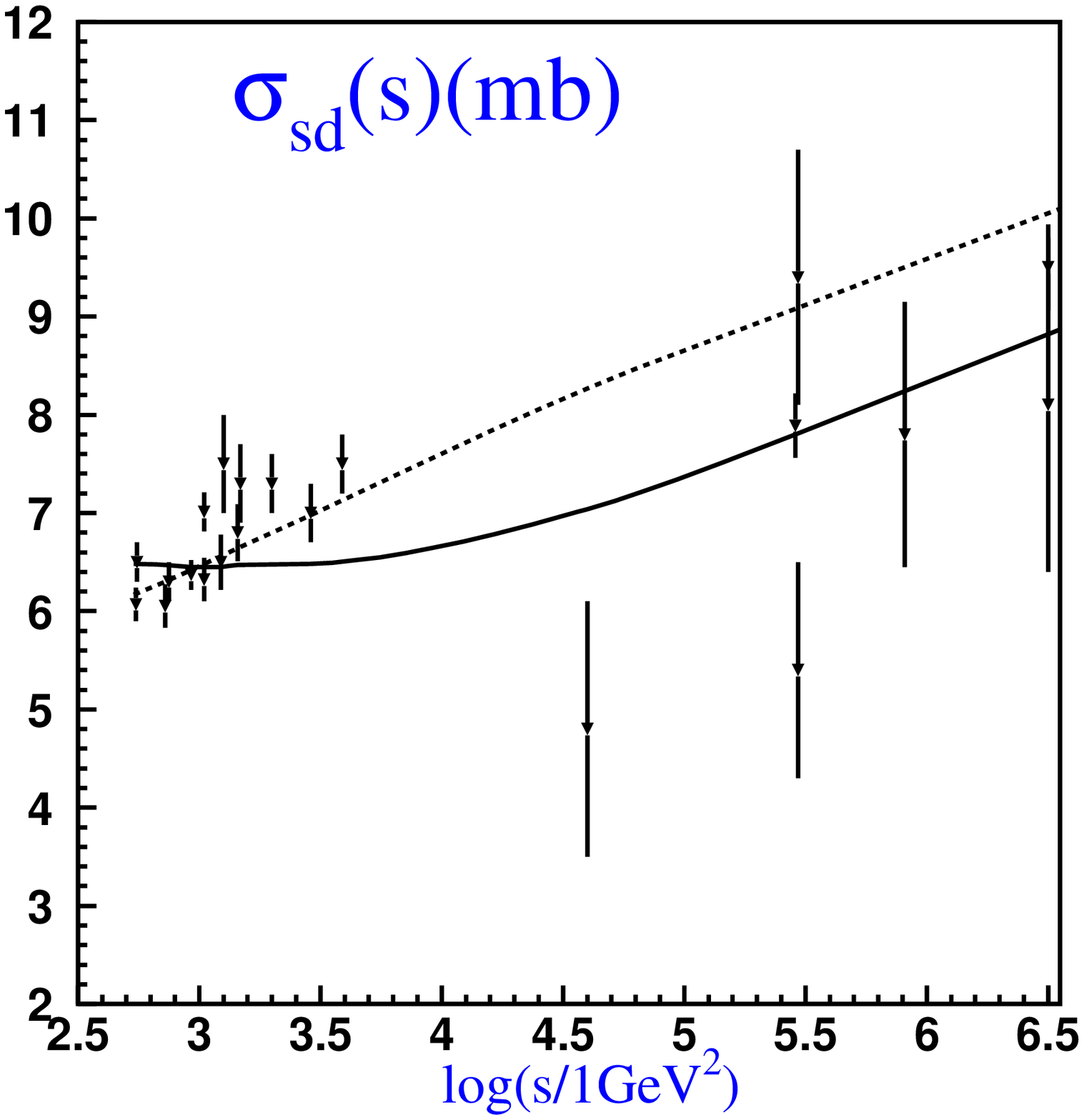,height=50mm,width=65mm}
 &\epsfig{file=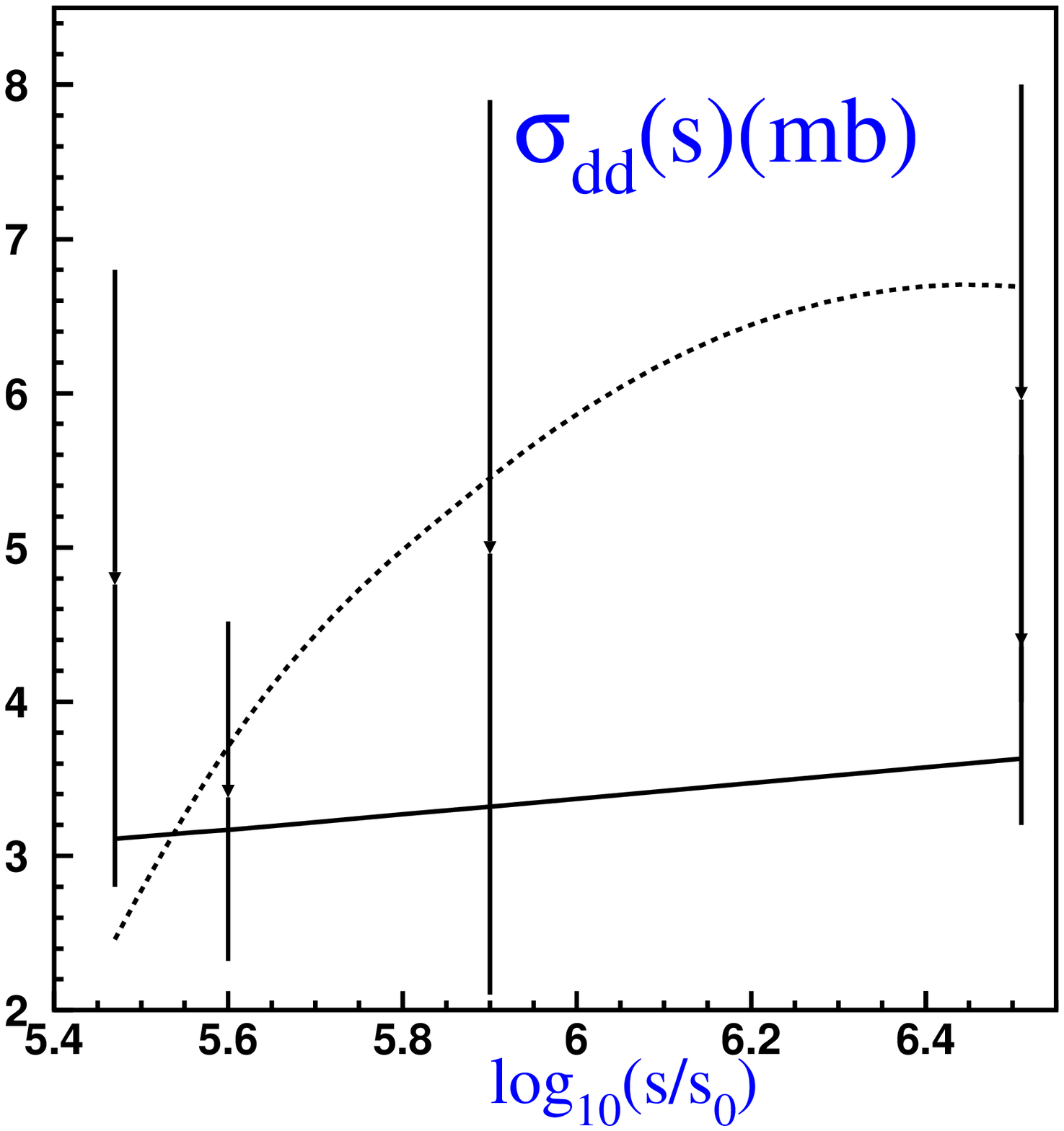,height=50mm,width=65mm}\\
\fig{fit}-c & \fig{fit}-d\\
\epsfig{file=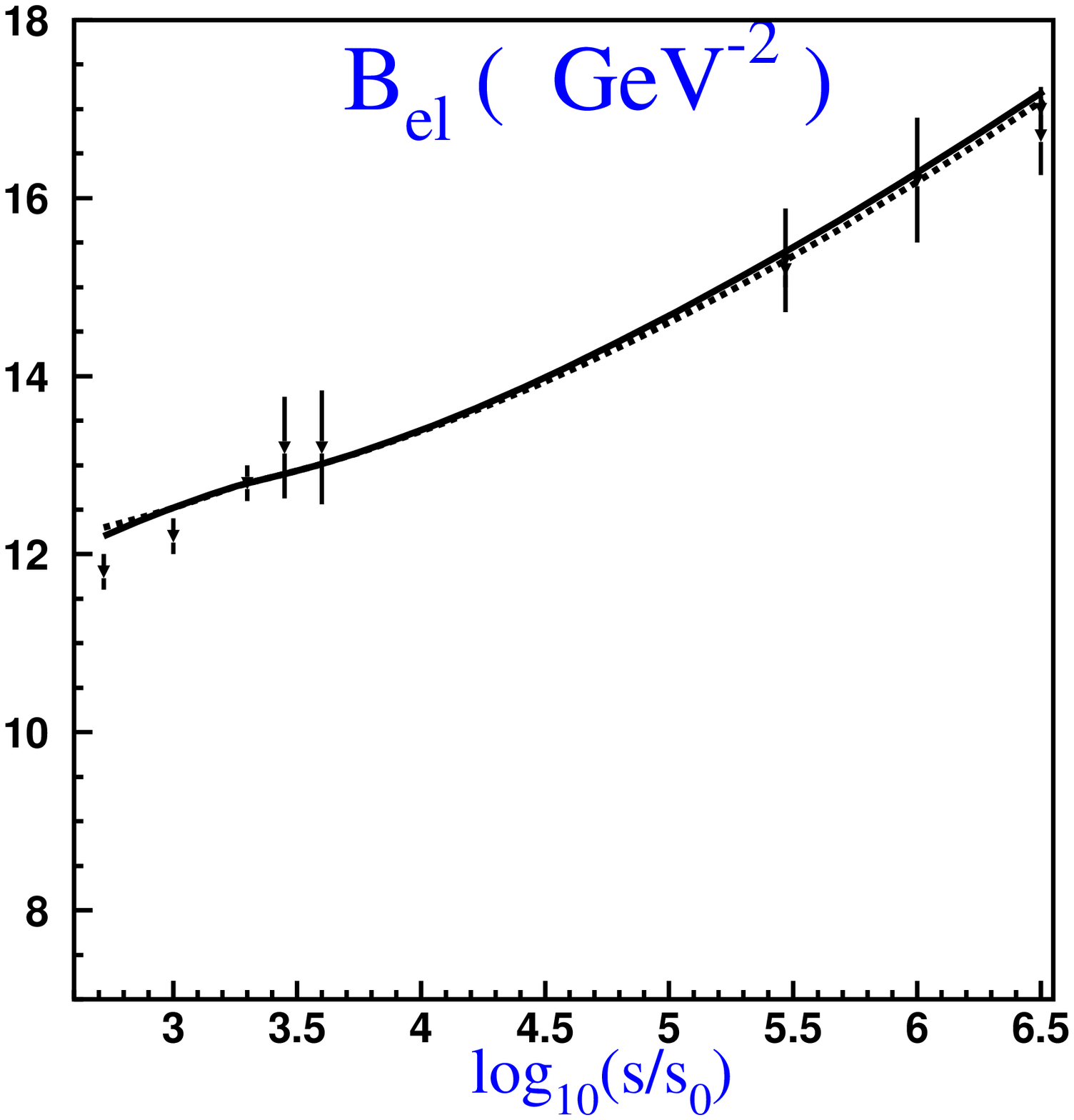,height=50mm,width=65mm}
 &\epsfig{file=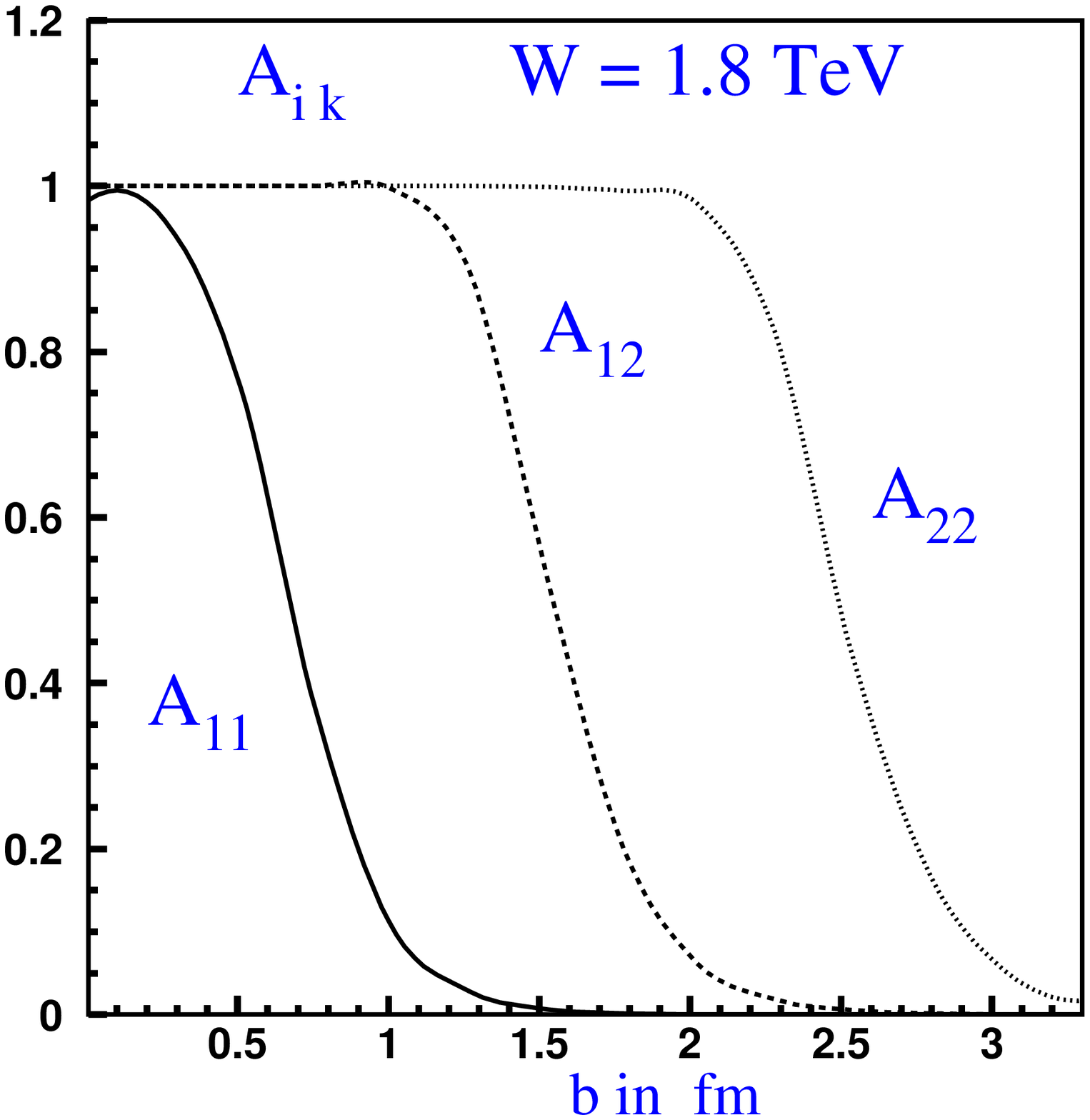,height=50mm,width=65mm}\\
\fig{fit}-e & \fig{fit}-f\\
\end{tabular}
\caption{Comparison with the experimental data the energy behaviour
 of the total (\protect\fig{fit}-a),
elastic  (\protect\fig{fit}-b), single diffraction  (\fig{fit}-c) and
 double diffraction (\protect\fig{fit}-d) cross sections
 and elastic slope( \protect\fig{fit}-e) .
 The solid lines show this fit while the dashed lines correspond
 to the fit of Ref. \protect\cite{GLMM}.
\protect\fig{fit}-f shows the behaviour of the
 amplitude $A_{i,k}$ as function of the impact parameter $b$ for
 the Tevatron energy.
}
\label{fit}}

\subsection{ The results of the fit}

As stated above, our fit is based on 58 experimental data points. 
 The model gives a good reproduction
of the data, with a $\chi^2/d.o.f.$ = 1.56.
 However, a large contribution to the value of $\chi^2/d.o.f.$
 stems from 
 the uncertainty of the
value of two single diffraction cross sections,
 and of the CDF total cross section \cite{CDF} at the Tevatron (W =
1800 $ GeV$ ).
 Neglecting the contribution of these three points to the total $\chi^2$ 
we 
obtain $\chi^2/d.o.f. = 0.86$.
The quality of the description of the experimental data
 is demonstrated in \fig{fit}.
 The values of fitted parameters are listed in Table 1.
An important advantage of our fit is that it provides a good reproduction 
of $\sigma_{dd}$.

In \fig{am} the amplitudes $A_{i k}$ are plotted.
 One can see that in spite of the smallness of $\alpha'_\pom$ we reproduce 
the
growth of the radius of interaction  with energy.
The values of physical observables  for higher energies are shown in Table 
2.

\TABLE[ht]{
\begin{tabular}{|l|l|l|l|l|l|l|}
\hline
$\Delta_\pom $ & $\beta$ &  $\alpha^{\prime}_{\pom}$& $g_1$ &  $g_2$ & $m_1$ &
$m_2$  \\ \hline
0.2 & 0.388 & 0.020 $GeV^{-2}$ &2.53 $GeV^{-1}$ &
88.4 $GeV^{-1}$ & 2.648 $GeV$& 1.37 $GeV$
\\ \hline
$ \Delta_{\reg} $ & $\gamma $ &  $\alpha^{\prime}_{\reg}$& $g^{\reg}_1$ &
$g^{\reg}_2$ & $ R^2_{0,1}$& $G_{3\pom}$ \\ \hline
-\,0.466 & 0.0033 & 0.4 $GeV^{-2}$ & 14.5 $GeV^{-1}$ & 1343 $GeV^{-1}$ &
4.0 $GeV^{-2}$ & 0.0173$GeV^{-1}$ \\ \hline
\end{tabular}
\caption{Fitted parameters for our model. The quality of the fit is $\chi^2/d.o.f.$ = 0.86 (see the detailed explanation in the text)
}
\label{t1}}

\subsection{ Comments on the parameter values of the fit}

  An attractive feature of our fit is that the exceedingly small value of
$\alpha'_{\pom}$ found in the original fit \cite{GLMM}  is reproduced.
However, the values obtained for $\Delta_{\pom}$ and $\gamma$ are smaller 
than our previous values \cite{GLMM}). Our results suggest that the 
complete summation of the Pomeron interaction sector, presented in this 
paper, induces a weaker screeing that was found by the partial summation 
presented in Ref. \cite{GLMM}.
A consequence of this feature is that $S^2_{enh}$ calculated
in this paper is expected to be  larger than the
corresponding values obtained in \cite{GLMM} .

The small value obtained for $\gamma$ = 0.0033 is encouraging, since 
 $\gamma \propto \alpha_{s}^2 $ in QCD, supports our key supposition that 
rather short distances,   
 contribute to the soft interaction at high energy,
 in agreement with all approaches to the Pomeron
 structure considered above.  Note,
 that since $\gamma^2 \,=\,\int d^2 k
\,G^2_{3 \pom}$
we can evaluate the value of the  typical transverse momentum of the 
Pomeron in
 the triple Pomeron vertex, which turns out to be  $\approx$ $1 GeV$.

 $g_2$ is rather large, as in our
 previous approaches (see Ref.\cite{GLMM}).
  The consequence of 
$\frac{g_2}{g_1} \gg 1$  is seen 
in \fig{fit}-f, where the amplitude $A_{i,k}$ are shown at the Tevatron 
energy.
 One can see that amplitude 
$A_{22}$ is equal to one in  a wide region of $b$ 
 and, therefore, 
corresponds to black disc scattering.
 However, its contribution is proportional to $\beta^4$ in elastic
 amplitude and 
 to $\beta^2$ in single and double diffraction amplitudes, 
consequently, it's relative contribution is  small.

It should be stressed  that the value of our phenomenological
 parameters (see Table 1) are in
 agreement with the theoretical estimates of \eq{EST}. 
 Choosing the typical soft scale $\mu = 1\,GeV$ we can see
 that $g_i \mu \approx 1$ and 
\beq \label{FI3}
G_{3\pom}\mu  \approx \gamma \,\approx\,\Delta^2_\pom \,\ll\,\Delta_\pom \,\ll\,g_i \mu
\eeq

As we have discussed, in this paper we sum all diagrams in an
 approximation in which $g_i G\Big(T\Lb 
Y\Rb\Big) \,\geq\,1$ while $\Delta^2_\pom  G\Big(T\Lb Y\Rb\Big) \,\,\ll\,\,1$.
 The values of the fit parameters support
  the use of the   approximation.

In \fig{am} the amplitudes $A_{i k}$ are plotted.

\FIGURE[ht]{
\begin{tabular}{c c c}
\epsfig{file=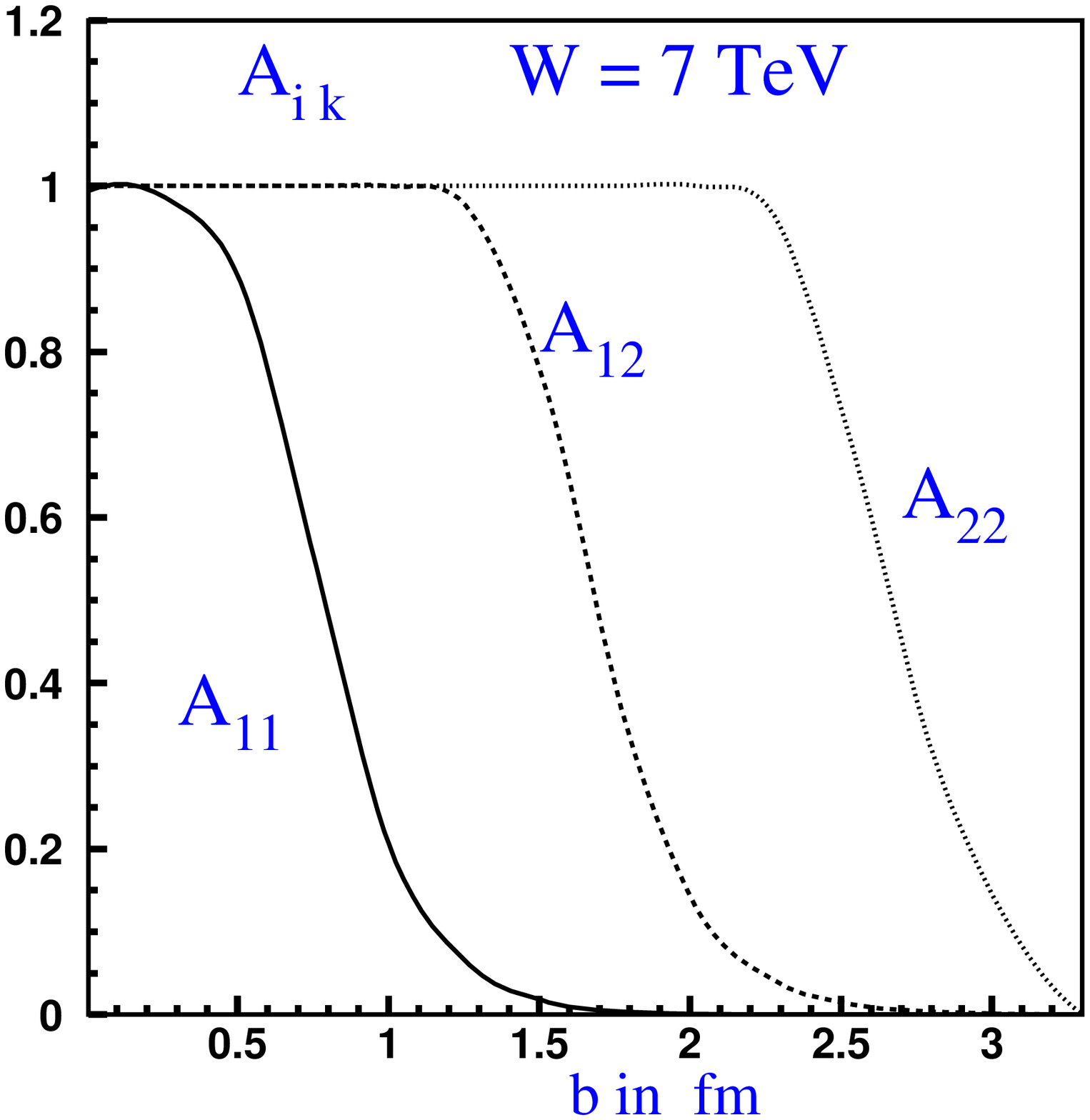,height=45mm,width=57mm} 
&\epsfig{file=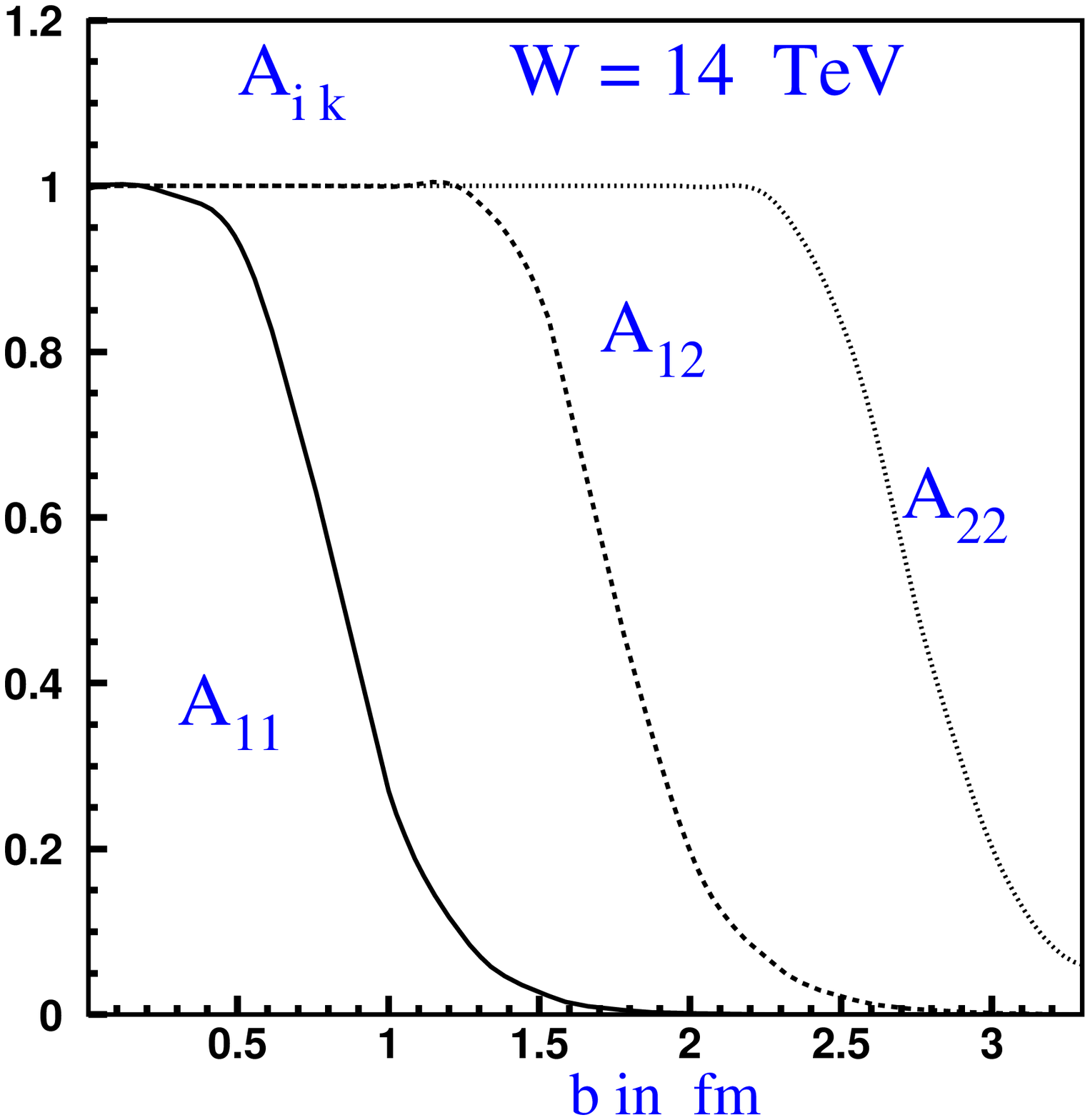,height=45mm,width=57mm}&
\epsfig{file=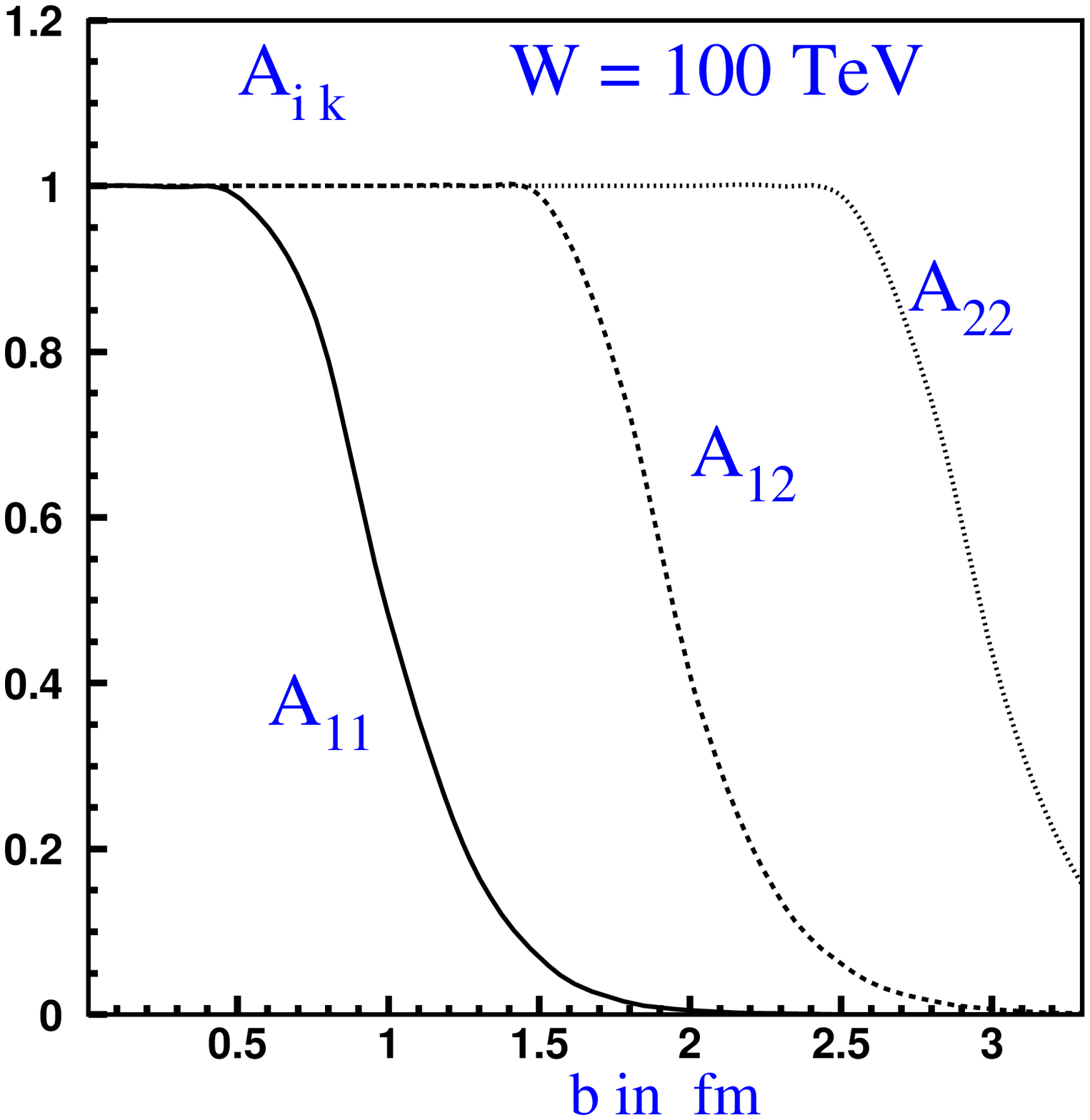,height=45mm,width=57mm}\\
\fig{am}-a & \fig{am}-b & \fig{am}-c \\
\end{tabular}
\caption{The  amplitudes $A_{i k}$ versus $b$ at  LHC energies.
}
\label{am}}

The values of physical observables  for higher energies are shown in 
Table 2.

\TABLE[ht]{
\begin{tabular}{|l|l|l|l|l|l|}
\hline
 W(TeV) & $\sigma_{tot}(mb)$ &  $\sigma_{el}(mb) $& $\sigma_{sd}(mb) $&  $\sigma_{dd}(mb)$ & $B_{el} (GeV^{-1})$  \\ \hline
0.9 & 66.3 & 15 & 8.24 &3.83 & 16.1\\ \hline
1.8 & 74.4 & 17.5 & 8.87 & 4.46 & 17.2\\ \hline
2.3 & 77.3 & 18.4 & 9.1& 4.59 & 17.5 \\ \hline
3.5 & 82.5 & 20.1 & 9.49 & 4.72& 18.2 \\ \hline
5    & 87   & 21.6 & 9.83 & 5.59& 18.8\\ \hline
7    & 91.3 & 23   & 10.2 & 6.46 & 19.3 \\ \hline
10  & 96    & 24.6 & 10.5 & 6.48 & 19.9  \\ \hline
14   & 101 & 26.1 & 10.8 &  6.5  & 20.5\\ \hline
100  & 128 & 35.6 & 12.7 & 7.79  & 29.9\\\hline
\end{tabular}
\caption{  Predictions for  energies that will be accessible at the LHC.}
\label{t2}}

\subsection{Comparison with other approaches.}

At present there are  three groups that are working on modelling 
the strong interaction at high energies: 
 Durham group\cite{KMRS}, Ostapchenko\cite{OS} and our group.
 The models have a lot in common: a rather large value of $\Delta_\pom$,
 small values of $\alpha'_\pom$,  a large contribution of the Good-Walker
 mechanism, 
 and a significant Pomeron-Pomeron interaction.
 The theories  mainly differ in the way  the Pomeron-Pomeron 
interaction is taken into account.
 In the Durham model as well as in the Ostapchenko one,
  all vertices of Pomeron-Pomeron interactions are 
 included,  using a slightly different phenomenological anzats.
 In our model we include only the triple Pomeron vertex, as has been 
discussed in section 1. 
In the Durham and in our models $\Delta_\pom \approx 0.2\div 0.3$ and 
$\alpha'_\pom \approx 0$, while
the  Ostapchenko model gives $ 
\Delta_\pom = 0.14$ and $\alpha'_\pom = 0.14\,GeV^{-1}$.
  All  models give more or less the same results at the Tevatron energy,
 and their results 
 at the highest LHC  energy: span the values
  $\sigma_{tot} = 86 \div 114 \,mb$,$\sigma_{el} = 20 \div  30\,mb$,
$\sigma_{sd} = 10 \div 16 mb$ and $\sigma_{dd} = 5 \div 13 mb$.

Note that the Ostapchenko model also includes a hard Pomeron, while our 
model has only a single Pomeron. The KMR model is more complicated as
 the Pomeron dependence on $k_t^2$ is described by three Pomerons.
 As one can see from  Table 3, the Ostapcheko model leads
 to the largest value for the total and elastic cross 
sections,
 and gives small values for the double diffractive cross section.
 The Durham model predicts the lowest value of the total cross section
 but a rather large double diffraction cross section.

\TABLE[ht]{
\begin{tabular}{|l|l|l|}
\hline
& \,\,\,\,\,\,\,\,\,\,\,\,\,\,\,\,\,\,\,Tevatron (1.8 TeV)
& \,\,\,\,\,\,\,\,\,\,\,\,\,\,\,\,\,\,\,\,\,\, LHC (14 TeV)
  \\
& GLMM\,\,GLM\,\,KMR(07)\,\,KMR(10)\,\,\,\,OS(C)
& GLMM\,\,GLM\,\,KMR(07)\,\,KMR(10)\,\,\,\,\,\,\,OS(C)
 \\
\hline
$\sigma_{tot}$(mb)
& \,\,73.29\,\,\,\,\,\,\,\,\,74.4 \,\,\,\,\,\,\,\,\,\,74.0\,\,\,\,\,\,\,\,\,\,\,\,\,\,73.9
\,\,\,\,\,\,\,\,\, 73.0 & \,\,\,\,92.1\,\,\,\,\,\,\,\,\,101
\,\,\,\,\,\,\,\,\,\,\,88.0\,\,\,\,\,\,\,\,\,\,\,\,\,\,\,86.3
\,\,\,\,\,\,\,\,\,\, 114.0 \\
\hline
$\sigma_{el}$(mb)
& \,\, 16.3\,\,\,\,\,\,\,\,\,\,17.5 \,\,\,\,\,\,\,\,\,\,16.3\,\,\,\,\,\,\,\,\,\,\,\,\,\,15.1
\,\,\,\,\,\,\,\,\,\, 16.8 & \,\,\,\,20.9  \,\,\,\,\,\,\,\,\,26.1
\,\,\,\,\,\,\,\,\,\,\,20.1\,\,\,\,\,\,\,\,\,\,\,\,\,18.1
\,\,\,\,\,\,\,\,\,\, 33.0
\\
\hline
$\sigma_{sd}$(mb) 
& \,\,\,\,\,9.76\,\,\,\,\,\,\,\,\,\, 8.87 \,\,\,\,\,\,\,\,10.9\,\,\,\,\,\,\,\,\,\,\,\,12.7
\,\,\,\,\,\,\,\,\,\,\,\,\,9.6 &\,\,\,\,11.8\,\,\,\,\,\,\,\,\,\,10.8
\,\,\,\,\,\,\,\,\,\,\,13.3\,\,\,\,\,\,\,\,\,\,\,\,\,\,\,16.1
\,\,\,\,\,\,\,\,\,\, 11.0  \\
\hline
$\sigma_{dd}$(mb) & \,\,\,\,\,5.36\,\,\,\,\,\,\,\,\,\, 3.53\,\,\,\,\,\,\,\,\,\,\,\,\,\,7.2
\,\,\,\,\,\,\,\,\,\,13.3\,\,\,\,\,\,\,\,\,\,\,\,\,\,3.93
&
\,\,\,\,\,\,\,6.1\,\,\,\,\,\,\,\,\,\,6.5
\,\,\,\,\,\,\,\,\,\,\,\,13.4\,\,\,\,\,\,\,\,\,\,\,\,\,\,\,\,12.9
\,\,\,\,\,\,\,\,\,\,\,\,\,\,\,4.83 \\
\hline 
\end{tabular}
\caption{  Comparison with the other models: GLMM is our model in which 
we summed only enhanced diagrams\cite{GLMM},
 GLM is the model  described in this paper, 
  KMR(07) and KMR(10) are two models of Durham group
 ((Ref. \cite{KMRS}) and OS(C)
is the model developed in Ref.\cite{OS}.
The predictions of the fit of 2010 is preliminary
 and are taken from the talk of A.Martin at Diffraction'10.}
\label{t3}}

~

~

~
\section{Conclusions}

In this paper we have constructed  a model to describe strong
 interactions at high energy, based on two main 
theoretical  criteria:
 it should include the main result of N=4 SYM,
 which is the only theory that is able to deal with  a
large coupling 
constant; and it should provide the natural matching with high enerrgy QCD.
  In accord with these ideas we assumed
 that $\Delta_\pom$ is relatively large and $\alpha'_\pom \to 0$.
  Using these assumptions  in this paper we sum  all 
enhanced and semi-enhanced diagrams.
 The enhanced diagrams have been calculated in our previous paper \cite{GLMM}.
 For 
the first time we obtain   analytical formulae for the scattering 
amplitude both for elastic
 scattering and for diffractive
 production.

Using these formulae we made a fit to the available experimental data
 and predict the main soft observables at  
LHC energies (see Tables 2 and 3).

We study the accuracy of our approach related to the corrections 
induced by small (both not equal to zero 
$\alpha'_\pom$) and by the fact that the `bare' Pomeron cannot
 be a Regge pole, but  a cut.

 In this paper we  have completed the formulation of a theoretical
 self consistent approach.
 In the future we intend  to apply this approach to
 the numerous practical problems,
 such as survival probability for di-jet and Higgs production,
 inclusive cross sections,
 rapidity and multiplicity correlations.

~

~

~

\section* {Acknowledgements}
We thank all participants of the conference ``Diffraction 2010" for 
fruitful discussion on the subject.
The research of one of us (E.L.)  was supported
in part  by the  Fondecyt (Chile) grant  \# 1100648.



\begin{thebibliography}{99}
\bibitem{COL}
P.D.B. Collins, {\it "An introduction to Regge theory and high energy physics"}, 
Cambridge University Press 1977.
\bibitem{SOFT}
Luca Caneschi (editor), {\it "Regge Theory of Low -$p_T$ Hadronic Interaction"}, 
North-Holland 1989.
\bibitem{LEREG}
E. Levin,
{\it "An introduction to Pomerons"}, arXiv:hep-ph/9808486; 
{\it "Everything about Reggeons. I: Reggeons in *soft* interaction"}, 
arXiv:hep-ph/9710546.


\bibitem{ATMP}
V. A. Khoze, A. D. Martin and M. G. Ryskin,
Eur. Phys. J. {\bf C18} (2000) 167;
Phys. Lett. {\bf B643} (2006) 93;\\
A. B. Kaidalov, V. A. Khoze, A. D. Martin and M. G. Ryskin,
Eur. Phys. J. {\bf C31} (2003) 387; {\bf C33} (2004) 261.
E. Gotsman, E. Levin and U. Maor,
Phys. Lett. {\bf B452} (1999) 387; {\bf B309} (1993) 199;
Phys. Rev. {\bf D49} (1994) R4321;\,\,\,
  A.~Capella, U.~Sukhatme, C-I~Tan {\it et al.}\,\,\,
  Phys.\ Rept.\  {\bf 236}, 225-329 (1994).
  \,\,\,A.~B.~Kaidalov,
  Phys.\ Rept.\  {\bf 50} (1979) 157;
\,\,\, A.~B.~Kaidalov, L.~A.~Ponomarev and K.~A.~Ter-Martirosian,
  Yad.\ Fiz.\  {\bf 44} (1986) 722
  [Sov.\ J.\ Nucl.\ Phys.\  {\bf 44} (1986) 468];
\,\,\, A.~B.~Kaidalov and K.~A.~Ter-Martirosyan,
  Nucl.\ Phys.\  B {\bf 75} (1974) 471.

\bibitem{2CH}
E. Gotsman, E. Levin and U. Maor,
Phys. Lett. {\bf B452} (1999) 387; {\bf B309} (1993) 199;
Phys. Rev. {\bf D49} (1994) R4321.
\bibitem{GLMLAST}
 E.~Gotsman, E.~Levin, U.~Maor,
  Phys.\ Rev.\  {\bf D81 } (2010)  051501  [arXiv:1001.5157 [hep-ph]];\,\,\,
  Braz.\ J.\ Phys.\  {\bf 38 } (2008)  431-436[arXiv:0805.0418 [hep-ph];\,\,\,
{\it "A Soft Interaction Model at Ultra High Energies: Amplitudes, Cross Sections
and Survival Probabilities"}, arXiv:0708.1506 [hep-ph];\,\,
\,
 E.~Gotsman, A.~Kormilitzin, E.~Levin {\it et al.},
  Eur.\ Phys.\ J.\  {\bf C52 } (2007)  295-304
  [arXiv:1001.5157 [hep-ph]].

\bibitem{GLMM}
 E.~Gotsman, E.~Levin, U.~Maor and J.~S.~Miller,
  Eur.\ Phys.\ J.\  {\bf C57 } (2008)  689-709.
  [arXiv:0805.2799 [hep-ph]].
\bibitem{GLMA}
E.~Gotsman, A.~Kormilitzin, E.~Levin and U.~Maor,
  Nucl.\ Phys.\  A {\bf 842} (2010) 82
  [arXiv:0912.4689 [hep-ph]].
\bibitem{KMRS}
 M.~G.~Ryskin, A.~D.~Martin, V.~A.~Khoze {\it et al.},
  J.\ Phys.\ G {\bf G36 } (2009)  093001
  [arXiv:0907.1374 [hep-ph]];\,\,\,
  Eur.\ Phys.\ J.\  {\bf C60 } (2009)  265-272.
  [arXiv:0812.2413 [hep-ph]];\,\,\,
  Eur.\ Phys.\ J.\  {\bf C60 } (2009)  249-264.
  [arXiv:0812.2407 [hep-ph]];\,\,\,
  AIP Conf.\ Proc.\  {\bf 1105 } (2009)  252-257.
  [arXiv:0811.1481 [hep-ph]];\,\,\,
  {\it ``Soft Diffraction at the LHC,''}
  arXiv:0810.3324 [hep-ph];\,\,{\it ``Rapidity gap survival probability and total cross sections,''}
  arXiv:0810.3560 [hep-ph];\,\,
Eur. Phys. J. {\bf C54} (2008) 199 [arXiv:0710.2494 [hep-ph]];\,\,\,

\bibitem{OS}
S.~Ostapchenko,
  Phys.\ Rev.\  D {\bf 81} (2010) 114028
  [arXiv:1003.0196 [hep-ph]];\,\,\,
  Phys.\ Rev.\  D {\bf 77} (2008) 034009
  [arXiv:hep-ph/0612175];\,\,\,
  Phys.\ Lett.\  B {\bf 636} (2006) 40
  [arXiv:hep-ph/0602139].
\bibitem{AdS-CFT}
 J.~M.~Maldacena,
  Adv.\ Theor.\ Math.\ Phys.\  {\bf 2} (1998) 231
  [Int.\ J.\ Theor.\ Phys.\  {\bf 38} (1999) 1113]
  [arXiv:hep-th/9711200];\,\,\,
S.~S.~Gubser, I.~R.~Klebanov and A.~M.~Polyakov,
  Phys.\ Lett.\  B {\bf 428} (1998) 105
  [arXiv:hep-th/9802109];\,\,\,
E.~Witten,
  Adv.\ Theor.\ Math.\ Phys.\  {\bf 2} (1998) 505
  [arXiv:hep-th/9803131].

\bibitem{POST}
 J.~Polchinski and M.~J.~Strassler,
  JHEP {\bf 0305} (2003) 012
  [arXiv:hep-th/0209211];\,\,
  Phys.\ Rev.\ Lett.\  {\bf 88} (2002) 031601
  [arXiv:hep-th/0109174].
\bibitem{BFKL4}
 A.~V.~Kotikov, L.~N.~Lipatov, A.~I.~Onishchenko and V.~N.~Velizhanin,
  Phys.\ Lett.\  B {\bf 595} (2004) 521
  [Erratum-ibid.\  B {\bf 632} (2006) 754]
  [arXiv:hep-th/0404092];\,\,
A.~V.~Kotikov and L.~N.~Lipatov,
  Nucl.\ Phys.\  B {\bf 661} (2003) 19
  [Erratum-ibid.\  B {\bf 685} (2004) 405]
  [arXiv:hep-ph/0208220];\,\,
  A.~V.~Kotikov and L.~N.~Lipatov,
  Nucl.\ Phys.\  B {\bf 582} (2000) 19
  [arXiv:hep-ph/0004008].

\bibitem{BST}
 R.~C.~Brower, J.~Polchinski, M.~J.~Strassler and C.~I.~Tan,
  JHEP {\bf 0712} (2007) 005
  [arXiv:hep-th/0603115];\,\,
R. C. Brower, M. J. Strassler and C. I. Tan,
  JHEP {\bf 0903 } (2009)  092
  [arXiv:0710.4378 [hep-th]];\,\,\, 
  JHEP {\bf 0903 } (2009)  050.
  [arXiv:0707.2408 [hep-th]].

\bibitem{HIM}
Y.~Hatta, E.~Iancu and A.~H.~Mueller,
  JHEP {\bf 0801} (2008) 026
  [arXiv:0710.2148 [hep-th]].
  \bibitem{COCO}
 L.~Cornalba and M.~S.~Costa,
 Phys. Rev. {\bf D 78}, (2008) 09010,
  arXiv:0804.1562 [hep-ph];\,\,\,
  L.~Cornalba, M.~S.~Costa and J.~Penedones,
  JHEP {\bf 0806} (2008) 048
  [arXiv:0801.3002 [hep-th]].
\bibitem{BEPI}
B.~Pire, C.~Roiesnel, L.~Szymanowski and S.~Wallon,
  Phys.\ Lett.\  B {\bf 670}, 84 (2008)
  [arXiv:0805.4346 [hep-ph]].
\bibitem{LMKS}
E.~Levin, J.~Miller, B.~Z.~Kopeliovich and I.~Schmidt,
  JHEP {\bf 0902}, 048 (2009)
  [arXiv:0811.3586 [hep-ph]].

\bibitem{FEYN}
R. P. Feynman,
Phys. Rev. Lett. {\bf 23} (1969) 1415;
{\it "Photon-Hadron Interactions"},
Reading 1972, p282
\bibitem{GRIB}
V. N. Gribov,
{\it "Space-time description of hadron interactions at high energies"}, 
arXiv:hep-ph/0006158;\\
Sov. J. Nucl. Phys. {\bf 9} (1969) 369 
[Yad. Fiz. {\bf 9} (1969) 640].
\bibitem{BJ}
J.~D.~Bjorken and E.~A.~Paschos,
  Phys.\ Rev.\  {\bf 185}, (1969) 1975.


\bibitem{LONU}
F. E. Low,
Phys. Rev. {\bf D12} (1975) 163.
S. Nussinov,
Phys. Rev. Lett. {\bf 34} (1975) 1286.
\bibitem{BFKL}
 E. A. Kuraev, L. N. Lipatov, and F. S. Fadin, {\it  Sov. Phys.
JETP}
                {\bf 45}, 199 (1977); \,\,\,
Ya. Ya. Balitsky and L. N. Lipatov,
               {\it   Sov. J. Nucl. Phys.}\, {\bf 28}, 22 (1978).

\bibitem{LI} 
L. N. Lipatov,
Phys. Rep. {\bf 286} (1997) 131; Sov. Phys. JETP {\bf 63} (1986) 904 
and references therein. 

\bibitem{GLR}
L. V. Gribov, E. M. Levin and M. G. Ryskin, 
Phys. Rep. {\bf 100} (1983) 1. 
\bibitem{MUQI}
A. H. Mueller and J. Qiu, 
Nucl. Phys. {\bf B268} (1986) 427.
\bibitem{MV}
L. McLerran and R. Venugopalan, 
Phys. Rev. {\bf D49} (1994) 2233, 3352; {\bf D50} (1994) 2225; 
{\bf D53} (1996) 458;\\ {\bf D59} (1999) 09400. 

\bibitem{B}
I.~Balitsky,
[arXiv:hep-ph/9509348];\,\,
{\it Phys.\ Rev.} {\bf D60}, 014020 (1999)
[arXiv:hep-ph/9812311]\,\,\,\,


\bibitem{K}
Y.~V.~Kovchegov,
{\it Phys.\ Rev.}  {\bf D60}, 034008  (1999),
[arXiv:hep-ph/9901281].
\bibitem{JIMWLK}
~J.~Jalilian-Marian, A.~Kovner, A.~Leonidov and H.~Weigert,
{\it  Phys.\ Rev.}\,  {\bf D59}, 014014 (1999),
[arXiv:hep-ph/9706377];\,\,  {\it Nucl.\ Phys.}\,{\bf B504}, 415
(1997),
[arXiv:hep-ph/9701284]; \,\,\,
J.~Jalilian-Marian, A.~Kovner and H.~Weigert,
  {\it Phys.\ Rev.}  {\bf D59}, 014015 (1999),
  [arXiv:hep-ph/9709432];\,\,\,
 A.~Kovner, J.~G.~Milhano and H.~Weigert,
 {\it  Phys.\ Rev.}  {\bf D62}, 114005 (2000),
  [arXiv:hep-ph/0004014]\,; \,\,\,
E.~Iancu, A.~Leonidov and L.~D.~McLerran,
{\it  Phys.\ Lett.}\,  {\bf B510}, 133 (2001);
[arXiv:hep-ph/0102009];\,\, {\it  Nucl.\ Phys.}\,  {\bf A692}, 583
(2001),
[arXiv:hep-ph/0011241];\,\,\,
E.~Ferreiro, E.~Iancu, A.~Leonidov and L.~McLerran,
 {\it  Nucl.\ Phys.}\  {\bf A703}, 489 (2002),
  [arXiv:hep-ph/0109115];\,\,\,
H.~Weigert,
{\it  Nucl.\ Phys.}  {\bf A703}, 823 (2002),
[arXiv:hep-ph/0004044].
\bibitem{BRN}
M. A. Braun,
Phys. Lett. {\bf B632} (2006) 297 [arXiv:hep-ph/0512057]; 
Eur. Phys. J. {\bf C16} (2000) 337 [arXiv:hep-ph/0001268]; 
Phys. Lett. {\bf B483} (2000) 115 [arXiv:hep-ph/0003004]; 
Eur. Phys. J. {\bf C33} (2004) 113 [arXiv:hep-ph/0309293]; 
{\bf C6}, 321 (1999) [arXiv:hep-ph/9706373]. 
M. A. Braun and G. P. Vacca,
Eur. Phys. J. {\bf C6} (1999) 147 [arXiv:hep-ph/9711486].
\bibitem{BART}
J. Bartels, M. Braun and G. P. Vacca,
Eur. Phys. J. {\bf C40} (2005) 419 [arXiv:hep-ph/0412218].
J. Bartels and C. Ewerz,
JHEP {\bf 9909} 026 (1999) [arXiv:hep-ph/9908454]. 
J. Bartels and M. Wusthoff,
Z. Phys. {\bf C6}, (1995) 157. 
A. H. Mueller and B. Patel,
Nucl. Phys. {\bf B425} (1994) 471 [arXiv:hep-ph/9403256].
J. Bartels,
Z. Phys. {\bf C60} (1993) 471. 
\bibitem{LEPO}
E.~Levin and I.~Potashnikova,
  JHEP {\bf 1008 } (2010)  112;
  [arXiv:1007.0306 [hep-ph]].
\bibitem{MUCD}
A. H. Mueller,
Nucl. Phys. {\bf B415} (1994) 373; {\bf B437} (1995) 107.
\bibitem{LALE}
E. Laenen and E. Levin,
Nucl. Phys. {\bf B451} (1995) 207.
\bibitem{LELU}
E. Levin and M. Lublinsky,
  Nucl.\ Phys.\  A {\bf 763} (2005) 172
  [arXiv:hep-ph/0501173];\,\,
  Phys.\ Lett.\  B {\bf 607} (2005) 131
  [arXiv:hep-ph/0411121];\,\, 
  Nucl.\ Phys.\  A {\bf 730} (2004) 191
  [arXiv:hep-ph/0308279].
\bibitem{KLP}
M. Kozlov, E. Levin and A. Prygarin,
Nucl. Phys. {\bf A792} (2007) 122 [arXiv:0704.2124 [hep-ph]].


\bibitem{LT}
E.~Levin and K.~Tuchin,
{\it Nucl.\ Phys.}\  {\bf A693} (2001) 787
[arXiv:hep-ph/0101275];\,\,\,{\bf A691} (2001) 779
[arXiv:hep-ph/0012167];\,\,\,{\bf B573} (2000) 833
[arXiv:hep-ph/9908317].

\bibitem{NS}
N.~Armesto and M.~A.~Braun,
  {\it Eur.\ Phys.\ J.}\  {\bf C20}, 517 (2001)
  [arXiv:hep-ph/0104038];\,\,
M.~Lublinsky,
  {\it Eur.\ Phys.\ J.}\  {\bf C21}, 513 (2001)
  [arXiv:hep-ph/0106112];\,\,\,\,
E.~Levin and M.~Lublinsky,
 {\it   Nucl.\ Phys.}  {\bf A712}, 95 (2002)
  [arXiv:hep-ph/0207374];\,\,
  {\it Nucl.\ Phys.}  {\bf A712}, 95 (2002)
  [arXiv:hep-ph/0207374];\,\,
{\it Eur.\ Phys.\ J.}\, {\bf C22}, 647 (2002)
  [arXiv:hep-ph/0108239];\,\,\,\,
M.~Lublinsky, E.~Gotsman, E.~Levin and U.~Maor,
 {\it   Nucl.\ Phys.}\,  {\bf A696}, 851 (2001)
  [arXiv:hep-ph/0102321];\,\,
 {\it  Eur.\ Phys.\ J.}\,  {\bf C27}, 411 (2003)
  [arXiv:hep-ph/0209074];\,\,\,\,
K.~Golec-Biernat, L.~Motyka and A.Stasto,
 {\it  Phys.\ Rev.} {\bf D65}, 074037 (2002)
  [arXiv:hep-ph/0110325];\,\,\,
E.~Iancu, K.~Itakura and S.~Munier, {\it
  Phys.\ Lett.}\, {\bf B590} (2004) 199
  [arXiv:hep-ph/0310338].
K.~Rummukainen and H.~Weigert,
{\it   Nucl.\ Phys.} \, {\bf A739}, 183 (2004)
  [arXiv:hep-ph/0309306];\,\,K.~Golec-Biernat and A.~M.~Stasto,
 {\it  Nucl.\ Phys.} {\bf B668}, 345 (2003)
  [arXiv:hep-ph/0306279];\,\,\,\,E.~Gotsman, M.~Kozlov, E.~Levin, U.~Maor and E.~Naftali,
 {\it   Nucl.\ Phys.}\, {\bf A742}, 55 (2004)
  [arXiv:hep-ph/0401021];\,\,\,\,K.~Kutak and A.~M.~Stasto,
 {\it  Eur.\ Phys.\ J.}\,  {\bf C41}, 343 (2005)
  [arXiv:hep-ph/0408117];\,\,\,\,G.~Chachamis, M.~Lublinsky and A.~Sabio Vera,
{\it   Nucl.\ Phys.}  {\bf A748}, 649 (2005)
  [arXiv:hep-ph/0408333];\,\,\,\,
 J.~L.~Albacete, N.~Armesto, J.~G.~Milhano, C.~A.~Salgado and U.~A.~Wiedemann,
 {\it  Phys.\ Rev.} {\bf D71}, 014003 (2005)
  [arXiv:hep-ph/0408216];\,\,\,\,E.~Gotsman, E.~Levin, U.~Maor and E.~Naftali,
{\it   Nucl.\ Phys.} {\bf A750} (2005) 391
  [arXiv:hep-ph/0411242];
Y.~V.~Kovchegov, J.~Kuokkanen, K.~Rummukainen and H.~Weigert,
Nucl.\ Phys.\  {\bf A823 } (2009)  47-82.
  [arXiv:0812.3238 [hep-ph]];\,\,\,
\,J.~L.~Albacete, N.~Armesto, J.~G.~Milhano and C.~A.~Salgado,
Phys.\ Rev.\  {\bf D80 } (2009)  034031.
  [arXiv:0902.1112 [hep-ph]].






\bibitem{BALE}
 J.~Bartels and E.~Levin,
  Nucl.\ Phys.\   {\bf B387} (1992) 617.
\bibitem{KHLE}
D.~Kharzeev and E.~Levin,
  Nucl.\ Phys.\  B {\bf 578} (2000) 351
  [arXiv:hep-ph/9912216].
\bibitem{KKL}
D.~E.~Kharzeev, Y.~V.~Kovchegov and E.~Levin,
  Nucl.\ Phys.\  A {\bf 690} (2001) 621
  [arXiv:hep-ph/0007182].






\bibitem{DL}
A. Donnachie and P.V. Landshoff,
Nucl. Phys. {\bf B231}, (1984) 189; 
Phys. Lett. {\bf B296}, (1992) 227; 
Zeit. Phys. {\bf C61}, (1994) 139.
\bibitem{GRIBMOV}
V.~N.~Gribov,
  Sov.\ Phys.\ JETP {\bf 15} (1962) 873
  [Zh.\ Eksp.\ Teor.\ Fiz.\  {\bf 42} (1962\ NUPHA,40,107.1963) 1260].
\bibitem{GW}
M. L. Good and W. D. Walker,
Phys. Rev. {\bf 120} (1960) 1857.

\bibitem{GRIBRT}
 V.~N.~Gribov,
  Sov.\ Phys.\ JETP {\bf 26} (1968) 414
  [Zh.\ Eksp.\ Teor.\ Fiz.\  {\bf 53} (1967) 654].

\bibitem{BORY}
K. G. Boreskov, A. B.~Kaidalov, V. A. Khoze, A. D. Martin and M. G. Ryskin,
Eur. Phys. J. {\bf C44} (2005) 523 [arXiv:hep-ph/0506211].





\bibitem{GRPO}
P. Grassberger and K. Sundermeyer,
Phys. Lett. {\bf B77} (1978) 220. 
E. Levin,
Phys. Rev. {\bf D49} (1994) 4469. 
K. G. Boreskov,
{\it "Probabilistic model of Reggeon field theory"}, 
arXiv:hep-ph/0112325 and reference therein.

\bibitem{AMCP}
D. Amati, M. Le Bellac, G. Marchesini and M. Ciafaloni, 
Nucl. Phys. {\bf B112} (1976) 107;
D. Amati, G. Marchesini, M.Ciafaloni and G. Parisi, 
Nucl. Phys. {\bf B114} (1976) 483.
\bibitem{KOLE}
M. Kozlov and E. Levin,
Nucl. Phys. A {\bf A779} (2006) 142 [arXiv:hep-ph/0604039].
\bibitem{KOLE1}
M.~Kozlov, E.~Levin, V.~Khachatryan and J.~Miller,
  Nucl.\ Phys.\  A {\bf 791} (2007) 382
  [arXiv:hep-ph/0610084].
\bibitem{FROI}
M.~Froissart,
{\it Phys.\, Rev.} \,  {\bf 123} (1961) 1053; \\
~A. ~Martin, {``Scattering Theory: Unitarity, Analitysity and Crossing."}
Lecture Notes in Physics, Springer-Verlag,  Berlin-Heidelberg-New-York,
1969.

\bibitem{MPSI}
A. H. Mueller and B. Patel,
Nucl. Phys. {\bf B425} (1994) 471. 
A. H. Mueller and G. P. Salam,
Nucl. Phys. {\bf B475}, (1996) 293. [arXiv:hep-ph/9605302]. 
G. P. Salam,
Nucl. Phys. {\bf B461} (1996) 512; 
E. Iancu and A. H. Mueller,
Nucl. Phys. {\bf A730} (2004) 460 [arXiv:hep-ph/0308315];            
494 [arXiv:hep-ph/0309276].

\bibitem{LEPR}
E. Levin and A. Prygarin,
Eur. Phys. J. {\bf C53} (2008) 385 [arXiv:hep-ph/0701178].
\bibitem{LMP}
E. Levin, J. Miller and A. Prygarin,
E.~Levin, J.~Miller, A.~Prygarin,
  Nucl.\ Phys.\  {\bf A806 } (2008)  245-286.
  [arXiv:0706.2944 [hep-ph]].
\bibitem{DG1}
A.~D.~Martin, M.~G.~Ryskin and V.~A.~Khoze,
  {\it ``Forward Physics at the LHC,''}
  arXiv:0903.2980 [hep-ph].
\bibitem{CDF}
CDF Collaboration, 
Phys. Rev.{\bf D50} (1994) 5535.







\bibitem{RY}
I. Gradstein and I. Ryzhik, 
{\it "Tables of Series, Products, and Integrals"}, Verlag MIR, Moskau,1981.

\bibitem{KL}
Y. V. Kovchegov and E. Levin,
Nucl. Phys. {\bf B577} (2000) 221 [arXiv:hep-ph/9911523].



\end{thebibliography}
\end{document}